\documentclass{aa}
\usepackage{graphicx}
\usepackage{natbib}
\usepackage{scalerel}
\usepackage{amsmath}	
\usepackage{multicol}       
\usepackage{bm}		
\usepackage{pdflscape}
\usepackage[usenames,dvipsnames,table]{xcolor}
\usepackage{tabularx}
\usepackage{multirow}
\usepackage{multicol}
\usepackage{tablefootnote}
\usepackage{threeparttable}
\usepackage{subcaption} 
\usepackage[normalem]{ulem}
\usepackage{booktabs}
\usepackage{footmisc}

\newcommand{\camels}{$\textsc{camels}$}

\usepackage{txfonts}
\usepackage[pdfencoding=auto,psdextra]{hyperref}
\hypersetup{
    colorlinks=true,
    linkcolor=blue,
    filecolor=magenta,      
    urlcolor=blue,
    citecolor=blue
}
\makeatletter
\renewcommand*\aa@pageof{, page \thepage{} of \pageref*{LastPage}}
\makeatother

\usepackage[utf8]{inputenc}

\usepackage[switch, modulo]{lineno}
\def\lsim{\mathrel{\rlap{\lower3.5pt\hbox{\hskip0.5pt$\sim$}}
    \raise0.5pt\hbox{$<$}}}                
\def\gsim{~\rlap{$>$}{\lower 1.0ex\hbox{$\sim$}}}

\def\App{\mbox{Appendix~}}
\def\Apps{\mbox{Appendices~}}

\begin{document}

\title{CASCO: Cosmological and AStrophysical parameters from Cosmological simulations and Observations}
\subtitle{IV.  Testing warm dark matter cosmologies with galaxy scaling relations: A joint simulation–observation study using DREAMS simulations
}

\newcommand{\orcid}[1]{} 
\newcommand{\ms}[1]{{\textcolor{red}{\sf{#1}}}}
\newcommand{\msbis}[1]{{\textcolor{cyan}{\sf{#1}}}}

\author{M. Silvestrini$^{1,2}$\thanks{E-mail: michela.silvestrini@unina.it},
C. Tortora$^{1}$,
V. Busillo$^{1,2,3}$,
Alyson M. Brooks$^{4}$,
A. Farahi$^{5,6}$,
A. M. Garcia$^{7}$,
N. Kallivayalil$^{7}$,
N. R. Napolitano$^{2}$,
J. C. Rose$^{8,9}$,
P. Torrey$^{5,7,10}$,
F. Villaescusa-Navarro$^{9,11}$,
M. Vogelsberger$^{12,13}$
}

\institute{$^{1}$INAF -- Osservatorio Astronomico di Capodimonte, Salita Moiariello 16, I-80131, Napoli, Italy\\
$^{2}$Dipartimento di Fisica “E. Pancini”, Universit\`{a} degli studi di Napoli Federico II, Compl. Univ. di Monte S. Angelo, Via Cintia, I-80126 Napoli, Italy\\
$^{3}$INFN, Sez. di Napoli, Compl. Univ. di Monte S. Angelo, Via Cintia, I-80126 Napoli, Italy\\
$^{4}$Department of Physics $\&$ Astronomy, Rutgers, the State University of New Jersey, 136 Frelinghuysen Rd, Piscataway, NJ 08854\\
$^{5}$The NSF-Simons AI Institute for Cosmic Origins, University of Texas at Austin, Austin, TX 78712,
 USA\\
 $^6$ Departments of Statistics and Data Sciences, University of Texas at Austin, Austin, TX 78712, USA\\
$^{7}$Department of Astronomy, University of Virginia, 530 McCormick Road, Charlottesville, VA 22904, USA\\
$^8$Department of Astronomy, University of Florida, Gainesville, FL 32611, USA\\
$^{9}$Center for Computational Astrophysics, Flatiron Institute, 162 5th Avenue, New York, NY 10010, USA\\
$^{10}$Virginia Institute for Theoretical Astronomy, University of Virginia, Charlottesville, VA 22904, USA\\
$^{11}$Department of Astrophysical Sciences, Princeton University, Peyton Hall, Princeton, NJ 08544, USA\\
$^{12}$Department of Physics $\&$ Kavli Institute for Astrophysics and Space Research, Massachusetts Institute of Technology, Cambridge, MA 02139, USA \\
$^{13}$The NSF AI Institute for Artificial Intelligence and Fundamental Interactions, Massachusetts Institute of Technology, Cambridge, MA 02139, USA\\
}

\abstract{Small-scale discrepancies in the standard Lamda cold dark matter paradigm have motivated the exploration of alternative dark matter (DM) models, such as warm dark matter (WDM).
In our work, we  investigate the constraining power of galaxy scaling relations on cosmological, astrophysical, and WDM parameters using a joint analysis of multiresolution hydrodynamic simulations and observational data. Our study is based on the DREAMS  project and combines large-volume uniform-box simulations with high-resolution Milky Way (MW) zoom-in runs exploring a $\Lambda$WDM cosmology.  To ensure consistency between the different simulation sets, we applied calibrations to account for resolution effects, which allowed us to better exploit the complementary strengths of the two suites. We compared the simulated relations, such as stellar size, DM mass, and fraction, within the stellar half-mass radius and the total-to-stellar mass ratio with two complementary galaxy samples: the Spitzer Photometry and Accurate Rotation Curves catalog, providing resolved kinematics for nearby spirals, and the Local Volume Database catalog, which includes structural and dynamical measurements for dwarf galaxies in the Local Volume. By applying a bootstrap-based fitting procedure, we show that key cosmological parameters ($\Omega_{\rm m}$, $\sigma_8$) and supernova feedback strength can be recovered with good accuracy, particularly from the uniform-box simulations. Although the WDM particle mass remains unconstrained, the MW zoom-in simulations reveal subtle WDM-induced trends, especially at low stellar masses, in the scaling relations of both the DM mass and the total-to-stellar mass ratio within the stellar half-mass radius. Additionally, we find that the galaxy abundance as a function of total stellar mass shows a measurable dependence on WDM particle mass, with a suppression at $\log_{10}\rm M_*/M_\odot\lsim$ 8 that appears separable from the impact of feedback, suggesting this observable is a valuable complementary probe. Our results highlight the importance of combining simulations at multiple resolutions with diverse observational catalogs to jointly constrain baryonic processes and DM properties. In particular, future low-mass galaxy surveys such as Euclid will play a crucial role in tightening the constraints on alternative DM scenarios through joint structural and statistical analyses. At the same time, higher-resolution simulations will be essential to fully capturing the small-scale features and improving the discriminatory power of such analyses, especially in the context of WDM.}

  \titlerunning{Constraining WDM mass with dwarf galaxies}
  \authorrunning{M. Silvestrini et al.}
   
  \maketitle
%
%-------------------------------------------------------------------
%

\section{Introduction}
The Universe, with its galaxies organized into clusters and filaments, is the result of approximately 13.8 billion years of evolution. Dark energy and dark matter (DM) shape spacetime and weave a cosmic web that connects astrophysical objects through structures governed by gravitational effects.
The nature and existence of DM remain among the most significant unresolved questions in astrophysics. According to the prevailing Lambda cold dark matter ($\Lambda$CDM) model, DM consists of cold—moving at velocities much lower than the speed of light—and collisionless particles (e.g., \citealt{Blumenthal:1984bp,1985ApJ...292..371D,1988ApJ...327..507F}).
The  $\Lambda$CDM model, despite its simplicity, has achieved significant success. For example, it explains the power spectrum of the cosmic microwave background and accurately reproduces the large-scale distribution of matter throughout the Universe. When it comes to smaller scales, such as galaxies and subgalactic scales, the $\Lambda$CDM model encounters several challenges, including the core-cusp (\citealt{Moore:1994yx,10.1093/mnras/290.3.533}), the too-big-to-fail (\citealt{10.1111/j.1745-3933.2011.01074.x}), the missing satellites (\citealt{1999ApJ...524L..19M,1999ApJ...522...82K}), and the diversity (\citealt{10.1093/mnras/stv1504}) problems. 
Recent advancements in simulations incorporating hydrodynamics and galaxy formation physics (\citealt{Vogelsberger2020}) have addressed some of these discrepancies within the $\Lambda$CDM paradigm, such as the core-cusp and the missing satellites problems (\citealt{Pontzen2012,Governato2012,Vogelsberger2012,2014Natur.509..177V,Brooks2013,2015MNRAS.454.2981C, Wetzel2016,10.1093/mnras/stab2437}). Moreover, baryonic feedback and tidal effects can alleviate the too-big-to-fail  problem in satellite galaxies (\citealt{Zolotov2012,Brooks2014}). However, in isolated dwarfs, where such processes are less effective, the too-big-to-fail and diversity problems remain significant challenges for the $\Lambda$CDM model (\citealt{Papastergis2015,10.1093/mnras/staa1089}).
Warm dark matter (WDM; e.g., \citealt{Bode_2001}) stands out as an alternative to CDM, proposing particles with non-negligible peculiar velocities. Recent hydrodynamic simulations have shown that WDM models can delay star formation in low-mass haloes and lead to observable differences in stellar ages and central densities of dwarf galaxies \citep{Herpich2014,Maccio2019,Shen2024}. Other compelling alternative candidates include self-interacting dark matter (e.g., \citealt{PhysRevLett.84.3760,burkert2000selfinteractingcolddarkmatter,Zavala2013}), where particles interact significantly with one another, and scalar field or fuzzy dark matter (e.g., \citealt{PhysRevD.95.043541,10.3389/fspas.2018.00048,NIEMEYER2020103787}), which involve ultralight bosons.

In this work, we focus on WDM models by analyzing cosmological simulations from the DREAMS project \citep[DaRk mattEr with AI and siMulationS,][]{rose2024introducingdreamsprojectdark}. The DREAMS project integrates astrophysics, particle physics, and machine learning to investigate the nature of DM, with a particular emphasis on small-scale physics in the Universe. The simulations model our Universe across various volumes and resolutions, ranging from large-scale uniform-boxes and Milky Way (MW) zoom-ins to dwarfs zoom-in configurations, tailored for specific DM models. Despite the existence of various experimental constraints on the lower mass limit of the WDM particle—up to 9.7 keV, as indicated by analyses of gravitationally lensed systems and MW satellites (\citealt{PhysRevLett.126.091101})—WDM remains a crucial benchmark for testing alternative DM  models in cosmological simulations. In fact, WDM models represent a simple alternative to CDM and serve as a natural starting point for exploring scenarios beyond the $\Lambda$CDM paradigm, as demonstrated in the DREAMS project.

In this study, we focus on key galactic scaling relations—specifically those linking galaxy size, DM content, and stellar mass—and investigate how varying WDM particle masses affect these relations, with particular attention given to low-mass galaxies. Indeed, galactic scaling relations arise from the fundamental physics governing galaxy formation and evolution, making them inherently sensitive to the nature of DM. Our goal is to evaluate how well WDM models can be constrained through these relations. In addition to the nature of DM, astrophysical feedback 
processes—such as those from supernovae (SNe), active galactic nuclei (AGNe), or gravitational heating—can significantly modulate galactic scaling relations \citep{dek_birn06,Governato2007,Moster+10,Tortora+19_LTGs_DM_and_slopes,CASCOIII,BusilloCascoI,CASCOII,vogels2014,Torrey2014,Weinberger2017,Pillepich2018,Roca2021,Ni2023}. To this end, we analyzed simulations at different resolutions and volumes, including both large-scale uniform-box simulations and high-resolution MW zoom-in simulations, and we compared the scaling relations obtained from these simulations with observational trends inferred from nearby dwarf galaxies in the Local Volume Database (LVDB; \citealt{pace2024localvolumedatabaselibrary}) and from the Spitzer Photometry and Accurate Rotation Curves (SPARC; \citealt{Lelli2016}).

This paper is the fourth in the CASCO project: Cosmological and Astrophysical parameters from Cosmological simulations and Observations. 
\citet[hereafter Paper~I]{BusilloCascoI} and \citet[hereafter Paper~II]{CASCOII} used the \camels\ simulations within a $\Lambda$CDM cosmology to constrain cosmological and astrophysical parameters—such as SN and AGN feedback—by comparing simulated and observed scaling relations, considering both early- and late-type galaxies. \citet[Paper~III]{CASCOIII} focused on the emergence of the so-called golden mass, a characteristic stellar mass scale where star formation efficiency peaks, and studied how its value evolves across cosmic time.
In this work (Paper~IV), we extend the CASCO framework to a WDM cosmology by using DREAMS simulations to investigate how feedback and the nature of DM shape the internal structure of galaxies.

In Section \ref{methodology}, we present the DREAMS simulations; the observational datasets used for comparison, namely the LVDB and SPARC catalogs; and the fitting procedure adopted to constrain cosmological, astrophysical, and WDM model parameters based on the agreement between simulations and observations. In Section \ref{results}, we discuss our results regarding the impact of these parameters on galaxy scaling relations and galaxy counts, and we assess the effectiveness and limitations of the fitting approach. We then compare our findings with observational data. Finally, in Section \ref{conclusions}, we summarize our conclusions and outline directions for future research.
%%%%%%%%%%%%%%%%%%%%%%%%%%%%%%%%%%%%%%%%%%%%%%%%%%%%%%%%%%%%%%%%%%%%%%%%%%%%%%
%%%%%%%%%%%%%%%%%%%%%%%%%%%%%%%%%%%%%%%%%%%%%%%%%%%%%%%%%%%%%%%%%%%%%%%%%%%%%%

\section{Data and methodology\label{methodology}}

%%%%%%%%%%%%%%%%%%%%%%%%%%%%%%%%%%%%%%%%%%%%%%%%%%%%%%%%%%%%%%%%%%%%%%%%%%%%%%

\subsection{WDM DREAMS simulations}

The DREAMS project\footnote{\url{https://dreams-project.readthedocs.io/en/latest/}} consists of an extensive suite of hydrodynamic simulations, complemented by N-body simulations that serve as their purely gravitational counterparts. In this section, we provide a description of the simulations used in this study, specifically those based on the WDM model, but we refer the reader to \cite{rose2024introducingdreamsprojectdark} for a more detailed description of the DREAMS project and the simulation setup.

Depending on the simulated environment, the simulations are organized into two different suites, each conducted in regions of different spatial scales and resolutions: cosmological boxes and MW zoom-in. 
The uniform-box simulations span a volume of (25 $h^{-1}$ Mpc)$^3$, adopt periodic boundary conditions, and include a $2\times 256^3$ grid of DM and gas particles. 
The DM mass resolution depends on the matter density parameter $\Omega_{\rm m}$ and is given by $7.81\times \left(\Omega_{\rm m}/0.302\right)\times10^7\; h^{-1}\;\rm M_\odot$, while the baryonic mass resolution is fixed at $1.27\times10^7\;h^{-1}\;\rm M_\odot$. The spatial resolution, set by the gravitational softening length, reaches 1.0 $h^{-1}$ kpc at $z = 0$. These simulations offer good statistical sampling thanks to their volume. 
However, the relatively low resolution of these runs, comparable to TNG300-1, limits the efficiency of baryonic feedback processes, especially those linked to star fformation and SNe. As a result, the stellar-DM relation tends to deviate from the results of higher-resolution simulations such as TNG100-1 and TNG50-1 \citep{CASCOII}, where feedback is better resolved and more effective.
To mitigate these resolution-dependent biases, we calibrate the stellar-DM mass relation of the uniform-box simulations by aligning them with the high-resolution TNG100-1 simulation, which serves as our reference. 
This empirical calibration improves the agreement with observational data and enhances the predictive reliability of the simulated galaxy populations. The calibration introduces a stellar-mass shift of approximately $c \sim 0.27$ dex in $\log_{10} \rm M_*/M_\odot$. A detailed description of the calibration procedure, including a discussion of its limitations, is provided in \App\ref{calibrazione}. 

Although the relatively low mass resolution of the uniform-box runs limits their ability to fully resolve small-scale baryonic processes, recent analyses have demonstrated that simulations at comparable resolution can still retain sensitivity to the WDM particle mass. In particular, \citet{Rose2023}  showed that, for fixed cosmology, convolutional neural networks trained on DM density fields can recover WDM masses up to $\sim$7 keV from N-body simulations with $256^3$  particles in a  (25 $h^{-1}$ Mpc)$^3$  box—identical to the DREAMS setup. When cosmological parameters are varied but the simulation resolution is kept fixed, their model remains accurate for WDM masses in the range 3.5–5 keV, though with slightly larger uncertainties. Furthermore, \citet{costanza2025} used the DREAMS hydrodynamical uniform boxes to show that galaxy population statistics, such as stellar and gas mass distributions, preserve strong discriminatory power between WDM and CDM models up to $\sim$ 6 keV. These results motivate the inclusion of the uniform-box suite in our analysis as a statistically representative yet still physically sensitive component of the DREAMS framework.

Complementing this setup, the MW zoom-in simulations simulate MW-mass analogs, with no nearby massive neighbor, and target smaller regions with a characteristic scale of approximately 200 $h^{-1}$ kpc. These simulations achieve higher resolution, only a factor of two lower in mass resolution than TNG50-1, with DM particle masses of $1.2\times10^6\;h^{-1}\;\rm M_\odot$, average baryonic masses of $1.9\times10^5\;h^{-1}\;\rm M_\odot$, and a spatial resolution of 0.31 $h^{-1}$ kpc at $z$ = 0. For this reason, no empirical calibration is applied to the MW zoom-in simulations, which are expected to capture feedback processes more accurately. 

 The construction of the MW zoom-in simulations involves a multistep procedure (see Appendix A in \citealt{rose2024introducingdreamsprojectdark}), which includes intermediate DM-only runs at progressively increasing resolution. As a result of this process, the final high-resolution region can be surrounded by particles from lower-resolution region, and some sub-halos—particularly in the outskirts—can be partially or predominantly composed of low-resolution DM particles. However, by design, the high-resolution region extends to $\sim$5 times the virial radius of the MW at $z=0$, ensuring that the main halo and most of its satellites lie well within the high-resolution volume, and are therefore minimally affected by low-resolution contamination. In cases where contamination could be significant, key baryonic processes such as gas cooling and star formation cannot be reliably modeled. To ensure physical consistency, we exclude from our analysis all galaxies in which more than 5$\%$ of the total DM mass originates from low-resolution particles.

It is also important to highlight that, although the central galaxy in each MW zoom-in simulation resides within a halo carefully selected to have a virial mass comparable to that of the MW at $z = 0$, the resulting galaxy properties are not guaranteed to resemble those of the MW itself. The diversity in feedback strength, and therefore in star formation efficiency, can lead to significant variations in stellar mass and morphology across the sample (see \App\ref{App:Host}). Thus, the MW-like host galaxies should not be interpreted as strict analogs of the MW, but rather as realizations within the broader class of galaxies hosted by MW-mass halos.

In the simulations of interest, cosmological parameters can either be variable (for the uniform-box runs) or fixed (for MW zoom-in). In the first case, $\Omega_{\rm m}$ varies within the range [0.1, 0.5] and $\sigma_8$ within [0.6, 1.0],  both sampled from a uniform distribution.  For the MW zoom-in, the reference values from the Planck Collaboration \citep{Planck2015} are adopted: $\Omega_{\rm m}$ = 0.302 and $\sigma_8$ = 0.839. Other cosmological parameters, fixed in both suites according to the Planck Collaboration results, are $\Omega_\Lambda$ = 0.698 and $H_0$ = 100 $h$ km $s^{-1}$ with $h$ = 0.691. N-body simulations assume $\Omega_{\rm b}$ = 0, while in hydrodynamic simulations, $\Omega_{\rm b}$ = 0.046. This study focuses exclusively on hydrodynamic simulations. In this context, we note that because $\Omega_{\rm m}$ is varied while $\Omega_{\rm b}$ is fixed in the hydrodynamical runs, the ratio $f_{\rm b}=\Omega_{\rm b}/\Omega_{\rm m}$, which sets the cosmic baryon fraction, also varies considerably. This ratio is relatively well constrained by cosmological observations (e.g., from the cosmic microwave background), typically around 0.15. In the simulations, however, it can depart significantly from that value, depending on the choice of $\Omega_{\rm m}$. These variations can influence the overall efficiency of galaxy formation and lead to increased scatter in quantities such as the stellar-to-halo mass ratio, beyond what would be expected if $\Omega_{\rm b}$ and $\Omega_{\rm m}$ were varied consistently within observational bounds.

%%%%%%%%%%%%%%%%%%%%%%%%%%%%%%%%%%%%%%%%%%%%%%%%%%%%%%%%%%%%%%%%%%%%%%%%%%%%%%
\subsubsection{Astrophysical parameters and WDM mass\label{sec:parameters}}

The WDM simulations employed in this study are based on the TNG galaxy formation model \citep{Springel2018, Pillepich2018, Nelson2018, Marinacci2018, Naiman2018, Nelson2019, Pillepich2019}. They start from redshift $z = 127$ and evolve down to $z = 0$ using the moving-mesh code AREPO (\citealt{2010MNRAS.401..791S}, \citealt{2019ascl.soft09010S}, \citealt{Weinberger_2020}). These simulations adopt the fiducial parameters of the TNG model, except for those governing SN and AGN feedback, and the matter power spectrum.

The TNG model incorporates a feedback mechanism to describe the interactions between star formation, supermassive black holes, and the circumgalactic medium. To understand how these processes affect DM distribution and galaxy growth, three key parameters related to stellar winds ($A_{\rm SN1}$ and $A_{\rm SN2}$) and AGN activity ($\rm BH_{\rm FF}$) are varied in the simulations\footnote{In future papers, the DREAMS collaboration will adopt the original TNG notation ($\bar{e}_w$, $\kappa_w$,  $\epsilon_{\rm f,\,high}$), with these parameters defined in detail shortly. We retain the CAMELS-like parameters for continuity with CASCO Papers I–III.}.

The first parameter is the wind energy, $A_{\rm SN1}$, which enters the expression for the wind mass-loading factor, $\eta_{\rm w}$, via $e_w$, the specific energy available to driving winds:
\begin{equation}
    \eta_w=\frac{2}{{\nu_w}^2}e_w(1-\tau_w)\;.
\end{equation}
Here, $\nu_w$ denotes  the wind velocity (see Eq. \ref{windvel}) and $\tau_w$ is the fraction of energy released thermally, fixed at its fiducial value of 0.1. 
The parameter $e_w$ is expressed as follows:
\begin{align}
     e_w&=A_{\rm SN1}\times \bar e_w \left[ f_{ w,\,Z}+\frac{1-f_{ w,\, Z}}{1+(Z/Z_{w,\,\rm ref})^{\gamma_{w, \,Z}}}\right]\nonumber
\\
&\quad \times N_{\rm SNII}E_{\rm SNII,\,51}10^{51}\rm erg\;M^{-1}_\odot\;,
\end{align}
where $Z$ is the metallicity of gas cells, $\bar e_w$ is the wind energy factor, and $f_{w,\,Z}$ accounts for the metallicity-dependent reduction. The reference metallicity is $Z_{w,\,\rm ref}$, while $\gamma_{w,\,Z}$ sets the power of the metallicity-dependent reduction. Finally, $N_{\rm SNII}$ represents the number of Type II SNe per unit stellar mass formed, and $E_{\rm SNII,\,51}$ is the energy available per core-collapse SN in units of $10^{51}\rm erg$.\\

The second parameter is the galactic wind velocity, $A_{\rm SN2}$, which depends on the properties of the DM halo and redshift. It is governed by $\nu_w$, determining the efficiency of gas ejection:
\begin{equation}\label{windvel}
    \nu_w= \max\Big[A_{\rm SN2}\kappa_{w,\,\rm{ref}}\times \sigma_{\rm DM}\left(\frac{H_0}{H(z)}\right)^{1/3},\;\nu_{w,\,\rm min}\Big]\;.
\end{equation}
In this expression, the parameter $A_{\rm SN2} = \kappa_w / \kappa_{w,\,\rm ref}$, where $\kappa_w$ is a dimensionless normalization factor with fiducial TNG value
$\kappa_{w,\,\rm ref} = 7.4$. The remaining quantities include $\sigma_{\rm DM}$, the DM velocity dispersion, $H(z)$, the Hubble function, and $\nu_{\rm w,\, min}=350\;\text{km\;s}^{-1}$, the minimum allowed wind speed. Increasing $\nu_{\rm w}$ suppresses star formation at later cosmic times.

The third key parameter is the AGN feedback efficiency, $\mathrm{BH_{FF}}$. Supermassive black holes influence gas heating and ejection, regulating star formation in massive galaxies. The model includes two feedback modes: a low-accretion state, which heats gas more uniformly, and a high-accretion state, which drives energetic outflows. The WDM DREAMS simulations vary only the high-accretion mode. Here, $\mathrm{BH_{FF}}=\epsilon_{f,\,\rm high}/\epsilon_{f,\,\rm{high,\,ref}}$, where  $\epsilon_{f,\,\rm high}$ determines the fraction of accreted energy transferred to the surrounding gas, and $\epsilon_{f,\,\rm{high,\, ref}}=0.1$ is its reference value:
\begin{equation}
\Delta\dot{E}=\mathrm{BH_{FF}}\,\epsilon_{f,\,\rm{high,\,ref}}\times \epsilon_r \dot{M}_{\rm BH}c^2\;,
\end{equation}
with $\epsilon_r$ is the radiative efficiency and $ \dot{M}_{\rm BH}$ is the black hole accretion rate. Higher values of $\mathrm{BH_{FF}}$ enhance AGN thermal feedback, suppressing star formation in massive galaxies. This parameter, also known as BH feedback factor, is implemented—together with other parameters regulating AGN feedback—in the latest suite of \camels\ simulations \citep{Ni2023, CASCOIII}.

The TNG model assumes fiducial values for
$A_{\rm SN1}$, $A_{\rm SN2}$ and $\rm BH_{\rm FF}$, equal to 1, but DREAMS simulations explore a broader range to assess their impact on galaxy evolution.
 Specifically, the parameter $A_{\rm SN1}$ and $\rm BH_{\rm FF}$ varies within the interval [0.25, 4], while  $A_{\rm SN2}$ spans the range [0.5, 2].  The astrophysical parameters are selected from a uniform distribution in logarithmic space.

In the WDM DREAMS simulations, the WDM is treated as a collisionless and pressureless fluid, similarly to CDM, and is modeled by a single parameter, $\rm P_{\rm WDM}$, defined as the inverse of the WDM particle mass ($\rm{P_{\rm WDM}}=$$1/ M_{\rm WDM}$). In the case of uniform-box simulations,  $\rm P_{\rm WDM}$ is uniformly chosen in the interval [0.062, 0.556] keV$^{-1}$, corresponding to WDM particle masses between $\sim$1.8 and 16 keV. In the MW zoom-in simulations,  $\rm P_{\rm WDM}$ spans a broader range of [0.033, 0.556] keV$^{-1}$, meaning that the WDM mass varies between $\sim$1.8 and 30 keV. $\rm P_{\rm WDM}$, as well as the cosmological and astrophysical parameters, is sampled using a Sobol sequence \citep{SOBOL196786}, which ensures homogeneous coverage of the multidimensional parameter space.

The  parameter $\rm P_{\rm WDM}$ fully determines the distribution of the WDM thermal velocities. Operationally, in the simulations, the WDM models differ from the CDM models in the initial conditions. Specifically, following \cite{Bode_2001}, the particle mass parameter enters the linear power spectrum through a suppression factor $\beta(k)$. As a consequence, the WDM power spectrum can be written as $P_{\rm WDM}(k) = \beta(k) P_{\rm CDM}(k)$, which is damped at large comoving wave numbers, i.e., on small scales,
\begin{align}
   \beta(k)&=\Big(\left(1+(\alpha k)^{2.4}\right)^{-5.0/1.2}\Big)^2\;,\nonumber
\\
\alpha&=0.048\:h^{-1}\;\mathrm{ Mpc}\;\left(\frac{M_{\mathrm{ WDM}}}{\rm{keV}}\right)^{-1.15}\left(\frac{\Omega_{\mathrm{ m}}-\Omega_{\mathrm{b}}}{0.4}\right)^{0.15}\left(\frac{h}{0.65}\right)^{1.3} \;.
  \end{align}
It is worth noting that in the limit where $M_{\rm WDM}\rightarrow \infty$, $\beta(k)\rightarrow 1$, and the power spectrum of the WDM coincides with that of the CDM. In this way, the DREAMS simulations cover a wide range of WDM models using a single parameter.
%%%%%%%%%%%%%%%%%%%%%%%%%%%%%%%%%%%%%%%%%%%%%%%%%%%%%%%%%%%%%%%%%%%%%%%%%%%%%%
\subsubsection{Simulations employed, astrophysical quantities, and selection criteria\label{sec: sel simulations}}

The WDM DREAMS simulation suite provides 1024 distinct configurations of astrophysical, cosmological, and WDM mass parameters for the uniform-box runs. Additionally, an equivalent number of simulations is available for the MW zoom-in runs, where the cosmology is fixed. 

Sub-halo identification is performed using the friends-of-friends and SUBFIND algorithms (\citealt{2001MNRAS.328..726S}). In this framework, each sub-halo is defined as a gravitationally self-bound structure, and its total mass is computed as the sum of all bound particles, including DM, gas, stars, and black holes, without relying on a spherical overdensity threshold. In this work, we compare simulated galaxy scaling relations with observational counterparts by analyzing several key astrophysical quantities derived from SUBFIND at redshift $z=0$:

\begin{itemize}
    \item Stellar half-mass radius, $R_{*,1/2}$: The radius enclosing half of the total stellar mass of the galaxy.
    \item Total stellar mass, $M_*$: Obtained as the sum of the stellar particle masses within a given sub-halo.
    \item Total mass, $M_{\rm tot}$: The sum of all mass components (DM, stars, gas, and black holes) in a sub-halo.
    \item Mass components within the stellar half-mass radius: Stellar mass ($M_{*,1/2}$), DM mass ($M_{\rm DM,1/2}$), and total mass ($M_{\rm tot,1/2}$).
    \item The DM fraction within the stellar half-mass radius, $f_{\rm DM,1/2}\equiv f_{\rm DM}(<R_{*,1/2})\equiv  M_{\rm DM,1/2}/M_{\rm tot,1/2} $.
    \item Total-to-stellar mass ratio within the half-mass radius, $M_{\rm tot,1/2}/M_{*,1/2}$.
    \item Number of star particles within the stellar half-mass radius, $N_{*,1/2}$.
    \item Star formation rate (SFR).
    \item Number of galaxies, $N_{\rm Galaxy}$: The total number of sub-halos. 
    \end{itemize}
To ensure a robust comparison between simulations and observational datasets, we imposed resolution-dependent selection criteria:

\begin{itemize}
    \item The value of $R_{*,1/2}$ must exceed the spatial resolution, which is 1 kpc for uniform-box simulations and 0.305 kpc for MW zoom-in runs.
    \item The DM fraction within the half-mass radius, $f_{\rm DM,1/2}$, must be positive for both simulation suites.
    \item The number of stellar particles within the half-mass radius, $N_{*,1/2}$, must exceed 50 for both simulation suites, corresponding to stellar mass thresholds of approximately $\rm\log_{10} M_*/M_\odot \sim 9.3$ for the uniform-box simulations and $\sim 7.4$ for the MW zoom-in simulations. This selection is applied when comparing the simulations to observational catalogs.
\end{itemize}
These criteria ensure that only well-resolved galaxies are included in the analysis and that a consistent comparison with observations can be made across all simulation sets. 
WDM simulations, however, may still be affected by numerical artifacts known as ``beads-on-a-string,'' which arise from discreteness noise along filaments and can produce spurious small-scale clumps (e.g. \citealt{WangWhite2007}). Such effects become relevant near the cutoff scale of the WDM power spectrum, where the half-mode mass approaches the particle-mass resolution. As discussed by \citet{Rose2023}, numerical fragmentation becomes significant only for very warm models ($M_{\rm WDM}\lesssim1.8$~keV). Since this value marks the lower bound of our explored range, we expect that the vast majority of our simulations are unaffected by such artifacts and that any residual spurious clumping at the lowest WDM mass is minimal and does not impact our results. The scaling relations analyzed in this work are consistently measured for both the uniform-box and MW zoom-in simulations, focusing on well-resolved galaxies above our stellar particle-number threshold (i.e., $N_{*,1/2}>50)$. Spurious haloes generated by numerical fragmentation are typically dark and therefore excluded by these cuts. Moreover, the abundance of satellites—where such artifacts could, in principle, be more apparent—is examined only in the higher-resolution MW zoom-in suite, which further mitigates possible contamination. Together with the findings of \citet{Rose2023}, these considerations indicate that numerical fragmentation does not materially affect the trends discussed in this paper.

%%%%%%%%%%%%%%%%%%%%%%%%%%%%%%%%%%%%%%%%%%%%%%%%%%%%%%%%%%%%%%%%%%%%%%%%%%%%%%
%%%%%%%%%%%%%%%%%%%%%%%%%%%%%%%%%%%%%%%%%%%%%%%%%%%%%%%%%%%%%%%%%%%%%%%%%%%%%%
\subsection{Observational catalogs\label{ObsCat}}

In the following, we introduce the two observational catalogs, the SPARC and the dwarf galaxy catalog, which serve as benchmarks for comparing the trends obtained from simulations. We specify the astrophysical quantities extracted from these datasets that contribute to the analyzed scaling relations and detail the methods used to derive them.
%%%%%%%%%%%%%%%%%%%%%%%%%%%%%%%%%%%%%%%%%%%%%%%%%%%%%%%%%%%%%%%%%%%%%%%%%%%%%%
\subsubsection{SPARC catalog overview\label{subsec:SPARC}}

The SPARC catalog (\citealt{Lelli2016}) provides a dataset of 175 disk galaxies, serving as a valuable reference for studying the interplay between baryonic matter and DM in galaxy dynamics. While it does not constitute a volume-limited sample, it includes a wide variety of late-type galaxy morphologies, rotational properties, and sizes, offering a representative view of disk galaxies in the local Universe. The SPARC galaxies span a stellar mass range of $6.9\lsim\rm log_{10} M_*/M_\odot\lsim11.4$, covering a wide range of galaxy masses from low-mass disks to massive spirals.

The near-infrared photometry at 3.6 $\mu$m of SPARC was analyzed to estimate stellar masses, adopting a standard stellar mass-to-light ratio ($\Upsilon_*$ = 0.6 $\Upsilon_\odot$). Busillo et al. (Paper I) tested the impact of this assumption by repeating the fitting-like analysis -- which is introduced in Sect. \ref{sec:fit bootstrap} -- with stellar mass-to-light ratios in the range $\Upsilon_* = 0.5\, \Upsilon_\odot$ to $\Upsilon_* = 0.7 \,\Upsilon_\odot$. While this range leads to differences in stellar mass estimates of about 0.06–0.08 dex, the resulting variations in the astrophysical and cosmological parameters were found to be negligible, remaining well within the statistical uncertainties. The catalog’s high-resolution HI 21 cm rotation curves allow for the reconstruction of the radial distribution of mass within each galaxy. The total enclosed 3D mass at a given radius $r$ is calculated using the relation $M(r)=v^2r/G$, allowing a direct comparison between the observed baryonic components and the inferred DM contributions.

Selection criteria further refine our working sample: galaxies with inclinations below $30^\circ$ are excluded to minimize projection effects, and only systems with well-measured rotation curves extending beyond the effective radius are retained, leading to a final selection of 152 galaxies.

To compare our results with the SPARC dataset, we select simulated galaxies based on their specific star formation rate (sSFR). We classify galaxies with sSFR $>10^{-10.5}\;\rm yr^{-1}$ as star-forming galaxies (SFGs), while we consider those with sSFR  $<10^{-10.5}\;\rm yr^{-1}$ passive galaxies (PGs). This selection is applied consistently throughout Papers I, II, and III, providing a practical way to isolate actively star-forming systems in the simulations. Although the SPARC sample targets late-type galaxies, these are generally star-forming in the local Universe. Selecting simulated galaxies by sSFR thus offers a practical proxy to identify SPARC analogs. This allows a meaningful comparison between simulations and observations.

In this study, we focus on several key parameters, including the total stellar mass, $M_*$; the stellar half-mass radius, $R_{*,1/2}$, defined multiplying the respective effective radii ($R_{\rm e}$),  which encloses half of the total 3.6 $\mu$m luminosity,  by a constant factor of $\sim 1.35$ \citep{Wolf2010}; the stellar, gas and total mass within this radius, $M_{*,1/2}$, $M_{g,1/2}$ and $M_{\rm tot,1/2}$, respectively. From these, we derive the enclosed DM mass, $M_{\rm DM,1/2}$, and compute the DM fraction, $f_{\rm DM,1/2}$. To complement these calculations, we incorporate virial mass estimates, $M_{vir}$, from \cite{2019AA...626A..56P}, obtained by fitting rotation curves using a Navarro-Frenk-White (NFW) DM halo profile \citep{NFW1996}.

%%%%%%%%%%%%%%%%%%%%%%%%%%%%%%%%%%%%%%%%%%%%%%%%%%%%%%%%%%%%%%%%%%%%%%%%%%%%%%

\subsubsection{Dwarfs in the local volume\label{subsec:dwarf}}

We also exploited the LVDB (\citealt{pace2024localvolumedatabaselibrary}) sample that is publicly available as a GitHub repository.\footnote{\url{https://github.com/apace7/local_volume_database}} LVDB is an extensive compilation of observed properties for astrophysical objects within the Local Volume, encompassing dwarf galaxies, globular clusters, and other stellar systems. The database primarily focuses on resolved stellar systems within approximately 10 Mpc, though its completeness is highest for objects within 3 Mpc, and it provides a valuable resource for studying galaxy formation, structure, kinematics, and chemical evolution.  The catalog covers a stellar mass range of $2.7\lsim\rm log_{10} M_*/M_\odot\lsim9.7$, offering a broad view of low-mass galaxy populations in the nearby Universe. The LVDB incorporates data from multiple sources, including the \cite{McConnachie_2012} catalog, the Catalog and Atlas of Local Volume Galaxies (\citealt{Karachentsev_2013}), and the Extragalactic Distance Database (\citealt{Tully_2009}, \citealt{Anand_2021}) and it is regularly updated to include new discoveries.

The LVDB contains key astrophysical parameters for each object, including structural, kinematic, and dynamical properties. For this study, we extract specific quantities necessary for establishing scaling relations, which are subsequently compared to numerical simulations. The primary parameters used in our analysis are:
\begin{itemize}
    \item Stellar mass: The parameter $\mathtt{mass\_stellar}$ stored in the catalog is defined as the logarithm of the stellar mass, estimated assuming a mass-to-light ratio of two, and derived from the absolute V-band magnitude ($M_V$) [$\rm \log_{10} M_\odot$].
    \item Half-light radius: $\mathtt{rhalf\_sph\_physical}$ is the azimuthally averaged 2D half-light radius, computed as the geometric mean incorporating ellipticity corrections, expressed as $\mathtt{rhalf}\times \mathtt{distance} \times \sqrt {(1-\mathtt{ellipticity})}$ in parsecs. Here, \texttt{rhalf} represents the major axis of the half-light radius (or Plummer radius) in arcminutes, \texttt{distance} is the heliocentric distance, and \texttt{ellipticity} = 1 - minor/major axis.
    \item Dynamical mass within the half-light radius: $\mathtt{mass\_dynamical\_wolf}$ is the dynamical mass enclosed within the 3D half-light radius [$\rm M_\odot$].
\end{itemize}
The 3D half-mass radius is determined following the method outlined in \cite{Wolf2010}, as done for the SPARC catalog: 
\begin{equation}
    R_{*,1/2}[\rm kpc]=4/3\times \frac{\mathtt{rhalf\_sph\_physical}}{1000}\;.
\end{equation}
The DM mass enclosed within the half-mass radius is computed as:
\begin{equation}
    M_{\rm DM,1/2}[\rm M_\odot]=\mathtt{mass\_dynamical\_wolf}-0.5\times10^{\mathtt{mass\_stellar}}\;.
\end{equation}
Due to the lack of robust gas mass estimates for the majority of the dwarf galaxies in the LVDB, we do not include a gas mass contribution in our analysis. The fraction of DM within the half-light radius was subsequently derived as 
 \begin{equation}
f_{\rm DM,1/2}=\frac{M_{\rm DM,1/2}}{\mathtt{mass\_dynamical\_wolf}}\;. 
 \end{equation}
Total virial masses of the dwarf galaxies are not included in the catalog and therefore will not be included in the next analysis.

To ensure a meaningful comparison with simulations, we applied specific selection criteria to the LVDB. The subset of 14 dwarf galaxies used in this study satisfies the following conditions: 
\begin{itemize}
    \item The system must be classified as a confirmed physical galaxy ($\mathtt{confirmed\_real}$ = 1) and a dwarf galaxy ($\mathtt{confirmed\_dwarf}$ = 1).
    \item The stellar mass must satisfy $\mathtt{mass\_stellar} > 7.4$ [$\rm\log_{10} M_\odot$], ensuring consistency with the resolution limit ($N_{*,1/2} > 50$) adopted in the WDM zoom-in simulations.
    \item The stellar half-mass radius must satisfy $ R_{*,1/2}[\rm kpc]>0.305\;\rm{kpc}$, ensuring that the selected objects have spatial resolutions above the simulation limits.
    \item The dynamical mass estimate must be positive and exceed the enclosed stellar mass ($\mathtt{mass\_dynamical\_wolf}>0.5\times10^{\mathtt{mass\_stellar}}$).
    \item The catalog must simultaneously contain all the quantities described above for the selected galaxies to ensure a complete dataset for comparison with simulations.
\end{itemize}

%%%%%%%%%%%%%%%%%%%%%%%%%%%%%%%%%%%%%%%%%%%%%%%%%%%%%%%%%%%%%%%%%%%%%%%%%%%%%%

\subsection{Analysis approach\label{sec:fit bootstrap}}

To compare the simulated trends with observational data, we adopt a fitting-like procedure, implemented in Mathematica, and designed to account for statistical uncertainties and parameter degeneracies. This method was first introduced in Paper~I and Paper~II, and is adapted here to the current simulations and data. In this context, we define the distance estimator $D^2$ as in Papers~I and II, but with a minor change in notation. This modification is intentional: we aim to emphasize that our fitting-like method and the $D^2$ estimator do not correspond to the standard definitions of fitting or the classical $\chi^2$ statistic. Instead, they represent an alternative approach developed in the context of the CASCO paper series, specifically designed to assess how well the simulations reproduce the observed trends in selected galaxy catalogs. Despite the differences, our method shares several features with the classical 
$\chi^2$. Below, we outline the main points of our analysis.

In this work, we focus on four key scaling relations used in the fit: the size–mass relation ($R_{*,1/2}$ versus $M_*$); the DM mass and fraction within the stellar half-mass radius ($M_{\rm DM,1/2}$ versus $M_*$ and $f_{\rm DM,1/2}$ versus $M_*$, respectively); and the relation between total sub-halo mass and stellar mass ($M_{\rm tot}$ versus $M_*$). In addition, we considered one relation not included in the fit but used elsewhere in our analysis. Specifically, we included the total-to-stellar mass ratio within the half-mass radius ($M_{\rm tot,1/2} / M_{*,1/2}$) in our plots, as it offers improved visual clarity compared to $f_{\rm DM,1/2}$. Since both quantities carry essentially the same information, we use $f_{\rm DM,1/2}$ in the fit to remain consistent with our previous works, where this methodology was first implemented. 

For the four scaling relations included in the fit, we first derive the observational trends by binning stellar mass and computing the 16th, 50th, and 84th percentiles in each bin. This provides the median trend and the scatter of the observational distribution. As discussed in Paper II, different binning choices can slightly influence this step. We then interpolate these percentiles linearly to define the observational function $f_{\rm obs}(x)$.
For each simulation, we extract the corresponding scaling relations, assigning each data point coordinates ($x_{\rm sim,\,i},\;y_{\rm sim,\,i})$, where $i$ indexes the simulation data and $n$ 
is the total number of points in that simulation. We computed the distance estimator as 
\begin{equation}
    D^2=\sum^{n%N_{\rm sim}
    }_{i=1}\frac{[y_{\rm sim,\,i}-\mathcal{N}_{\rm i}(f_{\rm obs}(x_{\rm sim,\,i},\sigma_{\rm obs,\,i}))]^2}{\sigma_{\rm obs,\,i}^2}\;,
\end{equation}
where the numerator is the squared difference between the ordinate of the simulation point $i$, $y_{\rm sim,\,i}$,
and a randomly extracted point from a Gaussian distribution, $\mathcal{N}$, centered on the median of the observed scaling relation with a standard deviation equal to $\sigma_{\rm obs,i}$. 
The quantity $\sigma_{\rm obs,\,i}$  is defined as the average of $\sigma_+$ and $\sigma_-$, which are the absolute differences between the interpolated 16th and 84th percentile trends and the interpolated median trend at $x_{\rm sim,\,i}$. At this stage, a few clarifications are necessary. First, we emphasize that we used the Gaussian distribution to model observational uncertainties, as it offers a simple yet effective way to account for the observed scatter in the data, a choice that, as we demonstrate, proves useful later in the analysis. It should be noted, however, that this approach does not explicitly account for potential selection effects in the observational data. Specifically, it assumes that the data points are independent and identically distributed samples from the underlying galaxy population. A more accurate treatment of these biases would require advanced modeling of the selection effects, which we leave for future work. Second, the quantity $\sigma_{\rm obs,\,i}$ accounts for the total observed scatter, encompassing both measurement errors and intrinsic astrophysical dispersion. Rather than comparing simulations to a single median trend, our analysis assesses how well they reproduce the full observed distribution.

Since each simulation contains a different number of data points, it is essential to normalize the $D^2$  estimator by this number, ensuring a fair comparison across simulations. We therefore define the reduced distance estimator as 
$\tilde{D}^2=D^2/(n-1)$.
Furthermore, to avoid extrapolations, we exclude from the fit any simulation points where $x_{\rm sim,\,i}$ falls outside the range of the observational median trend. The best-fitting simulation is the one with the minimum reduced distance estimator.

A simple fitting procedure, such as that described above, does not provide any uncertainty estimates for the best-fit parameters. Moreover, parameter degeneracy complicates the analysis: different parameter combinations can yield 
similarly good fits or produce very similar trends, making it difficult to identify a unique optimal configuration. This issue, which we discuss further in our work, motivates the use of a more robust approach: bootstrap resampling.
By applying the bootstrap method to the observational dataset, the simulations, and the simulation output, 
 we obtain a clearer statistical interpretation of the fit. Notably, since we have introduced a degree of randomness in the observational trends—by sampling from Gaussian distributions around the observed percentiles—this approach also allows us to estimate an uncertainty on the $D^2$ values themselves at the end of the bootstrap procedure.

We followed the steps described below. Using Mathematica \texttt{ResourceFunction["BootstrapStatistics"]},\footnote{\url{https://resources.wolframcloud.com/FunctionRepository/resources/BootstrapStatistics/}} we generated bootstrap samples, subsets of the original data with replacement. We performed resampling with 
$N\approx 100$ iterations, creating a larger statistical sample. Each resampled dataset undergoes the same $D^2$ analysis, producing a distribution of $N$ best-fit simulations. This distribution allows us to estimate the best-fit parameters using the median and quantify their scatter as the absolute difference between the median and the 16th/84th percentiles. The statistical reliability of the bootstrap resampling procedure, and its potential impact on parameter recovery and uncertainty estimates, have been explicitly tested through a series of dedicated analyses presented in \App\ref{Appendix: boot10}. We note that owing to the discrete nature of the sample, the 16th and 84th percentiles may occasionally coincide with the median, especially for sharply peaked distributions. 
This analysis provides both estimates of parameter uncertainties and insights into possible degeneracies. By examining the distribution of best-fit simulations obtained from the bootstrap resamples, each associated with a specific set of model parameters, we can assess the stability of the inferred results. A broad distribution of best-fit simulations indicates that multiple parameter configurations yield comparably good fits to the data, revealing potential degeneracies. Conversely, a narrow distribution may suggest that the best-fit parameters are robustly constrained. This approach thus allows for a more reliable evaluation of the model’s sensitivity to the data and the significance of the resulting parameter estimates.

%%%%%%%%%%%%%%%%%%%%%%%%%%%%%%%%%%%%%%%%%%%%%%%%%%%%%%%%%%%%%%%%%%%%%%%%%%%%%%
%%%%%%%%%%%%%%%%%%%%%%%%%%%%%%%%%%%%%%%%%%%%%%%%%%%%%%%%%%%%%%%%%%%%%%%%%%%%%%
\section{Results\label{results}}
%%%%%%%%%%%%%%%%%%%%%%%%%%%%%%%%%%%%%%%%%%%%%%%%%%%%%%%%%%%%%%%%%%%%%%%%%%%%%%
%%%%%%%%%%%%%%%%%%%%%%%%%%%%%%%%%%%%%%%%%%%%%%%%%%%%%%%%%%%%%%%%%%%%%%%%%%%%%%

In this section, we present the main results of our study. In Section \ref{sec:parameters_median}, we show the impact of cosmological and astrophysical parameters, as well as the WDM mass, on the scaling relations. We perform this analysis using both uniform-box and MW zoom-in simulations.  In addition, for the MW zoom-in simulations, we also evaluate the effect of these parameters on galaxy abundance. In Section \ref{sec:evaluation fit}, we assess the ability of our fitting method, supported by bootstrap resampling, to accurately estimate the parameters  for both simulation suites. Finally, in Section \ref{sec:bootstrap cataloghi}, we compare the simulation trends with observational data from the SPARC and LVDB catalogs, determining the set of parameters that best reproduces the observations.
%%%%%%%%%%%%%%%%%%%%%%%%%%%%%%%%%%%%%%%%%%%%%%%%%%%%%%%%%%%%%%%%%%%%%%%%%%%%%%
%%%%%%%%%%%%%%%%%%%%%%%%%%%%%%%%%%%%%%%%%%%%%%%%%%%%%%%%%%%%%%%%%%%%%%%%%%%%%%%%%%%%%%%%%%%%%%%%%%%%%%%%%%%%%%%%%%%%%%%%%%%%%%%%%%%%%%%%%%%%%%%%%%%%%%%%%%%%
%%%%%%%%%%%%%%%%%%%%%%%%%%%%%%%%%%%%%%%%%%%%%%%%%%%%%%%%%%%%%%%%%%%%%%%%%%%%%%

\subsection{Galactic scaling relations in  WDM cosmology\label{sec:parameters_median}}

\begin{figure*}
    \centering
     \includegraphics[width=0.95\linewidth]{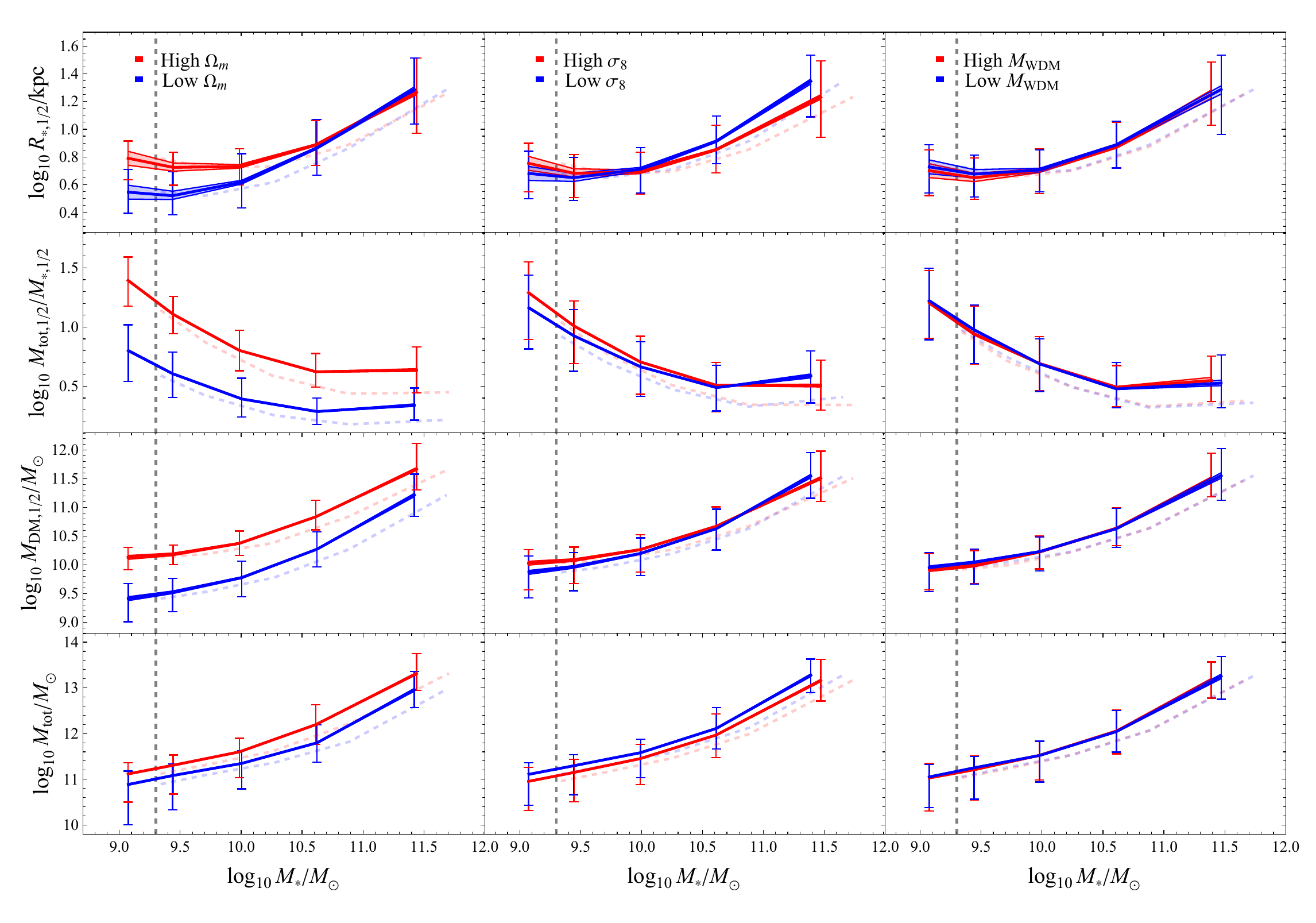}
        \caption{Median scaling relations as a function of total stellar mass illustrating the dependence on cosmological parameters. The relations for stellar half-mass radius, total-to-stellar mass ratio within the stellar half-mass radius, DM mass within the stellar half-mass radius, and total galaxy mass are shown for both uncalibrated (solid line) and calibrated (dashed line) cases. Trends were calculated using galaxies from 1024 uniform-box simulations divided into the following stellar mass intervals: $\rm log_{10} M_*/M_\odot<9.3$, $9.3\leq \rm log_{10} M_*/M_\odot<9.6$, $9.6\leq \rm log_{10} M_*/M_\odot<10.4$, $10.4\leq \rm log_{10} M_*/M_\odot<11.2$, and $\rm log_{10} M_*/M_\odot\geq 11.2$. Red and blue indicate trends for high and low parameter values, respectively, while the dashed gray line marks the threshold  $N_{*,1/2} =50$. The thresholds and median values for the parameters are provided in Table~\ref{tab:thresholds}. Error bars represent the scatter (16th–84th percentile range), while shaded areas indicate uncertainty estimates on the median values.}

  \label{fig: boxes cosm trend}
\end{figure*}

To investigate how cosmological and astrophysical parameters, including the WDM particle mass, influence galactic scaling relations, we use the full set of simulations. To clarify our methodology, we take the parameter $\Omega_{\rm m}$ as an example and describe our approach step by step; the same procedure is then applied to each of the other parameters. 

First, we aim to show how scaling relations vary for extreme values of the parameter. We select two threshold values for $\Omega_{\rm m}$, labeled "Low $\Omega_{\rm m}$"=0.25 and "High $\Omega_{\rm m}$"=0.35 (all threshold values are listed in Table \ref{tab:thresholds}).  From the full simulation set, we construct two subsamples containing simulations with $\Omega_{\rm m}$ values below the low threshold and above the high threshold, respectively. 
We then ensure that  
the distributions of the remaining parameters in each subsample remain similar to their  original distributions in the full sample and close to their fiducial values. 
Next, we divide the stellar mass range of the simulations into intervals and, for each interval, compute the median of the quantities of interest, along with the 16th and 84th percentiles. These values are then used to reconstruct the median scaling relations by combining the points from each mass interval. Median statistics are robust against the influence of outliers, and evaluating all simulations together within specific mass ranges allows us to assess the impact of the parameters on galaxies of different masses while maximizing the statistical sample size. 

On the other hand, the scatter does not represent an uncertainty on the median trend itself, but rather captures the statistical variation in the trends due to changes in the remaining parameters. As sources of uncertainty, we first consider the uncertainty on the medians, which is generally small and increases only moderately at higher stellar masses due to the reduced number of galaxies in those bins, reaching at most 0.08 dex. Second, we consider the error in estimating the stellar half-mass radius, which propagates into the measurements of the quantities computed within that radius. Details of this calculation are provided in \App\ref{app:ErrReff}. The uncertainty tied to the $R_{*,1/2}$ calculation depends on the number of stellar particles in the simulated galaxy, and can reach approximately 0.04–0.05 dex when $N_{*,1/2}=20$. For each scaling relation, we add in quadrature the sources of uncertainties on the y-axis: namely, the error of the median, and that on the plotted quantity  (discussed in \App\ref{app:ErrReff}), if the latter is — or is computed within — $R_{*,1/2}$. This procedure is applied consistently to both the low and high trends. We show the total uncertainties in the trends using shaded regions, while representing the scatter of the medians as error bars. In this section, we present the trends for galaxies with a minimum contribution of  $N_{*,1/2}=20$, along with the corresponding uncertainty estimates. However, when comparing with  observational catalogs, we adopted a more conservative cut of  $N_{*,1/2}>50$. Overall, the use of the full simulation set allows for precise estimates of median values, thanks to the large number of galaxies in each stellar mass bin.

%%%%%%%%%%%%%%%%%%%%%%%%%%%%%%%%%%%%%%%%%%%%%%%%%%%%%%%%%%%%%%%%%%%%%%%%%%%%%%
\subsubsection{Uniform-box suite \label{sec: Uniform-box parameter dep.}}
We begin by analyzing the uniform-box simulations, to which we apply the selection criteria described in Sect. \ref{sec: sel simulations}. 
The parameter thresholds are listed in Table~\ref{tab:thresholds}, which shows, for each parameter, the threshold ranges along with the corresponding median values computed within each subsample. These median values remain consistent across the different stellar mass intervals.

Figure \ref{fig: boxes cosm trend} illustrates the impact of cosmological parameters and the WDM particle mass on the galactic scaling relations discussed in Sect. \ref{sec:fit bootstrap}, and also shows the corresponding trends obtained from the calibrated models (dashed lines) for comparison. 
Figure~\ref{fig: boxes cosm trend} shows that $\Omega_{\rm m}$  has a significant impact on the total-to-stellar mass ratio and the DM mass within the stellar half-mass radius, shifting these relations toward higher DM fractions and greater DM content. Moreover, the total galaxy mass is higher for larger values of $\Omega_{\rm m}$, with this effect becoming more significant as the stellar mass increases and, in the lower stellar mass bins, higher values of $\Omega_{\rm m}$ correspond to larger $R_{*,1/2}$. The influence of $\sigma_8$ is generally weaker than that of  $\Omega_{\rm m}$. The DM content and the total-to-stellar half-mass ratio within $R_{*,1/2}$ show a mild increase at low stellar masses for higher values of $\sigma_8$, but this behavior reverses around $\rm log_{10} M_*/M_\odot\sim 10.6$. Conversely, the size–mass relation appears to flatten for elevated $\sigma_8$ values, starting from galaxies with $\rm log_{10} M_*/M_\odot\sim 10$. In addition, the total mass decreases with increasing parameter values, with little variation across stellar mass bins. Moreover, we find that the $\sigma_8$  also induces a slight horizontal shift in the highest mass bin, where higher values of $\sigma_8$  are associated with higher total stellar masses.
\begin{table}[h]
\centering
\caption{Thresholds and median values of cosmological and astrophysical parameters used in the scaling relation analyses.}
\label{tab:thresholds}
\begin{tabular}{lcccc}
\toprule
\midrule
\rm Parameter & \multicolumn{2}{c}{Low} & \multicolumn{2}{c}{High}\\
 & Range &  Median & Range & Median\\ 
\midrule
$\Omega_{\rm m}$         & $< 0.25$   & $ 0.19$   & $\geq 0.35$ &  $ 0.43$           \\
$\sigma_8$         & $< 0.75$   &$ 0.68$    & $\geq 0.85$ & $ 0.93$    \\
$A_{\rm SN1}$      & $< 1$      &  $ 0.6$ (0.5)  & $\geq 2$    & $ 2.8$ (2.8)   \\
$A_{\rm SN2}$      & $< 1$     & $ 0.7$ (0.7)    & $\geq 1$     & $ 1.4$ (1.4)     \\
$\rm BH_{\rm FF}$  & $< 1$     & $ 0.6$ (0.5)    & $\geq 2$     & $ 3.0$ (2.9)   \\
$M_{\rm WDM}$ [keV]& $< 2$     & $ 1.9$ (1.9)    & $\geq 10$   &   $ 12.3$ (15.4)     \\ 
\bottomrule
\hline
\end{tabular}
\tablefoot{Thresholds used in the scaling relation analyses shown in Figs.~\ref{fig: boxes cosm trend}, \ref{fig: boxes astro trend},
\ref{fig:MW zoom}, \ref{fig: diffmed}, and \ref{fig:Ngal_median}
for cosmological and astrophysical parameters and the WDM particle mass. Median values are computed for the selected subsamples in the uniform-box simulations, while values in parentheses refer to the MW zoom-in simulations.}
\end{table}
\begin{figure*}
    \centering
     \includegraphics[width=0.95\linewidth]{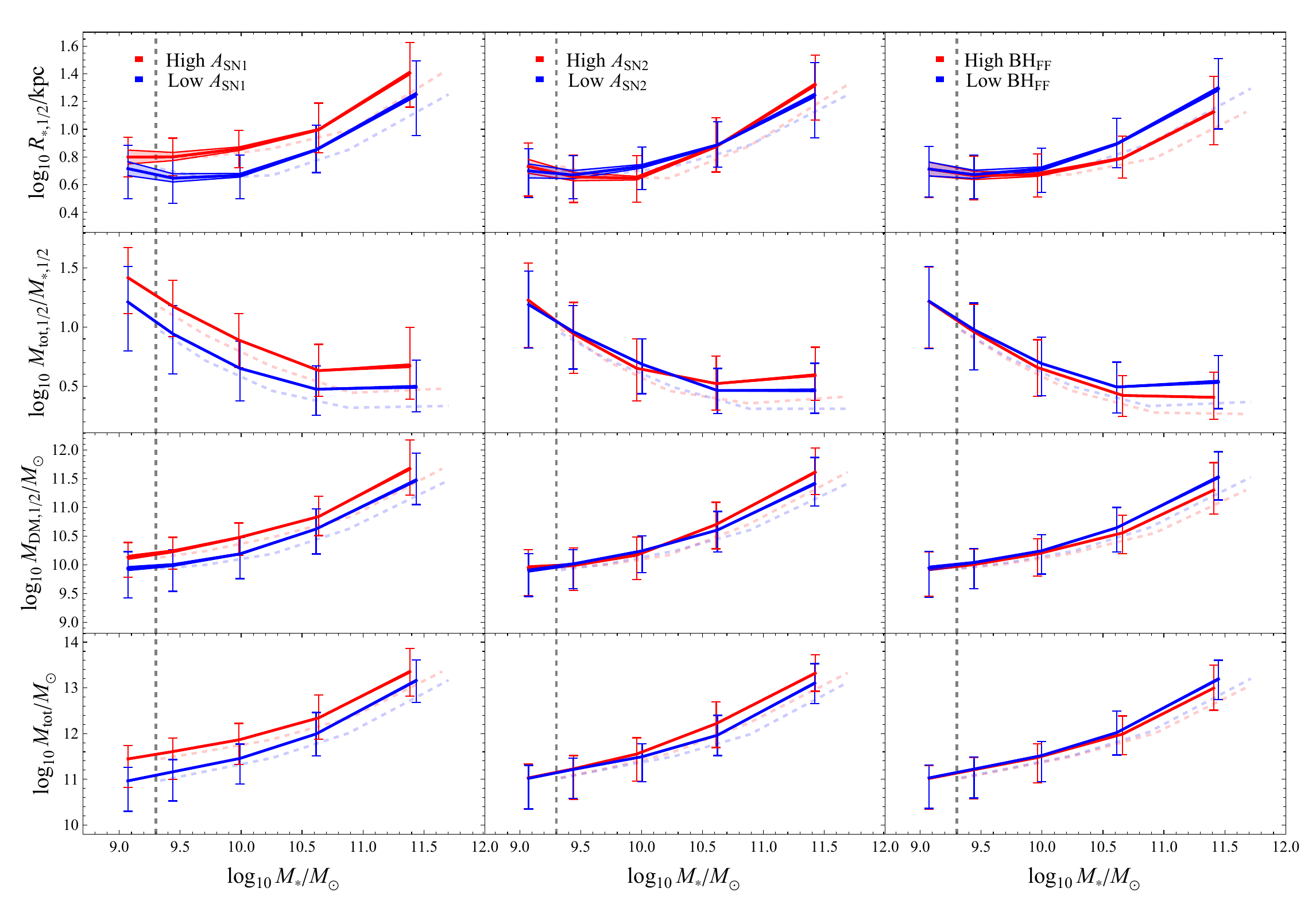}
        \caption{Same as Fig. \ref{fig: boxes cosm trend}, but for the astrophysical parameters $A_{\rm SN1}$,  $A_{\rm SN2}$, and $\rm BH_{\rm FF}$. }
  \label{fig: boxes astro trend}
\end{figure*}
These results are consistent with the findings of Papers I and II, where the effects of cosmological parameters in a $\Lambda$CDM framework were analyzed separately for SFGs and PGs, respectively. In our analysis, we do not differentiate between these morphological types. However, as discussed in Paper II, the influence of cosmological parameters is independent of galaxy type.  
 Moreover, we find that the WDM particle mass does not significantly alter the trends in the scaling relations. However, we do observe a horizontal shift, with higher WDM particle masses associated with slightly lower stellar masses in the most massive bin.
Finally, the comparison with the calibrated trends indicates that calibration effects become more pronounced at higher stellar masses, resulting in consistently lower values for all measured quantities compared to the uncalibrated case. This behavior is expected, and the same effect is naturally observed in Fig.~\ref{fig: boxes astro trend}, where we show how the scaling relations vary as a function of the astrophysical parameters.

Regarding the effect of astrophysical parameters on scaling relations, the wind energy--parameterized by $A_{\rm SN1}$--has the strongest impact, systematically shifting all four scaling relations upward (Fig. \ref{fig: boxes astro trend}).  
In addition, in the highest stellar mass bin, larger values of $A_{\rm SN1}$ suppress star formation, leading to a slight reduction in the median stellar mass of galaxies. Notably, the $A_{\rm SN1}$ parameter operates almost independently of the total stellar mass of the galaxy within the analyzed mass range, a result consistent with Papers I, II and III, where similar trends were observed for SFGs and PGs, respectively, using \camels\ simulations. 

The effects of wind velocity (i.e., $A_{\rm SN2}$) on stellar mass accretion are primarily evident at high stellar masses. In particular, higher values of the parameter increase, at fixed stellar mass, the DM content and the total-to-stellar mass ratio within the half-mass radius, as well as the total sub-halo mass. The size–mass relation, on the other hand, exhibits a turning point around $\log_{10} \rm M_*/M_\odot \sim 10.6$: above this threshold, larger $A_{\rm SN2}$ values are associated with larger sizes, whereas below it, the trend reverses and sizes decrease.
Since these effects are most pronounced at high masses, we first compare our results with Paper II, which focuses on PGs, typically more massive than SFGs. There, high $A_{\rm SN2}$ values consistently increase all four scaling relations, in agreement with our findings. The turning point we observe in the size–mass relation is also seen in Paper III, which reports a similar mass scale of $\log_{10}\rm M_*/M_\odot \sim 10.5$. Below this mass scale, the trend inverts: stronger feedback reduces galaxy sizes. This behavior aligns with Paper I, which examines SFGs, generally less massive systems, and finds that large $A_{\rm SN2}$ values lead to systematically smaller galaxy sizes.

Finally, the effects of the black hole feedback factor, $\rm BH_{\rm FF}$, are broadly consistent with the trends observed in the \camels\ simulations, as presented in Paper III, which showed that, at fixed stellar mass, higher values of the $\rm BH_{\rm FF}$ parameter reduce the total-to-stellar mass ratio already at stellar masses of $ \rm log_{10}\rm M_*/M_\odot \sim 10$. Here, we also examine other scaling relations and find that high $\rm BH_{\rm FF}$ values are associated with smaller galaxy sizes, lower DM fractions within $R_{*,1/2}$, and lower total galaxy masses.
As observed for the $A_{\rm SN1}$ parameter, the $\rm BH_{\rm FF}$   parameter also induces a slight reduction in the total stellar mass of galaxies within the highest stellar mass bin.

Among the scaling relations analyzed in this section, the ratio $M_{\rm tot,1/2}/M_{*,1/2}$ is particularly insightful, as it exhibits a minimum that defines the so-called golden mass \citep{dek_birn06, Moster+10, Tortora+10CG, Tortora+19_LTGs_DM_and_slopes, CASCOIII}. This characteristic mass scale appears as a bend or extremum in several scaling relations and is especially associated with the turnover in the stellar-to-halo mass relation (i.e., the $M_{\rm tot}$–$M_*$ relation), or equivalently with the peak in the correlation between star formation efficiency, defined as $\epsilon_{\rm SF} = M_*/(M_{\rm h} f_{\rm b})$, where $M_{\rm h}$ is the total halo mass,  and mass. Physically, the golden mass is thought to arise from the interplay of various processes—including SN and AGN feedback, virial shocks, and cold gas accretion via streams (see Paper III and references therein). A focused analysis of how this feature varies across different cosmological and feedback parameters, and between CDM and WDM models, is presented in \App\ref{sec:goldenmass}. This analysis, based on the DREAMS and \camels\ uniform-box simulations at comparable resolution, confirms the presence of a well-defined golden mass in both cosmologies. Notably, WDM simulations tend to exhibit more regular and pronounced minima in the $M_{\rm tot,1/2}/M_{*,1/2}$ ratio compared to CDM at the same resolution.
%%%%%%%%%%%%%%%%%%%%%%%%%
\subsubsection{MW zoom-in suite
\label{sec: MW parameter dep.}}
\begin{figure*}
    \centering     \includegraphics[width=0.93\linewidth]{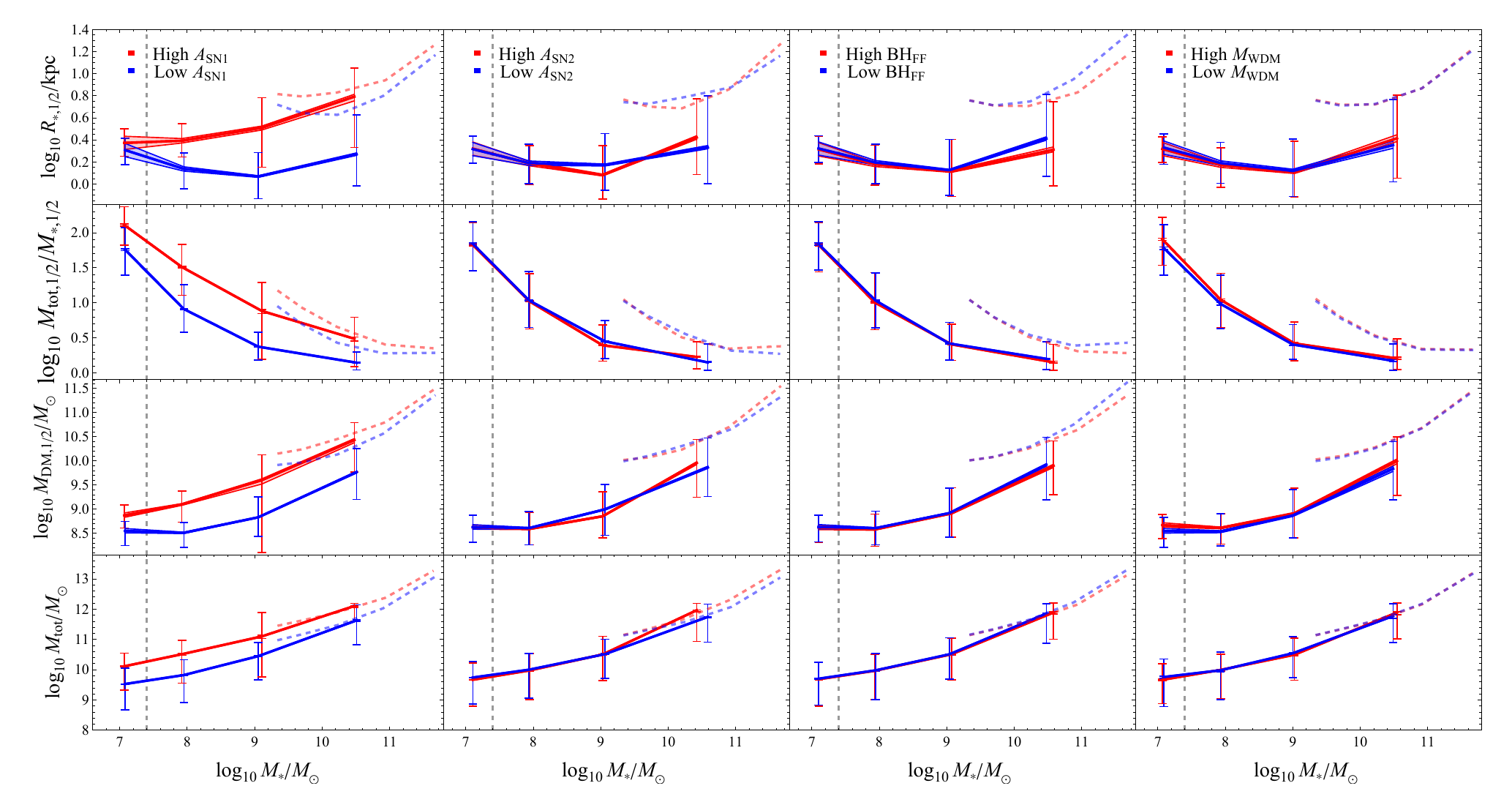}
        \caption{Median scaling relations as a function of total stellar mass showing the dependence on astrophysical parameters and WDM mass for both MW zoom-in simulations (solid line) and calibrated uniform-box simulations (dashed line) for comparison. Trends are derived from 1024 MW zoom-in simulations divided into the following stellar mass intervals: $\rm log_{10} M_*/M_\odot<7.5$,  $7.5\leq\rm log_{10} M_*/M_\odot<8.5$, $8.5\leq \rm log_{10} M_*/M_\odot<10$, and $\rm log_{10} M_*/M_\odot\geq10$. Red and blue indicate trends for high and low parameter values, respectively. The  dashed gray line indicates $N_{*,1/2} = 50$.}

\label{fig:MW zoom}
\end{figure*}
\begin{figure}
    \centering
     \includegraphics[width=0.8
     \linewidth]{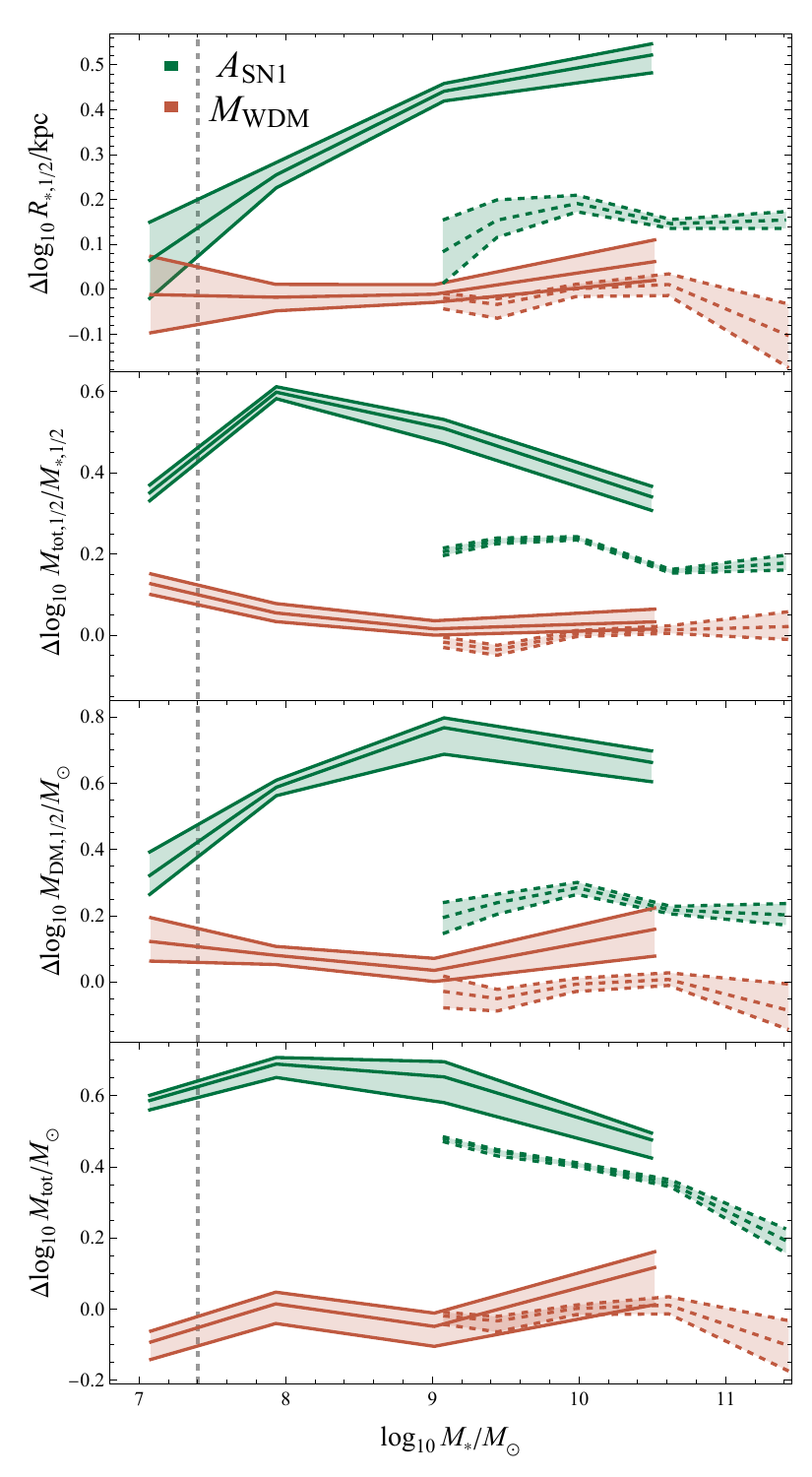}
        \caption{Differences between the median trends corresponding to the high and low values of two parameters: $A_{\rm SN1}$ (green) and $M_{\rm WDM}$ (brown). Solid lines represent MW zoom-in simulations; dashed lines represent uniform-box runs. Shaded regions show propagated uncertainties. The dashed gray line indicates $N_{*,1/2} = 50$.}
  \label{fig: diffmed}
\end{figure}
\begin{figure}
    \centering
     \includegraphics[width=0.4
    \textwidth]{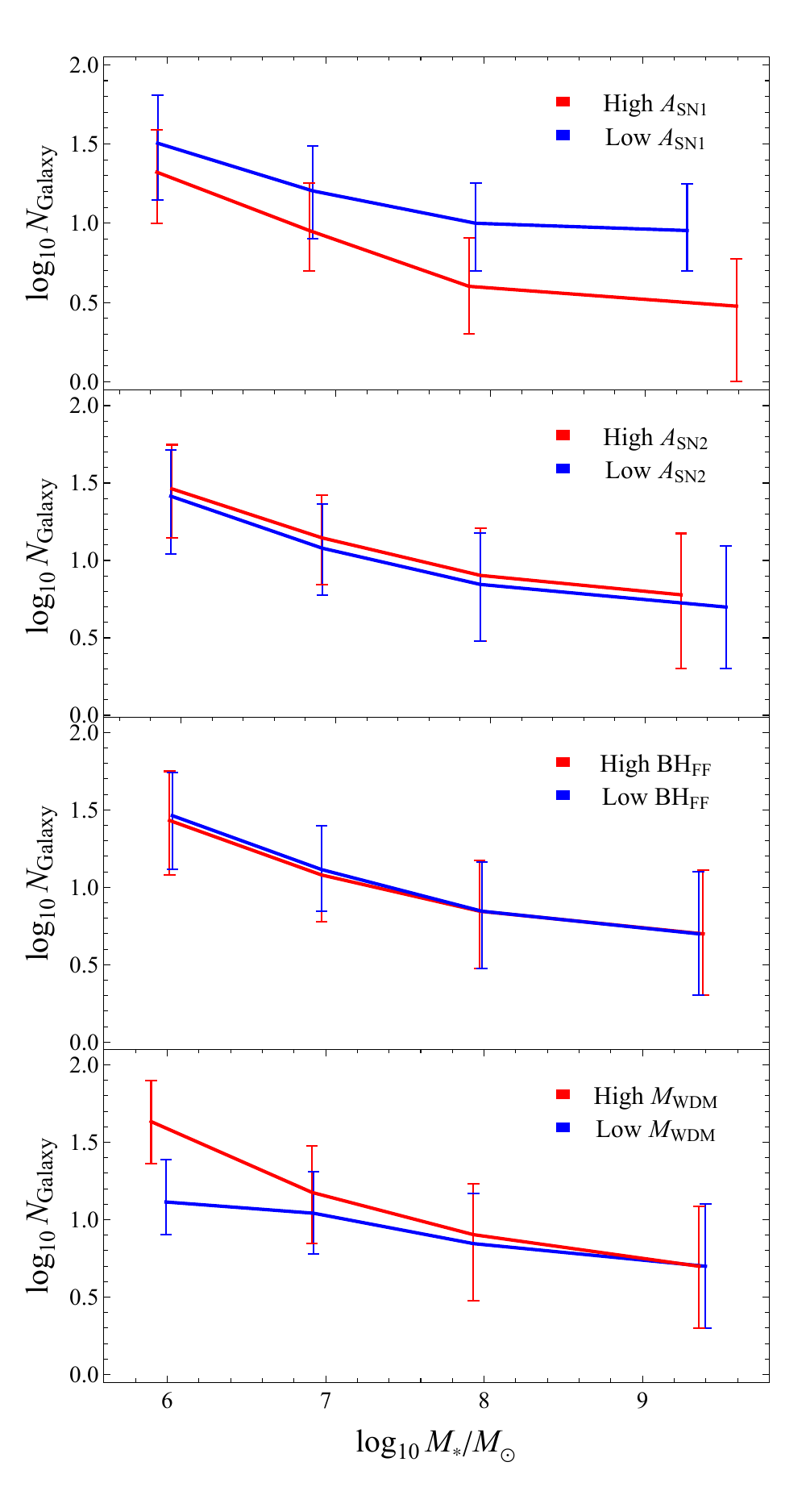}
        \caption{Median number of galaxies as a function of median total stellar mass computed in the stellar mass bins $\rm log_{10} M_*/M_\odot < 6.5$, $6.5 \leq\rm log_{10} M_*/M_\odot < 7.5$, $7.5\leq \rm log_{10} M_*/M_\odot < 8.5$, and $\rm log_{10} M_*/M_\odot \geq 8.5$. The four panels, from top to bottom, correspond to variations in $A_{\rm SN1}$, $A_{\rm SN2}$, $\rm BH_{\rm FF}$, and $M_{\rm WDM}$, respectively. In each case, the red and blue curves represent parameter values above and below the thresholds listed in Table~\ref{tab:thresholds}, respectively. The plots are based on the 1024 MW zoom-in simulations from DREAMS.}
  \label{fig:Ngal_median}
  \end{figure}
\begin{figure*}
    \centering
     \includegraphics[width=0.95\linewidth]{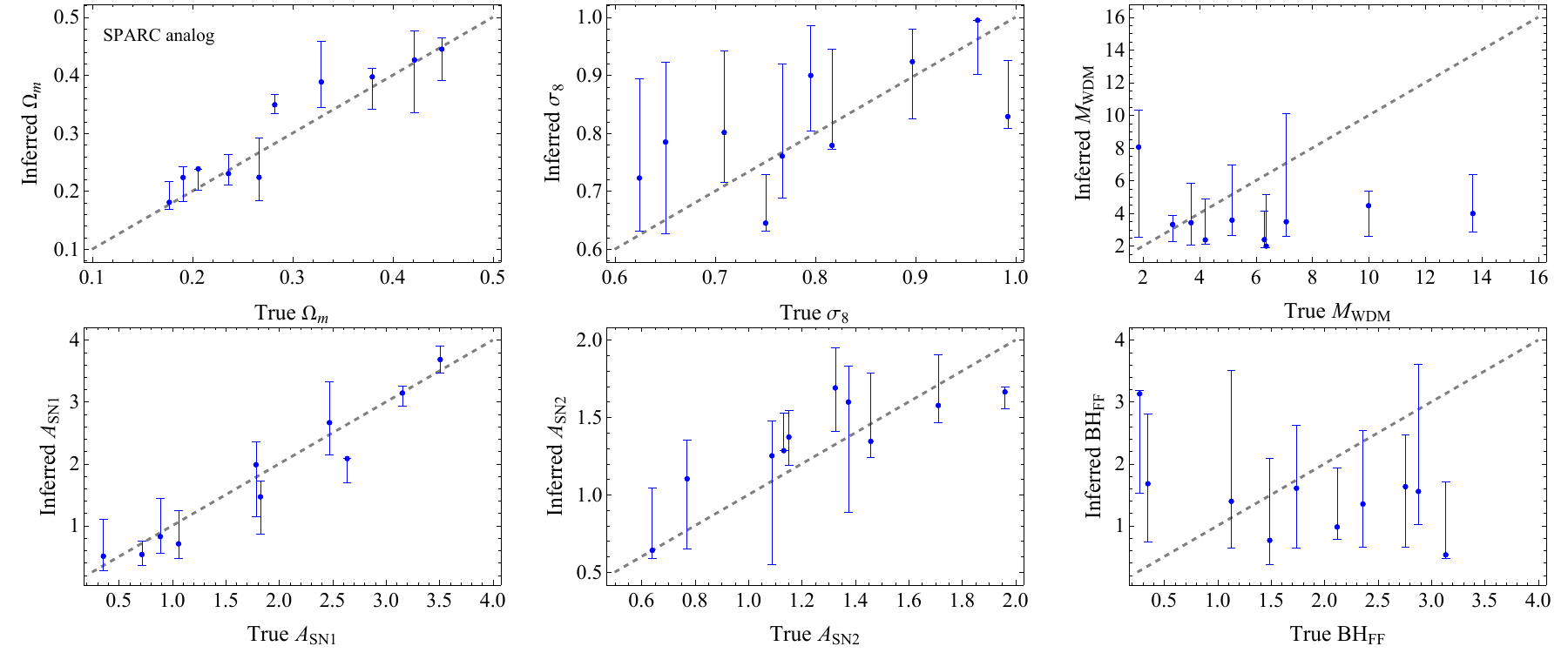}
        \caption{Recovery plot for the uniform-box simulations showing the bootstrap results with uncertainties for the estimated parameters: $\Omega_{\rm m}$, $\sigma_{8}$, $M_{\rm WDM}$, $A_{\rm SN1}$, $A_{\rm SN2}$, and $\rm BH_{\rm FF}$ against true values. These parameters were obtained by comparing the simulations with the simulation outputs mimicking the SPARC catalog. The dashed gray line indicates the ideal one-to-one relation between estimated and true parameter values.}
  \label{fig: recovery UB}
\end{figure*}
Figure \ref{fig:MW zoom} shows the dependence of the median scaling relations for astrophysical parameters and $M_{\rm WDM}$, derived from MW zoom-in simulations, alongside those from the calibrated uniform-boxes, for comparison. The consistency between the two sets of simulations is confirmed as they share the same parameter thresholds  (Table \ref{tab:thresholds}), and the median values are very similar, except for the WDM mass when $M_{\rm WDM}$ exceeds 10 keV.  In this case, the WDM mass is approximately 12.3 keV in the uniform-boxes compared to 15.4 keV in the MW zoom-in simulations; however, this discrepancy does not impact our analysis, as the scaling relations are largely insensitive to variations in $M_{\rm WDM}$ at high values of this parameter.
From the comparison between the two simulation suites, we observe discrepancies in some of the proposed scaling relations, most notably in the size–mass relation.
In particular, higher-resolution simulations tend to yield significantly smaller stellar half-mass radii compared to the uniform-box runs. This likely contributes to the lower trends observed in other relations measured within $R_{*,1/2}$. By contrast, the total stellar mass is broadly consistent with the trends seen in the lower-resolution suite. 

These discrepancies can be attributed to a combination of factors. A first and obvious contribution arises from resolution differences between the uniform-box and the MW zoom-in suites, which cannot be fully compensated by a simple stellar-mass calibration. Indeed, such a calibration only corrects for global offsets, without accounting for more complex effects related to gravitational softening, energy transfer between baryons and DM, or the detailed structure of the interstellar medium.
As shown in \Apps\ref{calibrazione} and \ref{app:cal_MW}, calibrating the 1P uniform-box run against TNG100-1 or TNG50-1 significantly improves the agreement of the scaling relations with the reference runs, leaving only moderate residual offsets. It is worth noting that TNG50-1 has a resolution comparable to that of the MW zoom-in simulations, while TNG100-1 is coarser. By contrast, when the same 1P simulation is calibrated against the satellite population extracted from the MW zoom-in suite, the resulting scaling relations remain systematically above the MW satellite trends, despite exhibiting a stellar-mass shift comparable to that obtained when calibrating against TNG100-1 or TNG50-1. This behavior indicates that resolution alone cannot explain the observed offsets: differences of this magnitude are not seen when comparing to TNG50-1 or TNG100-1, which suggests that environmental effects specific to the MW zoom-in simulations play a major role. We verified in Appendix \ref{APP:cal_R} that an additional calibration based on the stellar half-mass radius, $R_{*,1/2}$, further reconciles the two datasets, substantially improving the match with the MW zoom-in trends, however we do not pursue this approach here. The MW zoom-in galaxies, by construction, probe a highly specific environment — that of satellites within MW-mass haloes — and therefore cannot be considered representative of the general galaxy population at similar stellar masses.

Additional factors and motivations should also be considered. In particular, part of the discrepancy between the trends of the two simulation suites may be attributed to 
cosmic variance — i.e., the natural statistical variations caused by changing the random seed used to generate the initial conditions in cosmological simulations. Although our DREAMS simulations do not include multiple fiducial realizations (where only the seed is varied) to directly measure this effect, we expect its impact to be small. As a reference, Paper I provides an estimate of the effect of cosmic variance on the \camels\  simulations, finding the scatter in relevant galaxy properties to be at the percent level\footnote{See Paper I, Appendix A3, for a detailed assessment of cosmic variance on scaling relation parameters, reporting standard deviations around a few percent ($\sim 10^{-2}$).}. 

Moreover, we observe a flattening of the size–mass relation that amplifies the discrepancies between the two simulation suites. This flattening appears in the uniform-box simulations at total sub-halo stellar masses below $\log_{10}\rm M_*/M_\odot \sim 10.5$, while the MW zoom-in simulations also show some flattening at stellar masses below $\log_{10}\rm M_*/M_\odot \sim 9$, maintaining a relatively shallow trend over the full mass range. Such flattening can be attributed to a combination of physical and numerical factors. In particular, \citet{Ludlow2019} demonstrate that numerical artifacts may arise due to resolution limits, notably two-body scattering effects. In cosmological simulations, the spurious transfer of kinetic energy from massive DM particles to lighter stellar particles leads to artificial size inflation over time. This phenomenon becomes significant when the particle mass ratio $\mu=m_{\rm DM}/m_{\rm gas}$ exceeds unity, where $m_{\rm DM}$   and $m_{\rm gas}$  are the masses of individual DM and gas particles, respectively. Specifically, \citet{Ludlow2019} demonstrate that simulations from EAGLE project (\citealt{EAGLE2015}) with $\mu \gsim $ 5.4 show an artificial increase in galaxy sizes, which occurs at stellar masses below $\log_{10}\rm M_*/M_\odot \sim 9.5$, which leads to a systematic flattening of the size–mass relation\footnote{The convergence radius, below which numerical artifacts become prominent, can be approximated as $r_{\rm conv}=0.174 \kappa_{\rm rel}^{2/3}l$, where $\kappa_{\rm rel}=0.18$ (\citealt{Ludlow_Schaye2019}), and $l= L_{\rm box}/ N_{\rm part}^{1/3}$   is the mean interparticle spacing in physical units. Here, $L_{\rm box}$ denotes the linear size of the simulation box and $N_{\rm part}$ the total number of particles. For example, in the uniform box simulations with $  L_{\rm box}=25\;h^{-1}$ Mpc and $N_{\rm part}=2\times256^3$, this expression predicts a convergence radius of $r_{\rm conv}\sim6.2$ kpc (i.e., $\log_{10}R_{*,1/2}/\rm kpc\sim0.8$). Estimating $ r_{\rm conv}$  for the MW zoom-in simulations is more complex due to their nonuniform resolution of the DM particles and nested volume structure.}.
In DREAMS uniform box simulations, $\Omega_{\rm m}$  varies from 0.1 to 0.5, causing the particle mass ratio $\mu$  to range approximately from 2.0 to 10.2. In contrast, the MW zoom-in simulations have a roughly constant $\mu\sim 6.3$. Although these numerical effects likely contribute to the observed flattening, they do not fully account for the discrepancy between the two simulation suites. Even if the size–mass relation were extrapolated without flattening to lower stellar masses, a significant offset would remain.

Trends in Fig. \ref{fig:MW zoom} confirm that $A_{\rm SN1}$ plays a dominant role down to $\log_{10} \rm M_{*}/M_\odot \sim 7$, driving an increase in $R_{*,1/2}$ and shifting the total-to-stellar mass ratio and DM mass upward. In other words, at fixed stellar mass, higher values of $A_{\rm SN1}$ lead to larger galaxies with lower stellar fractions and a higher DM content, highlighting the efficiency of SN feedback in regulating star formation.  On the other hand, the effects of $A_{\rm SN2}$ and $\rm BH_{\rm FF}$ appear to be weak. For $\rm BH_{\rm FF}$ this is expected in MW zoom-in simulations, which primarily probe lower-mass galaxies, as they tend to regulate star formation and black hole growth more significantly in higher-mass systems.  Nevertheless, we find that the trends of $R_{*,1/2}$ with respect to $A_{\rm SN2}$ and $\rm BH_{\rm FF}$ are qualitatively consistent with those observed in uniform-box simulations.  Notably, the turning point where the size–mass relation inverts for $A_{\rm SN2}$ now occurs at $\log_{10}\rm M_*/M_\odot \sim 9.6$. Moreover, we observe that an increase in $A_{\rm SN2}$ leads to a reduction in the median stellar mass within the most massive bin, consistent with the expected effects of intensified stellar feedback. Conversely, higher values of $\rm BH_{\rm FF}$ correlate with increased stellar masses even in lower-mass bins. This trend likely reflects the fact that AGN feedback predominantly impacts galaxies exceeding a certain mass threshold, rather than indicating a direct causal enhancement of stellar mass. Notably, the turning point where the size–mass relation inverts for $A_{\rm SN2}$ now occurs at $\log_{10}\rm M_*/M_\odot \sim 9.6$. Finally, $M_{\rm WDM}$ shows a weak influence on the total-to-stellar half-mass ratio within the half-mass radius at low masses. This effect suggests that higher-resolution simulations could help clarify the impact of this parameter.%%%%%%%%%%%%%%%%%%%%%%%%%%%%%%%%%%%%%%%%%%%%%%%%%%%%%%%%%%%%%%%%%%%%%%%%%%%%%%
%%%%%%%%%%%%%%%%%%%%%%%%%%%%%%%%%%%%%%%%%%%%%%%%%%

In Fig. \ref{fig: diffmed}, we show the difference between the median trends obtained for the high and low settings of two representative parameters: $A_{\rm SN1}$ and $M_{\rm WDM}$. 
The associated uncertainties are computed through standard error propagation, based on the total uncertainties estimated for the individual high and low trends, and are displayed  as shaded regions. The figure shows results from both the uniform-box simulations (dashed lines) and the MW zoom-in runs (thick lines). Overall, the impact of these parameters is noticeably stronger in the MW zoom-ins. For $A_{\rm SN1}$, the difference between the extreme configurations reaches 0.5–0.7 dex, while in the uniform-box simulations the effect remains more modest, around 0.2–0.4 dex. This suggests that feedback processes regulated by $A_{\rm SN1}$ are more effective or more easily captured in the zoom-in setup. Regarding the WDM mass, the resulting variations are smaller in magnitude—typically on the order of 0.10-0.15 dex—but still detectable, particularly in quantities such as the ratio $M_{\rm tot,1/2}/M_{*,1/2}$  and  $M_{\rm DM,1/2}$. While subtle, these shifts imply that WDM mass can leave a non-negligible imprint on the inner structure of galaxies.  
%%%%%%%%%%%%%%%%%%%%%%%%%%%%%%%%%%%%%%%%%%%%%%%%%%%%%%%%%%%%%%%%%%%%%%%%%%%%%%%%%%%%%%%%%%%%%%%%%%%%%%%%%%%%%%%%%%%%%%%%%%%%%%%%%%%%%%%%%%%%%%%%%%%%%%%%%%%%%%%%%%%%%%%%%%%%%%%%%%%%%%%%
Although the WDM particle mass exhibits a weak signature on the galaxy scaling relations analyzed here, the high-resolution MW zoom-in simulations reveal a clear additional effect: a systematic reduction in the number of galaxies as a function of total stellar mass with decreasing WDM particle mass.
This trend reflects the expectation that WDM suppresses the formation of low-mass sub-haloes (e.g., \citealt{Lovell2014,Read2017}), thereby limiting galaxy formation at small scales. DREAMS simulations confirm this behavior: the total number of galaxies decreases as the WDM particle mass decreases, with the suppression becoming increasingly pronounced at lower stellar masses.

Figure \ref{fig:Ngal_median} illustrates the mass-dependent suppression by showing the median number of galaxies as a function of stellar mass, comparing models with $M_{\rm WDM}$<2 keV and $M_{\rm WDM}$>10 keV. The figure also reports the median trends for different values of the other feedback parameters. The threshold values used to define the parameter subsets are the same as those adopted in this section for the other scaling relations (see Table \ref{tab:thresholds}). Among all feedback parameters, $A_{\rm SN1}$ has the strongest influence on galaxy abundance, as expected. Its effect manifests primarily as a mass-independent vertical shift in the number of galaxies. In contrast, the WDM particle mass introduces a mass-dependent suppression: the number of galaxies is strongly reduced at $\rm log_{10}M_*/M_\odot\lsim 8$, with the effect gradually weakening and becoming negligible at $\rm log_{10}M_*/M_\odot\sim 9$. Regarding the other parameters, $\rm{BH_{FF}}$ has no measurable impact on galaxy abundance in this stellar mass range, consistent with AGN feedback becoming relevant only in more massive systems. $A_{\rm SN2}$ produces a minor increase in the number of galaxies. The overall sensitivity of galaxy abundance to the feedback parameters and on the WDM mass is qualitatively consistent with the findings of \citet{rose2024introducingdreamsprojectdark}.
While the suppression of the stellar mass function by WDM is theoretically expected, our results provide a direct, quantitative realization of this effect within a fully hydrodynamical framework that systematically explores baryonic feedback. The stellar mass dependence of galaxy abundance thus emerges as a promising diagnostic to help disentangle DM properties from feedback processes.

In this section, we have focused on the cumulative median trends of all galaxies in the MW zoom-in simulation, including both the central (host) galaxy and its satellites, with particular attention given to the lowest-mass systems. However, it is also informative to isolate the host galaxies and analyze their properties separately. A detailed discussion of the host galaxy properties—as a function of the astrophysical feedback parameters and the WDM particle mass—is presented in \App\ref{App:Host}, where we find that their overall trends are qualitatively consistent with those of the full galaxy population.
%%%%%%%%%%%%%%%%%%%%%%%%%%%%%%%%%%%%%%%%%%%%%%%%%%%%%%%%%%%%%%%%%%%%%%%%%%%%%%
\subsection{Evaluation of the fitting procedure\label{sec:evaluation fit}} 
In this section we assess the effectiveness of the fitting procedure described in Sect. \ref{sec:fit bootstrap} in recovering astrophysical and cosmological parameters, including the mass of the WDM, through an internal validation process. This methodology closely follows the framework established by \cite{BusilloCascoI}. As a first step, we select a subset of simulations to serve as reference datasets for evaluation—that is, datasets with known input values used as benchmarks to test the quality of the fitting procedure. We then apply the fitting algorithm to infer astrophysical and cosmological parameters as well as the WDM mass. 
Since the true input values are known in these reference  
datasets, we can directly evaluate the accuracy of the inference by comparing the recovered parameters to the ground truth. To quantify the level of agreement, we compute the Pearson correlation coefficient and determine its statistical significance\footnote{
A correlation is considered significant when its p-value falls below the adopted significance level $\alpha$.}.
Once a significant correlation between true and inferred parameters is established, we assess whether the relationship is consistent with an ideal one-to-one correspondence, performing a weighted linear fit and computing the slope $m$ and intercept $c$. The recovery is considered successful if $m$ is statistically compatible with 1 and $c$ with 0.

%%%%%%%%%%%%%%%%%%%%%%%%%%%%%%%%%%%%%%%%%%%%%%%%%%%%%%%%%%%%%%%%%%%%%%%%%%%%%%
\subsubsection{Parameter recovery in uniform-box simulations\label{subsec:rec unif}}
We assess here the performance of the fitting procedure when applied to uniform-box simulations. To ensure consistency, we apply the same selection criteria, used for the comparison with the SPARC catalog and described in Sect. \ref{sec: sel simulations}, to all simulations, including the subset serving as the reference. 
Correlation coefficients and linear fit parameters, are reported in \App\ref{Appendix: table recovery} (Table \ref{tab:recovery p ub}).

Figure~\ref{fig: recovery UB} exhibits the recovery plots for cosmological and astrophysical parameters, as well as for the WDM mass. 
Strong positive correlations are observed for $\Omega_{\rm m}$,  $\sigma_8$, $A_{\rm SN1}$, and $A_{\rm SN2}$, with statistically significant results at the $\alpha = 0.05$ level. The linear fits support these findings: for $\Omega_{\rm m}$, $A_{\rm SN1}$, and $\sigma_8$, the slope and intercept are statistically consistent with a one-to-one relation, indicating a successful recovery. For $A_{\rm SN2}$, the slope is slightly underestimated and the intercept is overestimated. 

By contrast, $M_{\rm WDM}$ and $\rm BH_{\rm FF}$ exhibit negative correlations with the ground truth and large uncertainties, suggesting that the fitting procedure fails to constrain these parameters. The failure to recover $M_{\rm WDM}$ is expected, as the median trends show a lack of sensitivity to this parameter (see Fig. \ref{fig: boxes cosm trend}), whereas for $\rm BH_{\rm FF}$, it can be attributed to the sample selection.  Indeed, the median trends discussed in Sect. \ref{sec: Uniform-box parameter dep.} are computed over the full galaxy population, while here the fitting procedure is restricted to SFGs, which underrepresent massive galaxies where the effect of $\rm BH_{\rm FF}$ is more prominent. 
Interestingly, while our analysis does not constrain $M_{\rm WDM}$, other approaches have reported sensitivity to WDM masses of several keV \citep[e.g.][]{Rose2023, costanza2025}. These constraints rely on methodological choices that differ from ours—such as deep-learning models applied to DM-only density fields or machine-learning emulators trained on galaxy-population statistics—which extract information in ways complementary to the global scaling-relation framework adopted here.

These results are in agreement with Paper I, where a similar analysis on CDM \camels\ simulations identified $\Omega_{\rm m}$ and $A_{\rm SN1}$ as the most robustly recovered parameters. Moreover, the parameter $\sigma_8$ was reasonably well recovered, although  with larger associated uncertainties. In contrast, $A_{\rm SN2}$ exhibited less consistent behavior, passing the consistency test in some cases but failing in others, which indicates greater difficulty in obtaining reliable constraints.
The DREAMS and \camels\ simulations change different AGN feedback parameters, so a direct comparison is not possible.

%%%%%%%%%%%%%%%%%%%%%%%%%%%%%%%%%%%%%%%%%%%%%%%%%%%%%%%%%%%%%%%%%%%%%%%%%%%%%%
\subsubsection{Parameter recovery in MW zoom-in simulations\label{sec:recovery}}

  \begin{figure}
    \centering
     \includegraphics[width=0.48 \textwidth]{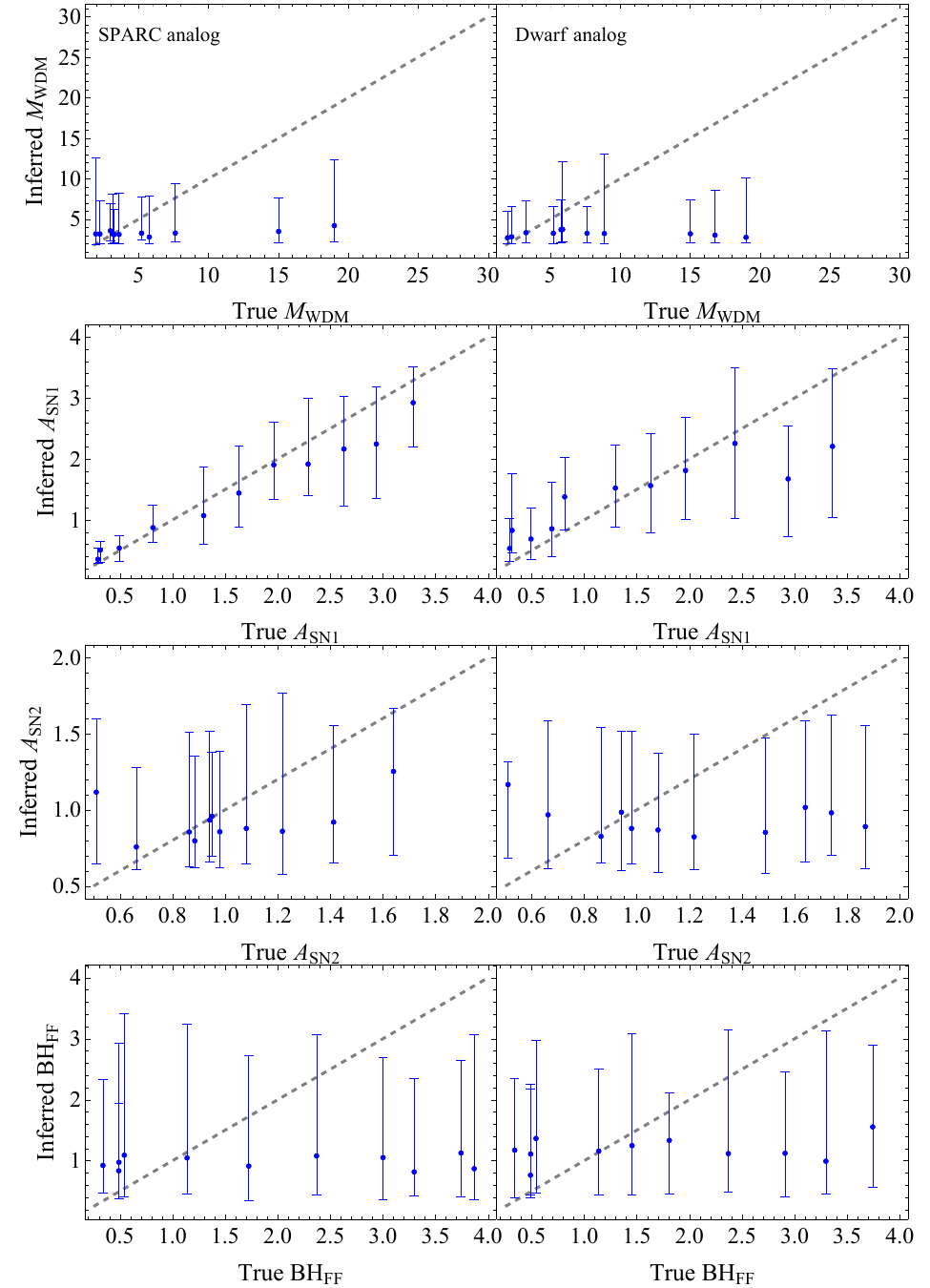}
        \caption{Recovery plot for the MW zoom-in simulations. The  figure shows the bootstrap results with uncertainties for the estimated parameters: $M_{\rm WDM}$, $A_{\rm SN1}$, $A_{\rm SN2}$, and $\rm BH_{\rm FF}$ (from top to bottom). These parameters were obtained by comparing the simulations with the  simulation outputs  
        mimicking SPARC and the dwarf catalog.  The dashed gray line indicates the ideal one-to-one relation between estimated and true parameter values.}
  \label{fig:recovery MW cumulative}
  \end{figure}

To assess how well the fitting procedure performs when applied to the MW zoom-in simulations, we used simulated galaxy samples filtered according to the selection criteria described in Sect. \ref{sec: sel simulations}, ensuring consistency with the SPARC and dwarf galaxy catalogs.

Figure \ref{fig:recovery MW cumulative} shows parameter estimates with uncertainties from scaling relation fits, compared to the ground truth provided by the simulation. 
In the comparison with the dwarf-like simulated samples,
we consider two cumulative fit cases: one including all the scaling relations, and one excluding the  $M_{\rm tot}$ versus $M_*$ relation. In the figure, we show the results obtained by considering all scaling relations. Correlation coefficients and fit parameters, are reported in Table \ref{tab:recovery p ub} in the \App\ref{Appendix: table recovery}.

The parameter $A_{\rm SN1}$ passes the correlation test at a significance level of $\alpha = 0.05$  for both the SPARC-matched and dwarf-matched simulation samples. 
In both samples, however, the linear fit systematically underestimates the slope and overestimates the intercept. For $M_{\rm WDM}$, a strong correlation is observed with the SPARC analog; however, the linear fit deviates significantly from the ideal one-to-one relation, with the slope close to zero,  in clear disagreement with the expected value of one.  Overall, in both cases considered here, the SPARC and dwarf analogs, these results indicate that we are not able to  constrain the WDM particle mass using scaling relations at the current resolution. 
Finally, both $A_{\rm SN2}$ and $\rm BH_{\rm FF}$  exhibit no significant correlation, consistent with the findings reported  in Sect. \ref{sec: MW parameter dep.}. 

%%%%%%%%%%%%%%%%%%%%%%%%%%%%%%%%%%%%%%%%%%%%%%%%%%%%%%%%%%%%%%%%%%%%%%%%%%%%%%
%%%%%%%%%%%%%%%%%%%%%%%%%%%%%%%%%%%%%%%%%%%%%%%%%%%%%%%%%%%%%%%%%%%%%%%%%%%%%%

\subsection{Comparing simulated and observed scaling relations\label{sec:bootstrap cataloghi}}

After discussing the impact of each parameter on the scaling relations and validating our fitting procedure using simulation output, 
we now compare the scaling relations derived from the SPARC and dwarf galaxy catalogs (see Sect. \ref{ObsCat}) with those obtained from the WDM simulations. We apply both a standard fitting procedure and the bootstrap method described in Sect. \ref{sec:fit bootstrap}, to constrain the variable parameters adopted in the simulations.
%%%%%%%%%%%%%%%%%%%%%%%%%%%%%%%%%%%%%%%%%%%%%%%%%%%%%%%%%%%%%%%%%%%%%%%%%%%%%%
\subsubsection{Fitting uniform-box simulations to observations\label{subsec:ub sparc fit}}
\begin{figure}[h!]
    \centering
     \includegraphics[width=0.38
    \textwidth]{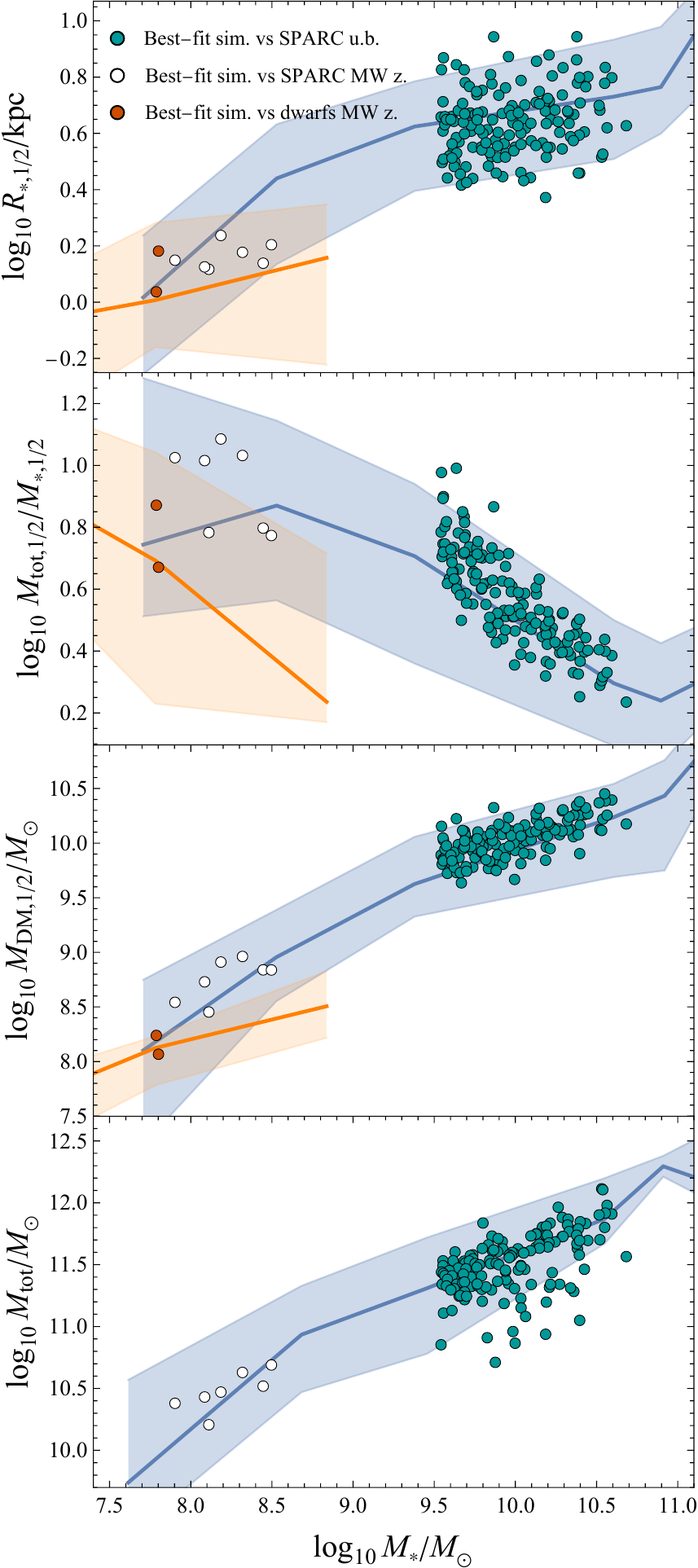}
        \caption{ Best-fit simulations (points) compared to median observational trends (shaded areas) from SPARC (green) and the dwarf catalog (orange). From top to bottom: Stellar half-mass radius, total-to-stellar mass ratio within $R_{*,1/2}$, DM mass within $R_{*,1/2}$, and total versus stellar mass. Green points indicate the best-fit uniform-box versus SPARC, white points represent the MW zoom-in best matching SPARC, and orange points show the MW zoom-in best matching the dwarf catalog.}
  \label{fig:fit all}
  \end{figure}
In Fig. \ref{fig:fit all} we present the simulations that best reproduce the observational catalogs, displaying both the trends from the calibrated uniform-box suite and from the MW zoom-in simulations within the same panel. Here, we focus exclusively on the uniform-box runs, leaving the discussion of the MW zoom-in simulations for the next section.

The simulated trends (green points) show good agreement with the SPARC observational data (green shaded area). The best-fit parameters are $\Omega_{\rm m} = 0.26$, $\sigma_8 = 0.99$, $S_8 = 0.93$, $A_{\rm SN1} = 1.75$, $A_{\rm SN2} = 1.42$, $\rm BH_{\rm FF} = 1.45$ and $\rm P_{\rm WDM}=0.38\;\rm keV^{-1}$ (i.e., $M_{\rm WDM} = 2.6$ keV). We include $S_8$ as a derived parameter, defined as $S_8 := \sigma_8 \sqrt{\Omega_{\rm m} / 0.3}$. The cumulative reduced D-square is $\tilde{D}^2 = 1.09$, while all individual and cumulative $D^2$ and $\tilde{D}^2$ values are reported in Table \ref{tab:chi2SPARCdwarf}.
The fitting results indicate significantly higher values for $A_{\rm SN1}$, $A_{\rm SN2}$, $\sigma_8$, and $S_8$ compared to those reported in Paper I, where $A_{\rm SN1} = 0.48$, $A_{\rm SN2} = 1.24$, $\sigma_8 = 0.83$, and $S_8 = 0.78$, while the matter density is similar: $\Omega_{\rm m} = 0.27$. We remind the reader that Paper I, which analyzes CDM \camels\ simulations, included the same cosmological and astrophysical parameters, except for AGN feedback and the WDM mass.

Bootstrap resampling analysis further supports the results of the previous fit;  the parameter estimates and their associated uncertainties
are reported in Table~\ref{tab:bootstrapSPARCub}. In the text below, we also report the corresponding percentage uncertainties, computed relative to the full prior range of each parameter — that is, each uncertainty is expressed as a fraction of the allowed variation range. The results are as follows: $\Omega_{\rm m} = 0.28^{+0.08}_{-0.04}$ ($\pm10/20\%$), $\sigma_8 = 0.89^{+0.10}_{-0.18}$ ($\pm25/45\%$), $S_8 = 0.85^{+0.09}_{-0.12}$ ($\pm 10/13\%$),  $A_{\rm SN1} = 1.67^{+0.49}_{-0.84}$ ($\pm13/22\%$), $A_{\rm SN2} = 1.46^{+0.32}_{-0.21}$ ($\pm14/21\%$), $\rm BH_{\rm FF} = 1.40^{+0.59}_{-0.66}$ ($\pm16/18\%$),  and $\rm P_{\rm WDM}=0.37^{+0.10}_{-0.13}\;keV^{-1}$ ($\pm 20/27\%$). Applying the inverse transformation, $M_{\rm WDM}=1/\rm P_{\rm WDM}$, we obtain a corresponding WDM particle mass of $M_{\rm WDM}=2.7^{+1.3}_{-0.6}$ keV. Since this transformation is nonlinear, the propagated uncertainties become asymmetric. In this case, it is more appropriate to express the uncertainty relative to the central value (i.e., $21/54\%$) rather than to the range of variation of the original parameter. The reduced D-square from the bootstrap is $\tilde{D}^2 = 5.08^{+0.32}_{-0.36}$.
Next, we compare these results with those from Paper I, noting that, at these resolutions, the WDM mass has no significant influence on the scaling relations analyzed in our study. 
While the cosmological parameters are consistent within $1 \sigma$ with those recovered in Paper I—$\Omega_{\rm m} = 0.27^{+0.01}_{-0.05}$, $\sigma_8 = 0.83^{+0.08}_{-0.11}$, and $S_8 = 0.78^{+0.03}_{-0.09}$—the supernova feedback parameters are considerably higher with respect to Paper I, as they found $A_{\rm SN1} = 0.48^{+0.25}_{-0.16}$ and $A_{\rm SN2} = 1.21^{+0.03}_{-0.34}$. Furthermore, the parameter estimates here are generally less precise with respect to Paper I.
These discrepancies are attributed to the introduction of the $\rm BH_{\rm FF}$ parameter, which has a more significant impact in this study compared to the AGN feedback parameters used in Paper I (see Sect. \ref{sec:parameters_median}). The higher inferred values of $A_{\rm SN1}$ and $A_{\rm SN2}$ compared to Paper I, and of $\rm BH_{\rm FF}$ relative to its fiducial value of 1, likely reflect their distinct effects on the model’s scaling relations: $A_{\rm SN1}$ tends to shift them upward, while $\rm BH_{\rm FF}$ generally shifts them downward, with a dependence on stellar mass. The effect of $A_{\rm SN2}$ is more intricate, but when focusing on SFGs, its impact resembles that of $\rm BH_{\rm FF}$: higher values of $A_{\rm SN2}$ are associated with smaller galaxy sizes and lower enclosed quantities, although this does not hold for the total stellar mass. Therefore, instead of converging to very low values of $A_{\rm SN1}$ to fit SPARC data, as seen in Paper I, the fit identifies a new combination of parameters that provides a good match to the observations. Moreover, the inclusion of the calibration procedure—introduced to mitigate the tendency of simulations to overestimate the scaling relations—prevents the fit from favoring artificially low values of $A_{\rm SN1}$, which would otherwise shift the predicted trends downward. Indeed, in the absence of calibration, the simulations yield cosmological parameter values and $\rm A_{SN1}$ that are systematically lower, as shown in Table \ref{tab:bootstrapSPARCuncub}: $\Omega_{\rm m} = 0.26^{+0.04}_{-0.04}$, $\sigma_8 = 0.87^{+0.11}_{-0.16}$, $S_8 = 0.80^{+0.11}_{-0.12}$, and $\rm A_{\rm SN1} = 1.04^{+0.78}_{-0.56}$. Compared to the calibrated runs, we also find a lower value of $\rm A_{\rm SN2} = 1.27^{+0.47}_{-0.48}$, and a higher value of $\rm BH_{\rm FF} = 1.87^{+1.38}_{-0.95}$. This helps explain the broader uncertainties seen in the results, which are likely due to stronger parameter degeneracies caused by the influence of $\rm BH_{\rm FF}$. In the future, a dedicated study incorporating PGs, where the impact of $\rm BH_{\rm FF}$ and $A_{\rm SN2}$ is more significant, will be performed, possibly helping to break these degeneracies and lead to tighter constraints on the model parameters.

%%%%%%%%%%%%%%%%%%%%%%%%%%%%%%%%%%%%%%%%%%%%%%%%%%%%%%%%%%%%%%%%%%%%%%%%%%%%%%
\subsubsection{Fitting MW zoom-in simulations to observations\label{subsec:fit zoomin}}
%%%%%%%%%%%%%%%%%%%%%%%%%%%%%%%%%%%%%%%%%%%%%%%%%%%%%%%%%%%%%%%%%%%%%%%%%%%%%%
%%%%%%%%%%%%%%%%%%%%%%%%%%%%%%%%%%%%%%%%%%%%%%%%%%%%%%%%%%%%%%%%%%%%%%%%%%%%%%
In this section, we compare the MW zoom-in simulations with the observational catalogs from SPARC and dwarf galaxies. Before going into the details of the analysis, a few considerations are in order.

The limited number of galaxies involved reduces the statistical robustness of the fit. We computed the median number of galaxies used in the comparison between simulations and observations, along with the 16th and 84th percentiles. Taking into account the stellar mass range covered by the catalogs and applying a possible cut in sSFR, we compare $12^{+14}_{-7}$ galaxies with SPARC and $10^{+11}_{-6}$ galaxies with the dwarf catalog. When the sample size is very small, some simulations may appear to agree well with the data due to statistical fluctuations rather than an accurate description of the underlying physics. Nevertheless, we retain such cases to avoid introducing selection biases.  Importantly, a low number of galaxies can sometimes be associated with strong supernova feedback—particularly high values of the $A_{\rm SN1}$ parameter, as shown in Fig.~\ref{fig:Ngal_median}. However, this is not a strict correlation: some simulations with weak feedback also yield few galaxies, and vice versa. Moreover, within the stellar mass range probed by the dwarf catalog (approximately 
$\rm log_{10}M_*/M_\odot\sim7.4$ to 8), the actual number of satellite galaxies observed in the MW is only about two. As such, models producing few galaxies in this regime remain physically plausible and cannot be excluded from the analysis on the basis of low statistics alone. 

As shown in Sects.~\ref{sec:parameters_median} and \ref{sec:evaluation fit}, our analysis of the MW zoom-in simulations allows us to constrain only the feedback parameter $A_{\rm SN1}$. Although these simulations contain a relatively small number of galaxies, the fixed cosmological background reduces parameter degeneracies, thereby improving the robustness of the constraints on 
$A_{\rm SN1}$. Next, we present our results, starting with the comparison with the SPARC catalog, and then compare them with those obtained from the uniform-box simulations. 

In Fig. \ref{fig:fit all} we show the trends from the best-fit simulation (white points) compared to the SPARC catalog. The best-fit simulation parameter configuration is the following: $A_{\rm SN1}$ = 0.85, $A_{\rm SN2}$ = 0.62, $\rm BH_{\rm FF}$ = 0.67, and $\rm P_{\rm WDM}=0.22\;\rm keV^{-1}$ (i.e., $M_{\rm WDM}$ = 4.50 keV), yielding a reduced D-squared value of $\tilde D^2$ = 1.27 (see Table \ref{tab:chi2SPARCdwarf}). The fit obtained via bootstrap resampling is consistent with our initial parameter estimates, with the exception of the
$\rm BH_{\rm FF}$. In fact, we find that (Table \ref{tab:bootstrapSPARC}) $A_{\rm SN1}= 1.31^{+0.83}_{-0.52} \; (\pm 14/22\%)$, $A_{\rm SN2}= 0.89^{+0.67}_{-0.32}  \; (\pm 21/45\%)$, $\rm BH_{\rm FF}= 1.08^{+1.23}_{-0.72}  \; (\pm 19/33 \%)$, and $\rm P_{\rm WDM}= 0.31^{+0.18}_{-0.20}\;\rm keV^{-1}\; (\pm 37/40 \%) $, which corresponds to $M_{\rm WDM}=3.56 $ keV $(\pm37/181\% $), with $\tilde{D}^2= 2.71^{+0.63}_{-0.70} $.
This estimate of $A_{\rm SN1}$ is consistent with that derived from the uniform-box analysis, with comparable uncertainties.

The best-fit simulation with respect to the dwarf catalog is shown in Fig. \ref{fig:fit all} (orange points). 
The parameters of the best-fit simulation are: $A_{\rm SN1}=0.74$, $A_{\rm SN2}=1.84$,  $\rm BH_{\rm FF}=0.56$ and $\rm P_{\rm WDM}=0.39$ keV$^{-1}$ (i.e., $M_{\rm WDM}=2.55\;\rm keV$), with a reduced D-squared value of $\tilde{D}^2$ = 0.96. As usual, the individual D-squared values for the different scaling relations are listed in Table \ref{tab:chi2SPARCdwarf}. The bootstrap analysis shows (Table \ref{tab:bootstrapSPARC}): $A_{\rm SN1}= 0.68_{-0.34}^{+0.50} \; (\pm 9/13\%) $, $A_{\rm SN2}= 0.91_{-0.34}^{+0.77}  \; (\pm 23/51\%)$, $\rm BH_{\rm FF}= 0.78_{-0.46}^{+1.14} \; (\pm 12/30\%) $, and $\rm P_{\rm WDM}=0.37_{-0.17}^{+0.10}\;\rm keV^{-1} \; (\pm 20/35\%)$, i.e., $M_{\rm WDM}= 7.04 \;\rm keV\; (\pm 22/85\%)$, with $\tilde{D}^2=2.8_{-0.9}^{+1.7}$. 
The estimate of $A_{\rm SN1}$   obtained from the comparison with the dwarf galaxy catalog is more precise than those derived from the SPARC catalog, uniform-box, and MW zoom-in simulations. However, the dwarf catalog tends to favor models with lower $A_{\rm SN1}$ values compared to the overall results from the SPARC analysis. Despite these differences, the estimates remain consistent within their respective uncertainties.

%%%%%%%%%%%%%%%%%%%%%%%%%%%%%%%%%%%%%%%%%%%%%%%%%%%%%%%%%%%%%%%%%%%%%%%%%%%%%%
%%%%%%%%%%%%%%%%%%%%%%%%%%%%%%%%%%%%%%%%%%%%%%%%%%%%%%%%%%%%%%%%%%%%%%%%%%%%%%

\section{Conclusions\label{conclusions}}
In this work, we extended the CASCO project by analyzing DREAMS hydrodynamic simulations in a WDM cosmology \citep{rose2024introducingdreamsprojectdark} and comparing them to observational data from the SPARC and LVDB catalogs. Our goal was to evaluate how cosmological and astrophysical parameters, especially the WDM particle mass, affect galactic scaling relations. We considered two complementary simulation suites—uniform-box and MW zoom-in runs—characterized by different numerical resolutions and volumes, focusing on the small-scale signatures of WDM. We began by analyzing how variations in simulation parameters affect the median trends of the scaling relations between stellar mass and four key quantities: the stellar half-mass radius, the enclosed DM mass, the enclosed total-to-stellar mass ratio, and the total mass. We then constrained simulation parameters by comparing DREAMS WDM outputs to observations from SPARC and LVDB by using a bootstrap-based fitting method developed in Papers I and II. This approach was validated across both simulation types for its ability to constrain cosmological, astrophysical, and WDM parameters. Below, we summarize our main results:
\begin{itemize}
    \item Simulated scaling relations in the uniform-box simulations, where cosmological parameters were varied, demonstrate a clear dependence on the matter density parameter. Higher values of $\Omega_{\rm m}$ systematically increase the median values across all four scaling relations. In contrast, $\sigma_8$ has milder effects: It reduces stellar sizes and total masses while increasing the DM content and total-to-stellar mass ratio below $\log_{10}\rm M_*/M_\odot \sim 10.8$, where the trend reverses. 
    \item Astrophysical parameters were varied in both simulation suites, allowing for a comparative analysis. The SN wind energy (i.e., $A_{\rm SN1}$) has the most significant impact: Increasing $A_{\rm SN1}$  leads to a systematic rise in galaxy sizes, higher total-to-stellar mass ratios, and DM masses within the stellar half-mass radius, as well as an overall increase in the total galaxy mass.  
    This effect is observed in both the uniform-box and MW zoom-in simulations, but it is more pronounced in the latter, where differences between the high and low median trends reach 0.5–0.7 dex, compared to 0.2–0.4 dex in the uniform-box suite. The SN wind velocity ($A_{\rm SN2}$) and black hole feedback factor ($\rm BH_{\rm FF}$) generally have weaker effects than $A_{\rm SN1}$ and primarily affect galaxies with higher stellar masses. As a result, the MW zoom-in simulations, which include mostly low-mass galaxies, are less sensitive to these parameters.
    The $\rm BH_{\rm FF}$ parameter is found to suppress all four scaling relations at the high-mass end, as consistently observed across both simulation suites.  
    \item Regarding the WDM particle mass,
    previous studies have shown that simulations at resolutions comparable to the DREAMS uniform boxes can retain constraining power on WDM models up to $\sim$6~keV, particularly when analyzed using population-level statistics or machine-learning techniques \citep[e.g.,][]{Rose2023, costanza2025}. In our analysis based on scaling relations that are expected to be most sensitive to the underlying DM physics, we find no significant differences in the uniform-box simulations across the explored WDM mass range. By contrast, slight trends in $M_{\rm tot,1/2}/M_{*,1/2}$  and $M_{\rm DM,1/2}$ emerge in the MW zoom-in simulations, suggesting that higher-resolution simulations could offer valuable insights and represent a promising direction for future study.  
    
    \item Moreover, although the MW zoom-in and uniform-box simulations exhibit similar overall behaviors, a direct comparison between the MW zoom-in trends and those of the calibrated uniform-box suite also indicates that high-resolution simulations produce galaxies with smaller stellar half-mass radii as well as lower values of $M_{\rm DM,1/2}$ and $M_{\rm tot,1/2}/M_{*,1/2}$. The total mass versus stellar mass trend, however, remains smoothly aligned across both simulation types. These systematic shifts are likely driven by resolution effects and environmental factors, among other influences, that can impact the underlying astrophysical processes.
    \item We also analyzed galaxy abundance as a function of total stellar mass, showing that this relation may help disentangle the effects of WDM particle mass from those of other parameters due to its strong dependence on total stellar mass. 
    \item We confirm the presence of a golden mass,
a characteristic stellar mass scale where star formation efficiency
peaks, in the uniform-box WDM simulations, which show more clearly defined minima compared to \camels\ CDM simulations at the same resolution.  
    \item Internal validation of the bootstrap-based fitting strategy was performed by accounting for the characteristics of the observational catalogs used for comparison. While this approach does not reflect the method’s full potential when applied to the two simulation suites, it offers a realistic estimate of its ability to constrain the model parameters. In the uniform-box suite, the most robustly constrained parameters are $\Omega_{\rm m}$ and $A_{\rm SN1}$. The parameter $\sigma_8$ is reasonably well recovered, though with larger uncertainties, while $A_{\rm SN2}$ passes the correlation test but fails to reproduce an ideal one-to-one relation with the ground truth. In the MW zoom-in simulations, $A_{\rm SN1}$ shows significant correlation with the target values in both the SPARC-analog and dwarf-analog cases, though the one-to-one relation is only closely approached in the former. This reduced performance is likely due to the limited statistics of the MW zoom-in sample, which hinders tighter constraints. Finally, the WDM particle mass cannot be constrained.

    \item Comparison between the simulations and observational catalogs yielded the following results. In the uniform-box simulations, $\Omega_{\rm m} = 0.28^{+0.08}_{-0.04}$ is 
   in agreement with the value reported for the \camels\ simulations analyzed  in Paper I ($0.27^{+0.01}_{-0.05}$), with uncertainties at the 10–20$\%$ level. The estimate for $\sigma_8 = 0.89^{+0.10}_{-0.18}$ is slightly higher than in the earlier analysis ($0.83^{+0.08}_{-0.11}$), though still compatible within 1$\sigma$, resulting in a comparable $S_8$. The feedback parameter $A_{\rm SN1}$ is significantly higher than in Paper I ($1.67^{+0.49}_{-0.84}$ versus $0.48^{+0.25}_{-0.16}$), although it is more loosely constrained. In contrast, $A_{\rm SN2} = 1.46^{+0.32}_{-0.21}$ shows a milder increase compared to the previous estimate ($1.21^{+0.03}_{-0.34}$). This discrepancy may stem from the fact that the \camels\ and DREAMS simulations change  different AGN feedback parameters—resulting in negligible effects in the former and more significant impacts in the latter—making a direct comparison unfeasible. For the MW zoom-in simulations, only $A_{\rm SN1}$ is meaningfully constrained. Its estimates—$1.31^{+0.83}_{-0.52}$ from the SPARC catalog and $0.68^{+0.50}_{-0.34}$ from the dwarf catalog—are consistent with the uniform-box result of $1.67^{+0.49}_{-0.84}$. The dwarf catalog provides the tightest constraint, although it tends to prefer lower values, closer to the estimate from Paper I.
\end{itemize}
Our analysis highlights the value of combining simulations with different resolutions and observational catalogs that probe complementary stellar mass ranges. While we find that the current constraints are insufficient to determine the WDM particle mass, our results in a WDM framework reveal subtle mass-dependent effects, particularly in the galaxy abundance as a function of stellar mass, that may help disentangle the influence of WDM from other parameters. This suggests that a multiscale approach, leveraging both zoom-in and uniform-box simulations, is a promising avenue for future studies aiming to constrain the physics of structure formation in nonstandard cosmologies. 
In future work, we plan to extend our analysis by examining the DM density profiles of different components and by quantifying enclosed masses within radii smaller than the effective radius ($r < R_{*,1/2}$). These inner regions are expected to be more sensitive to the suppression of small-scale power induced by WDM, especially  in low-mass (dwarf) galaxies. Furthermore, we will investigate whether the core-like structures predicted in DM-only WDM simulations \citep{Maccio2012,Maccio2013} are preserved or altered in the presence of baryonic feedback. We also aim to explore  the redshift and environmental dependence of these trends, which may further enhance the discriminatory power between WDM and CDM models. In this context, upcoming large-scale surveys, such as Euclid \citep{EuclidPerseus2024,EuclidQ12025}, with their ability to map the faint end of the luminosity function with unprecedented precision, will be instrumental. When combined with future improvements in hydrodynamic simulations—including enhanced resolution and refined baryonic physics—these observational datasets will allow us to better probe the role of WDM and its viability as an alternative to CDM.

\begin{acknowledgements}
All the calculations underlying this work have been performed via the use of Wolfram Mathematica ver. 13.1.
A.M.B. acknowledges support from grant FI-CCA-Research-00011826 from the Simons Foundation.
A.F. acknowledges support from the National Science Foundation under Cooperative Agreement 2421782 and the Simons Foundation award MPS-AI0001051. N.R.N. acknowledges support from the Guangdong Science Foundation grant (ID: 2022A1515012251).  P.T. acknowledges support from NSF-AST 2346977 and the NSF-Simons AI Institute for Cosmic Origins which is supported by the National Science Foundation under Cooperative Agreement 2421782 and the Simons Foundation award MPS-AI-00010515.
\end{acknowledgements}

\bibliography{bibliography}

\begin{appendix}    
\section{Calibration of the uniform-box simulation suite with respect to TNG100-1\label{calibrazione}}
In this section, we describe the calibration procedure adopted for the lower-resolution uniform-box DREAMS simulations, based on the fiducial TNG model. This calibration  is essential to reproduce observational trends more accurately.

Numerical resolution significantly impacts the effectiveness of astrophysical feedback processes. Lower resolution implies more massive gas particles, leading to a less accurate representation of the gas distribution. This limits the ability to resolve star formation, which occurs in dense, small-scale environments. Furthermore, feedback models, such as those associated with SN energy release, may become ineffective or poorly modeled, as the energy is distributed over excessively large masses. Higher resolution enables better physical modeling and more efficient feedback implementation.
To account for these limitations, we calibrated the WDM uniform-box simulations of DREAMS using the TNG100-1 fiducial run as a reference. TNG100-1 features a DM particle mass of $5.1\times 10^6 \;h^{-1}\;\rm M_\odot$, a baryonic mass resolution of $9.4\times 10^5 \;h^{-1}\;\rm M_\odot$, and a comoving box size of $75h^{-1}$Mpc per side. As no WDM uniform-box simulation is available for direct comparison, we use the 1P \footnote{We note that 1P simulations are only available for the uniform-box suite.} WDM 16 keV simulation as a proxy. These simulations are specifically designed to isolate the impact of individual parameters by varying one at a time while keeping all others fixed. In this case, the only difference from the reference setup is the WDM particle mass.
\begin{figure}[ht]
    \centering
     \includegraphics[width=0.48
    \textwidth]{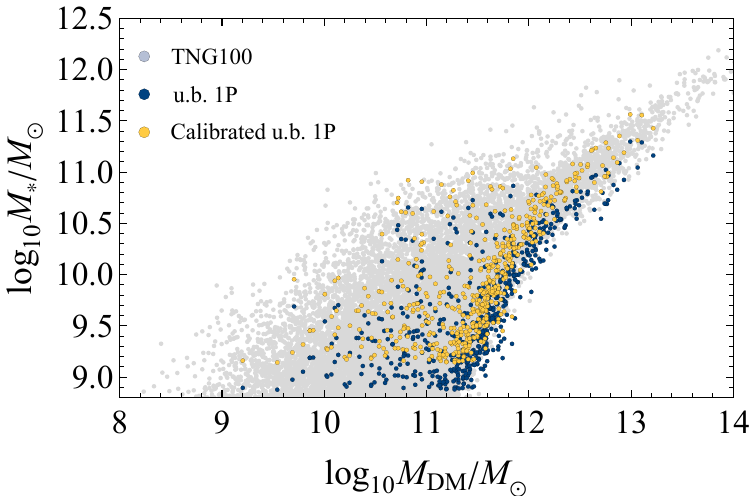}
        \caption{ Stellar-DM mass relation for the  WDM  uniform-box 1P DREAMS simulation with a 16 keV WDM particle mass (blue), its calibrated counterpart (yellow), and the TNG100-1 fiducial run (light gray), used as reference.}
  \label{fig:calibration}
  \end{figure}  
  At this resolution, we verified that the WDM mass parameter has a negligible impact on the relevant physical processes (see Sect. \ref{sec:parameters_median}). The calibration is based on the stellar-to-DM mass relation ($M_* $ versus $M_{\rm DM}$) and proceeds as follows:
 \begin{itemize}
     \item TNG100-1 interpolation: we compute the median, 16th, and 84th percentiles of the 
     $M_*$ versus $M_{\rm DM}$ relation from TNG100-1 and generate an interpolating function of $M_*$  as a function of $M_{\rm DM}$.
\item Cost function definition: for each object in the WDM simulation (1P, 16 keV), we evaluate the squared difference between the stellar mass (shifted by a constant $\rm c_{M_*}$ in log) and the interpolated TNG100-1 prediction, normalized by the TNG100-1 uncertainty (average between 16th and 84th percentiles) at the same $M_{\rm DM}$.
\item Filtering: objects outside the interpolation validity range are excluded.
\item Minimization: we determine the optimal logarithmic shift $\rm c_{M_*}$ by minimizing the sum of the normalized residuals. The resulting offset is $\rm c_{M_*}$ $\sim 0.27$ in $\rm log_{10} M_*/M_\odot$.
 \end{itemize}
%%%%%%%%%%%%%%%%%%%%%%%%%%%%%%%%%%%%%%%%%%%%%%%%%%%%%%%%%%%%%%%%%%%%%%%%%%%%%%%%%%%%%%%%%%%%%%%%%%%%%%%%%%%%%%%%%%%%%%%%%%%%%%%%%%%%%%%%%%%%%%%% 
Finally, we rescale stellar-mass-dependent astrophysical quantities as follows:
\begin{itemize}
    \item $\rm{log}_{10} M_*/M_\odot\rightarrow \mathrm{log}_{10} M_*/M_\odot+c_{M_*}$;
    \item $\rm{log}_{10} M_{\mathrm{tot}}/M_\odot \rightarrow \mathrm{ log}_{10} ((10^{c_{M_*}} M_*+M_{\mathrm{DM}}+M_{\mathrm{gas}}+M_{\mathrm{BH}})/M_\odot$);
    \item $f_{\rm{DM},1/2}\rightarrow \frac{M_{\rm {DM},1/2}}{ 10^{\mathrm{c_{M_*}}} M_{*,1/2}+ M_{\mathrm {DM},1/2}+ M_{\mathrm{BH},1/2}+ M_{\mathrm{ gas},1/2}}$.
\end{itemize}
%%%%%%%%%%%%%%%%%%%%%%%%%%%%%%%%%%%%%%%%%%%%%%%%%%%%%%%%%%%%%%%%%%%%%%%%%%%%%%%%%%%%%%%%%%%%%%%%%%%%%%%%%%%%%%%%%%%%%%%%%%%%%%%%%%%%%%%%%%%%%%%%
Figure \ref{fig:calibration} shows the stellar-DM mass relation before and after applying the calibration.  While this procedure effectively mitigates resolution effects on the $M_*$ versus $ M_{\rm DM}$ relation, it is important to note that the calibration is performed a posteriori and tailored to the TNG100-1 reference run. Nevertheless, it consists solely of a rigid shift in stellar mass and does not involve tuning of the simulation parameters themselves. Therefore, it does not trivially favor the reference configuration when analyzing other scaling relations. Indeed, the effects of physical parameters in the simulations are generally non linear and mass-dependent, whereas the applied correction is uniform across all stellar masses.
In Fig. \ref{fig:calibration_scalings}, we present the median values together with the 16th and 84th percentiles for the same set of scaling relations adopted in our analysis, for both the reference simulation TNG100-1 and the 1P run at 16 keV, before and after the stellar-mass calibration.
We note that the calibration described in this section brings the DREAMS uniform-box simulation into much closer agreement with the reference model, with a substantial overlap between the two. This indicates that a systematic shift in $M_*$ has a generally beneficial effect on the other reference scaling relations as well.
%%%%%%%%%%%%%%%%%%%%%%%%%%%%%%%%%%%%%%%%%%%%%%%%%%%%%%%%%%%%%%%%%%%%%%%%%%%%%%%%%%%%%%%%%%%%%%%%%%%%%%%%%%%%%%%%%%%%%%%%%%%%%%%%%%%%%%%%%%%%%%%%
\begin{figure}[ht]
    \centering
     \includegraphics[width=0.48
    \textwidth]{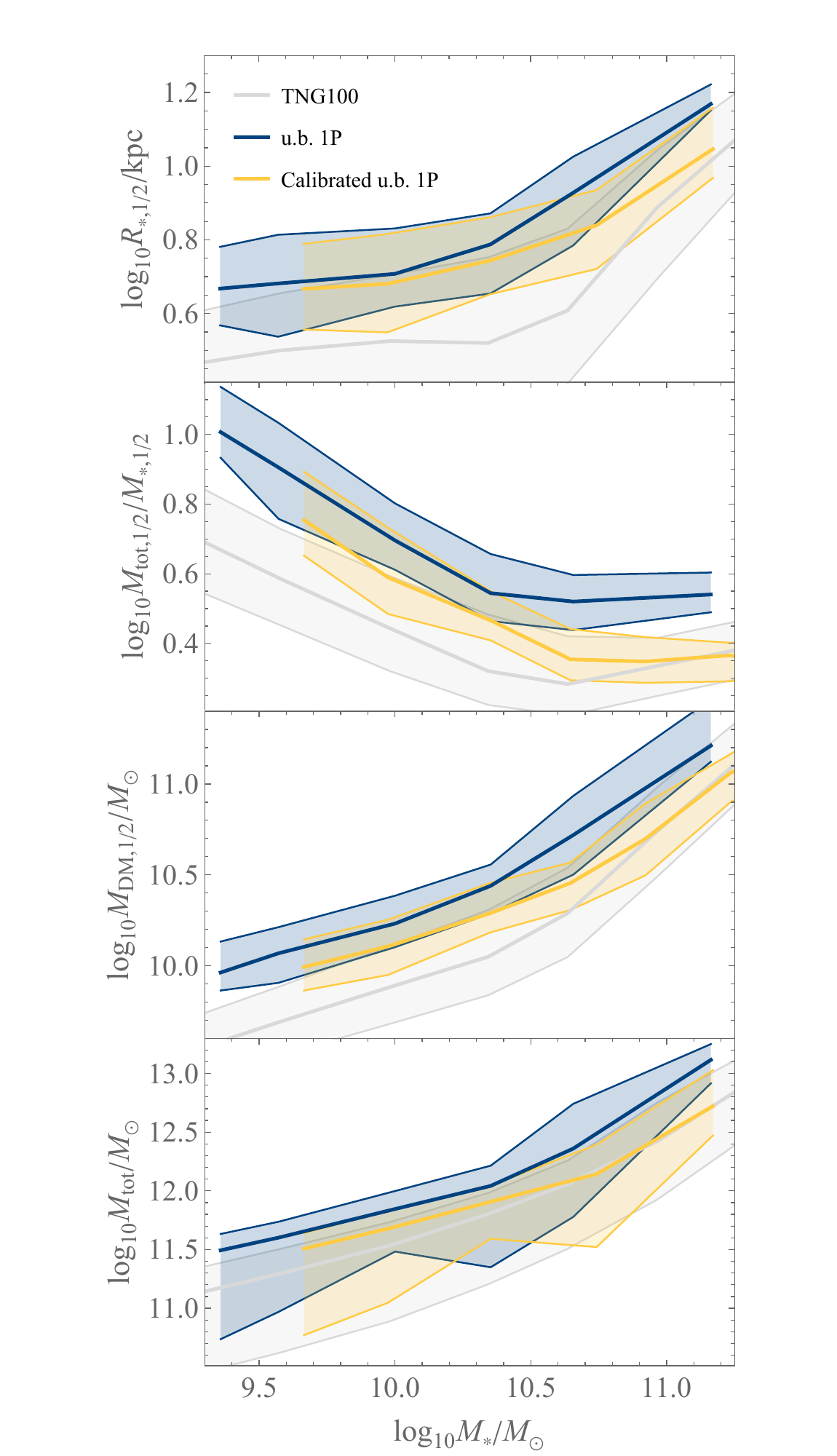}
        \caption{ Median scaling relations with 16th and 84th percentiles for the reference simulation TNG100-1 (light gray), the uncalibrated 1P run at 16 keV (blue), and the stellar-mass–calibrated 1P run (yellow).}
  \label{fig:calibration_scalings}
  \end{figure}  %%%%%%%%%%%%%%%%%%%%%%%%%%%%%%%%%%%%%%%%%%%%%%%%%%%%%%%%%%%%%%%%%%%%%%%%%%%%%%%%%%%%%%%%%%%%%%%%%%%%%%%%%%%%%%%%%%%%%%%%%%%%%%%%%%%%%%%%%%%%%%%%%%%%%%%%%%%%%%%%%%%%%%%%%%%%%%%%%%%%%%%%%%%%%%%%%%%%%%%%%%%%%%%%%%%%%%%%%%%%%%%%%%%%%%%%%%%%%%%%%%%
\section{Calibration of the uniform-box simulation suite with respect to the MW zoom-ins\label{app:cal_MW}}%%%%%%%%%%%%%%%%%%%%%%%%%%%%%%%%%%%%%%%%%%%%%%%%%%%%%%%%%%%%%%
%%%%%%%%%%%%%%%%%%%%%%%%%%%%%%%%%%%%%%%%%%%%%%%%%%%%%%%%%%%%%%
\begin{figure}[ht]
    \centering
     \includegraphics[width=0.48
    \textwidth]{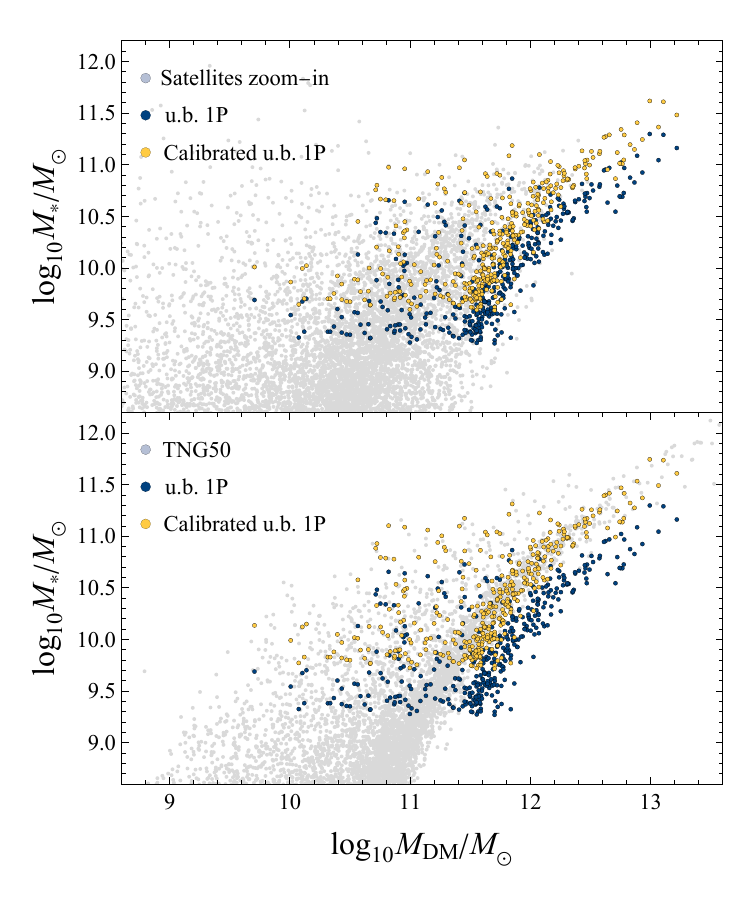}
        \caption{ Stellar–DM mass relation for the WDM uniform-box 1P DREAMS simulation with a 16 keV WDM particle mass (blue), its stellar-mass–calibrated counterpart (yellow), and the corresponding reference samples (gray). The upper panel shows the calibration relative to the satellite galaxies from the MW zoom-in simulations, while the lower panel shows the calibration relative to TNG50-1.}
  \label{CALMWTNG50}
  \end{figure}  %%%%%%%%%%%%%%%%%%%%%%%%%%%%%%%%%%%%%%%%%%%%%%%%%%%%%%%%%%%%  %%%%%%%%%%%%%%%%%%%%%%%%%%%%%%%%%%%%%%%%%%%%%%%%%%%%%%%%%%%%
\begin{figure}[ht]
    \centering
     \includegraphics[width=0.49
    \textwidth]{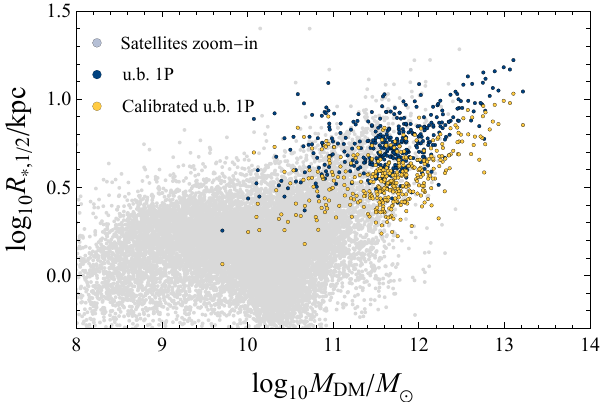}
        \caption{Stellar half-mass radius–DM mass relation for the WDM uniform-box 1P DREAMS simulation with a 16 keV WDM particle mass (blue), its calibrated counterpart (yellow), and the satellite galaxies from the MW zoom-in simulations used as reference (gray). The equivalent plot for the stellar-mass calibration is shown in the left panel of Fig.~\ref{CALMWTNG50}.}
  \label{CAL_RStar}
  \end{figure}
  %%%%%%%%%%%%%%%%%%%%%%%%%%%%%%%%%%%%%%%%%%%%%%%%%%%
  %%%%%%%%%%%%%%%%%%%%%%%%%%%%%%%%%%%%%%%%%%%%%%%%%%%
  \begin{figure}[ht]
    \centering
     \includegraphics[width=0.48
    \textwidth]{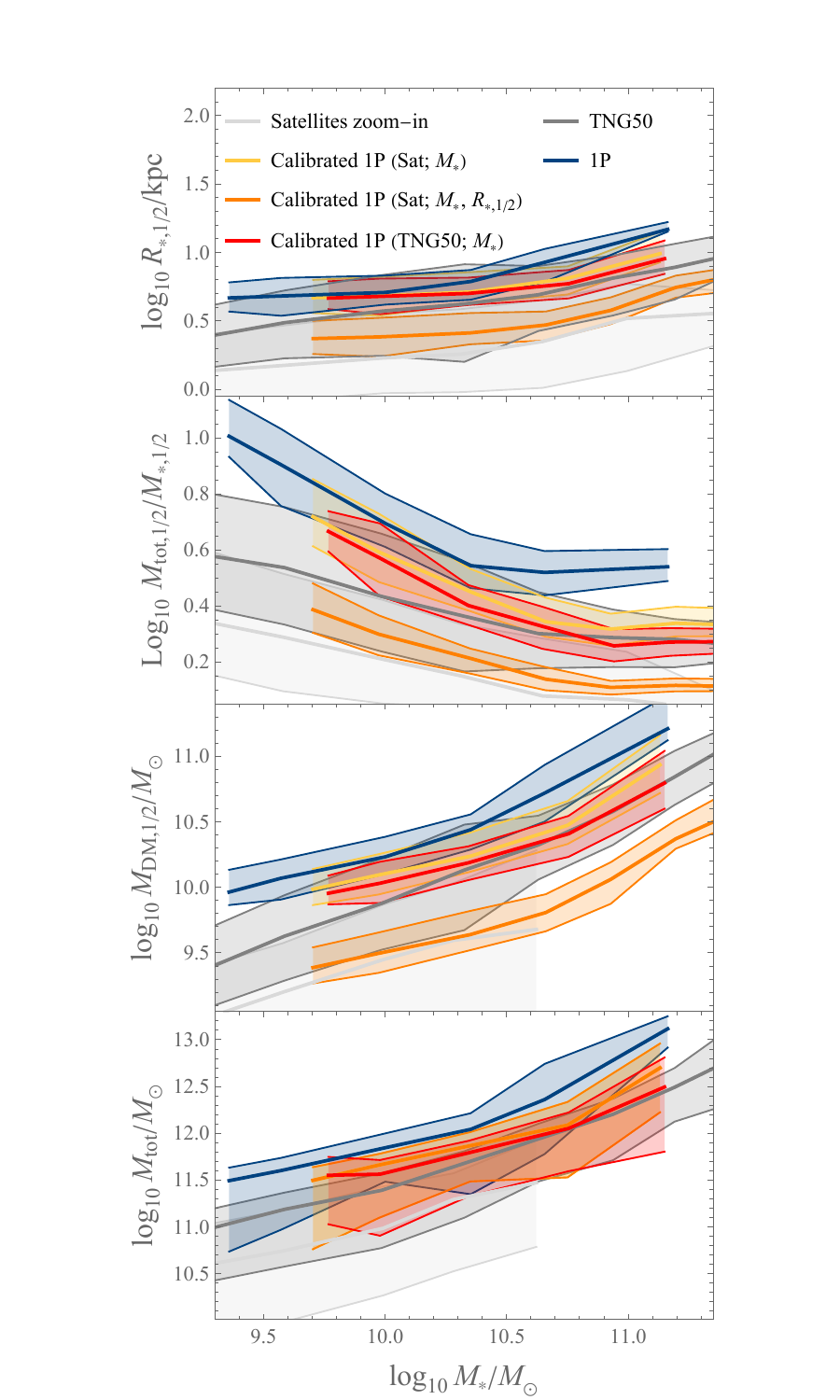}
        \caption{ Median scaling relations with 16th and 84th percentiles for the MW zoom-in satellites (light gray), TNG50-1 (dark gray), the uncalibrated 1P run (blue), the 1P run calibrated on stellar mass with respect to the MW zoom-ins (yellow), the same run further calibrated on stellar half-mass radius (orange), and the 1P run calibrated on stellar mass with respect to TNG50-1 (red).}
  \label{fig:calibration_scalings_MW_TNG50}
  \end{figure}
In this section we extend the analysis presented in \App\ref{calibrazione} by performing a two-step calibration of the standard uniform-box 1P simulation with a 16~keV WDM particle mass, using as reference the population of satellite galaxies extracted from the 1024 MW zoom-in simulations.\footnote{All simulated galaxies are selected according to the standard selection criteria (see Sect
. \ref{sec: sel simulations}).}
We refer to this procedure as a double calibration, as it is carried out sequentially on the stellar mass, $M_*$, and on the stellar half-mass radius, $R_{*,1/2}$.

  %%%%%%%%%%%%%%%%%%%%%%%%%%%%%%%%%%%%%%%%%%%%%%%%%%%%%%%%%%%%
  %%%%%%%%%%%%%%%%%%%%%%%%%%%%%%%%%%%%%%%%%%%%%%%%%%%%%%%%%%%%%

\subsection{Stellar-mass calibration \label{Mstarcal}}

We calibrate the stellar mass with respect to the satellites of the MW zoom-in simulations, following the same procedure outlined in \App\ref{calibrazione}.
We find a best-fitting offset of $\rm c_{M_*} = 0.32$ in $\log_{10}\rm M_*/M_\odot$ (see upper panel of Fig.~\ref{CALMWTNG50}), close to the value obtained in \App\ref{calibrazione} (i.e., $\rm c_{M_*}=0.27$). After applying this correction, the median scaling relations of the calibrated uniform-box simulation (yellow) remain systematically above those of the MW zoom-in satellites (light gray), especially for relations involving the stellar half-mass radius $R_{*,1/2}$ (see Fig.~\ref{CAL_RStar}).
This residual discrepancy may arise from a combination of factors, including resolution effects that can limit the modeling of small-scale baryonic structure, as well as physical or environmental differences between the two simulation suites.

To further test the impact of resolution, we repeated the calibration using the TNG50-1 run as reference.
TNG50-1 has a spatial and mass resolution comparable to that of the MW zoom-in simulations, and therefore provides an appropriate reference for this comparison.
Using TNG50-1 as reference, we calibrated the stellar mass of the uniform-box run (see lower panel of Fig.~\ref{CALMWTNG50}); the resulting shift, $\rm c_{M_*}=0.45$, is slightly higher but of the same order as those obtained for TNG100-1 and the MW zoom-ins.
The resulting scaling relations of the calibrated uniform-box simulation (red) closely follow those of TNG50-1 (dark gray; see Fig.~\ref{fig:calibration_scalings_MW_TNG50}).
It is plausible that environmental factors, inherent to the MW zoom-in sample of MW-mass haloes, also contribute to the residual offsets at fixed stellar mass.
To further explore this aspect, we examined the uniform-box runs and found that regions of higher local density—for example, spheres of 350~kpc radius containing at least ten neighboring galaxies—tend to host systems with lower total mass at fixed stellar mass. Although only indicative, this behavior points to a possible environmental contribution to the differences observed between the two simulation suites.

\subsection{Stellar half-mass radius calibration\label{APP:cal_R}}

Building on the stellar mass correction ($\rm c_{M_*}=0.32$; see Appendix~\ref{Mstarcal}), we identified residual offsets in the size–mass relation and in quantities measured within $R_{*,1/2}$. To mitigate these, we used the $R_{*,1/2}$ versus $M_{\rm DM}$ relation from the MW zoom-in satellites (see Fig.~\ref{CAL_RStar}), which yields a logarithmic shift of $c_{\rm R}=-0.30$, corresponding to:
\begin{equation}
\log_{10}R_{*,1/2}/ \mathrm{ kpc} \rightarrow \log_{10} R_{*,1/2}/\mathrm{ kpc}+ c_{\mathrm R}\,.
\end{equation}
A calibration on $R_{*,1/2}$ implies corresponding adjustments to all quantities defined within this radius.
The stellar mass within $R_{*,1/2}$ is easily corrected, since $M_{*,1/2}\equiv\tfrac{1}{2}M_*$ and therefore follows the same scaling as $M_*$.
For the DM mass within the stellar half-mass radius, we assume a simple power-law dependence around $R_e$, $M_{\rm DM}\propto r^{\eta}$, where $\eta=2$ corresponds to an NFW profile and $\eta=1.2$ to a contracted NFW profile \citep{Boylan2005,Tortora2018}.
Accordingly,
\begin{equation}
\log_{10}M_{\rm DM,1/2}/M_\odot\rightarrow \log_{10} M_{\rm DM,1/2}/M_\odot+ \eta c_R\,.
\end{equation}
Throughout the analysis we adopt $\eta=2$, noting that the differences with a contracted NFW assumption are marginal. Finally, when computing the total mass enclosed within $R_{*,1/2}$, we do not rescale the black hole and gas masses, as the resulting error is negligible. For completeness, we note that recalibrating these components would have only a minor effect, since their median contributions to the total mass within $R_{*,1/2}$ in the simulations are about $0.2\%$ for the black hole and $6\%$ for the gas. The 1P simulation calibrated on both $M_*$ and $R_{*,1/2}$ is represented by the orange trends in Fig. \ref{fig:calibration_scalings_MW_TNG50}, which closely follow those of the MW zoom-in satellites (light gray trends). For comparison with the other results in this paper, we performed a bootstrap fit of the uniform-box simulations to the SPARC catalog after the double calibration. In this case, the resulting $A_{\rm SN1}$ value is significantly higher, around 3.5. This occurs because the double calibration causes the uniform box trends to be systematically underestimated, requiring a higher $A_{\rm SN1}$ to raise the trends and match the SPARC data.

%%%%%%%%%%%%%%%%%%%%%%%%%%%%%%%%%%%%%%%%%%%%%%%%%%%%
%%%%%%%%%%%%%%%%%%%%%%%%%%%%%%%%%%%%%%%%%%%%%%%%%%%%
%%%%%%%%%%%%%%%%%%%%%%%%%%%%%%%%%%%%%%%%%%%%%%%%%%%%

%%%%%%%%%%%%%%%%%%%%%%%%%%%%%%%%%%%%%%%%%%%%%%%%%%%%
%%%%%%%%%%%%%%%%%%%%%%%%%%%%%%%%%%%%%%%%%%%%%%%%%%%%
%%%%%%%%%%%%%%%%%%%%%%%%%%%%%%%%%%%%%%%%%%%%%%%%%%%%

\section{Determination of uncertainties on simulated galaxy quantities\label{app:ErrReff}}

In this section, we evaluate the typical uncertainties of relevant simulated galaxy quantities (e.g., effective radius) as a function of galaxy stellar mass.  We apply this procedure for the galaxies in both MW zoom-in and uniform-box simulations, in order 
to evaluate quantitatively how the low particle number affects the knowledge of these quantities.

For this analysis, we both assume that light follows mass and spherical symmetry. We consider a galaxy as a collection of randomly extracted point particles from the following deprojected Sérsic distribution (e.g., \citealt{PS97}):
\begin{equation}
    L(<x)=L_{\textrm{tot}}\,\frac{\Gamma \left(n\,(3-p(n)),0,x^{1/n}\right)}{\Gamma (n\,(3-p(n)))}\,,
\end{equation}
where $p(n)\approxeq1-0.6\,n^{-1}+0.05\,n^{-2}$, $n$ is the Sérsic index, $L_{\textrm{tot}}$ is the total luminosity of the galaxy,

\begin{equation}
    x=\frac{r}{R_{\textrm{e}}}\,b^{n}(n)\label{eq:adimensional_radius_deproj_Sersic}\;,
\end{equation}
\begin{figure}[ht]
    \centering
     \includegraphics[width=0.48
    \textwidth]{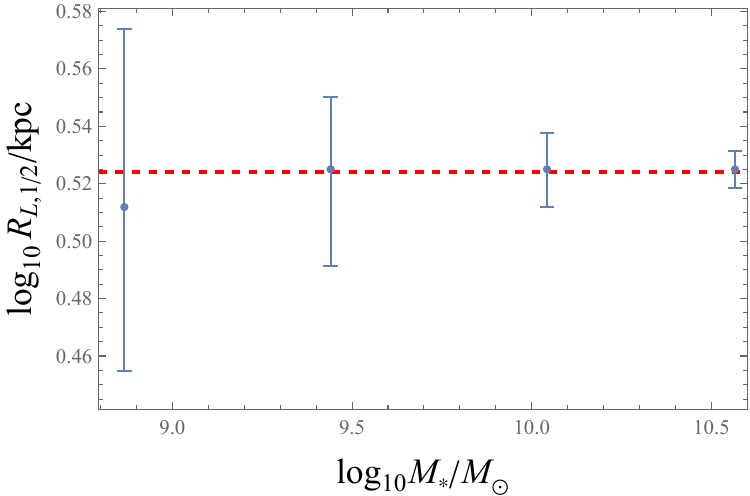}
        \caption{ Estimate and uncertainty of the half-light radius for galaxies in the DREAMS uniform-box simulations, computed for four different fixed total numbers of stellar mass particles ($N_{\textrm{par}} = [40, 150, 600, 2000]$). The dashed red line marks the true half-light radius. }
  \label{fig:errorReff}
  \end{figure}
is an adimensional radius (with $b(n)\approxeq2\,n-1/3+9.9\,\times10^{-3}\,n^{-1}$ and $R_{\textrm{e}}$ projected effective radius), $\Gamma(z_{0},z_{1},z_{2}) := \Gamma(z_{0},z_{1})-\Gamma(z_{0},z_{2})$ is the generalized incomplete Gamma function, $\Gamma(z_{0},z_{1})$ the incomplete Gamma function, and $\Gamma(z)$ is the Euler Gamma Function. The deprojected effective radius can thus be obtained by solving $L(<x)=0.5$ for $x$ and then substituting the obtained value in Eq. \eqref{eq:adimensional_radius_deproj_Sersic}.

To find the uncertainties on the deprojected effective radius, we start by considering fixed values for the projected effective radius and Sérsic index equal to $\log_{10}(R_{\textrm{e}}/\rm kpc) = 0.4$ (which is roughly equal to the median of the dwarfs' effective radii in Fig. \ref{fig:MW zoom}) and $n=1$ (noting that the Sérsic index of dwarf galaxies is typically between $0.5$ and $1.5$, see for example \citet[][Fig. 13]{Lazar2024} while for the SFGs is $\sim 1$, \citealt{Ward2024}).

In the case of the uniform-box  simulations, for example (the procedure is analogous for the MW zoom-in simulation), we select 
four  different fixed total numbers of stellar mass particles ($N_{\textrm{par}} = [40, 150, 600,2000]$), which correspond to a mass range between $\rm log_{10} M_*/M_\odot\sim 8.8$ and $10.6$. For each value of $N_{\textrm{par}}$, we produce $N_{\textrm{gal}} = 400$ random galaxies. For all these galaxies, we then compute the respective normalized growth curve, obtained by counting the number of particles within a generic radius $r$ and then dividing the count by the total number of particles. We can then estimate the median deprojected effective radius for a galaxy with number of particles equal to a given $N_{\textrm{par}}$ by finding the empirical deprojected effective radius (i.e., the nearest $r$ value such that the normalized growth curve is equal to $0.5$) for each of the $N_{\textrm{gal}}$ randomly produced galaxies, and evaluating the median and quantiles of these radii. The trend of the effective radii estimated from this procedure as a function of stellar mass is shown in Fig. \ref{fig:errorReff}. As expected, the uncertainties are larger when the number of stellar mass particles is lower. The maximum error on the effective radius, when the total number of stellar mass particles is 40, does not exceed 0.05 dex.

Given the uncertainty on the deprojected effective radius, one can determine the uncertainties on all other derivated quantities: for example, the uncertainty on the stellar mass within the deprojected effective radius can be obtained by considering for each simulated galaxy the number of particles within the empirical deprojected effective radius, and then converting to stellar mass by multiplying the results for the resolution of the star particles. Uncertainty on the DM mass within the deprojected effective radius can then be obtained by considering the $M_{\textrm{DM},1/2}/M_{*,1/2}$ values corresponding to a certain number of particles, and proceed by error propagation by multiplying this ratio for the estimated $M_{*,1/2}$. Similarly, the uncertainties on the total mass within the deprojected half-light radius can be obtained from the expected ratio, $M_{\rm tot,1/2}/M_{*,1/2}$,  for a given number of particles. Since we already have an estimate of the uncertainty on $M_{*,1/2}$, we can then propagate this to determine the uncertainty on the total-to-stellar half-mass ratio. We observe that the propagated uncertainties on $M_{\rm DM,1/2}$ and on $M_{\rm tot,1/2}/M_{*,1/2}$ are also very small and do not exceed 0.04 dex. The uncertainty estimates derived in this section are incorporated in the analysis presented in Sect.~\ref{sec:parameters_median}, where they are used to propagate errors on quantities computed within the stellar half-mass radius.  In that section, the resulting uncertainties are combined in quadrature with the statistical uncertainty on the medians and shown as shaded regions in Figs.~\ref{fig: boxes cosm trend}, \ref{fig: boxes astro trend}, and \ref{fig:MW zoom}.

\section{Analysis of the golden mass in DREAMS and \camels\ \label{sec:goldenmass}}

\begin{figure}[ht]
    \centering
     \includegraphics[width=0.49
    \textwidth]{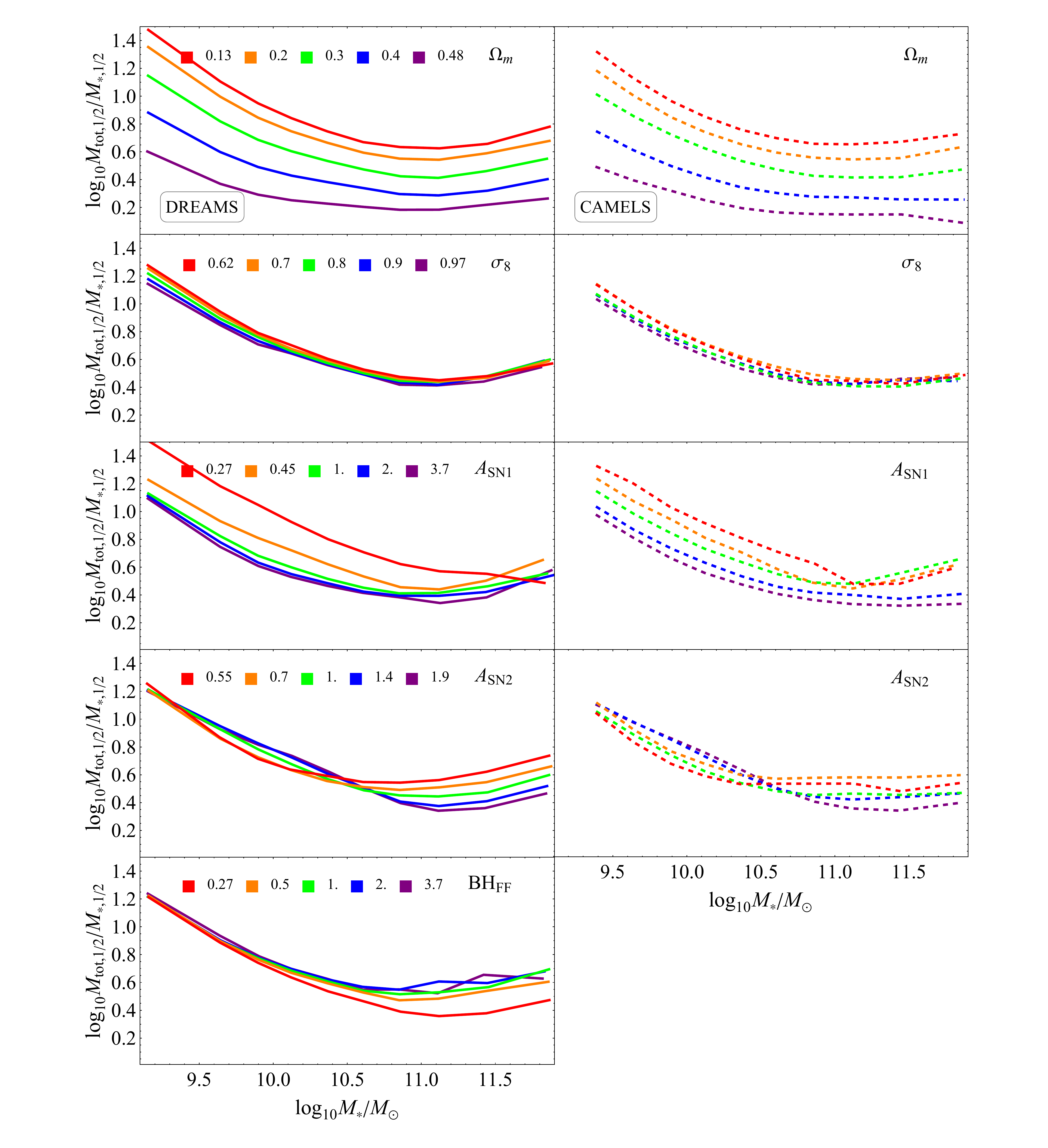}
        \caption{ Median trends of the total-to-stellar half-mass ratio as a function of total stellar mass, for different cosmological and astrophysical parameters in the DREAMS (left, thick lines) and \camels\ (right, dashed lines) simulations. Colored curves correspond to subsets of simulations grouped by parameter value; the legend indicates the median value for each group.} 
  \label{fig:goldenmassDC}
  \end{figure}

In this section, we investigate the so-called golden mass, the characteristic stellar mass scale at which the ratio $M_{\rm tot,1/2}/M_{*,1/2}$   reaches a minimum. This minimum marks the peak of star formation efficiency and a transition in galaxy formation physics \citep{CASCOIII}. 
The analysis presented in this work is based on the DREAMS and \camels\ uniform-box simulations, which have a resolution comparable to that of TNG300 within a volume of $(25\;h^{-1}\rm Mpc)^3$. These results are then compared to other \camels\ simulations with the same resolution but covering a larger volume of $(50\;h^{-1}\rm Mpc)^3$, as presented in Paper~III.

We note that, unlike Paper III, which defines the golden mass using the ratio $M_{\rm DM,1/2}/M_{*,1/2}$, we adopt $M_{\rm tot,1/2}/M_{*,1/2}$ throughout our analysis. This choice includes the contributions from baryons in the numerator, yielding slightly higher values for the ratio. However, as explicitly stated in footnote 4 of Paper III, the effect of using the total mass instead of the DM mass within the same aperture leads to an increase in the golden mass of at most 0.05–0.1 dex and does not affect the physical conclusions. 

Figure~\ref{fig:goldenmassDC} illustrates how the total-to-stellar half-mass ratio responds to variations in cosmological and astrophysical parameters, with results shown for DREAMS (left) and \camels\ (right).  Since the parameter distributions differ between DREAMS and \camels, the comparison between CDM and WDM is not one-to-one, and differences in the resulting curves are partly driven by sampling choices. Moreover, cosmic variance may also contribute to the observed discrepancies. 
The golden mass is identified by fitting a parabola to the median trend of the ratio as a function of stellar mass. However, not all parameter choices yield a clear minimum. In several cases, especially for extreme parameter values, the trend is either monotonic or only mildly curved, indicating that the golden mass is not always well-defined. When the minimum is identifiable, its location varies systematically with the physical or cosmological parameter under consideration. For the DREAMS simulations, minima are found around $\rm log_{10} M_*/M_\odot \sim$11.1 for variations in both $\Omega_{\rm m}$ and $\sigma_8$, and approximately 11.0 for moderate values of $A_{\rm SN1}$. Higher values of $A_{\rm SN1}$ often do not show a detectable minimum. $A_{\rm SN2}$ shows a gradual shift in the minimum, ranging from 11.2 (stronger winds) to 10.8 (weaker winds). AGN feedback, parametrized by $\rm BH_{\rm FF}$, produces minima between 10.8 and 11.2 when present. In the \camels\ simulations, the minima  are located at even higher stellar masses than in Paper III. The golden mass occurs around $\rm log_{10} M_*/M_\odot \sim$11.2 for varying $\Omega_{\rm m}$, and between  11.3 and 11.4 for changes in $\sigma_8$. For $A_{\rm SN2}$, the minimum shifts from 11.4 at weak feedback to 10.6 for the strongest winds. In the case of $A_{\rm SN1}$, when a minimum is present, it lies between 11.4 and 11.5.

Compared to Paper III, which consistently reports a golden mass near $\rm log_{10} M_*/M_\odot \sim$10.6, the values found in both DREAMS and \camels\ uniform boxes are systematically higher. 
Our analysis confirms the presence of a golden mass in both CDM and WDM scenarios, with trends that qualitatively agree with those reported in Paper~III. Interestingly, WDM models tend to show more regular and well-defined minima in $M_{\rm tot,1/2}/M_{*,1/2}$, possibly reflecting a smoother buildup of central mass in the absence of small-scale structure. Despite the differing parameter distributions in DREAMS and \camels\, the median trends consistently highlight the golden mass as a robust feature of galaxy formation, shaped by both feedback and the DM model. The higher golden masses observed in smaller-volume simulations relative to Paper III is likely due to volume-dependent sampling effects. Larger volumes sample intermediate- and high-mass galaxies more thoroughly, allowing a more accurate determination of the minimum in $M_{\rm tot,1/2}/M_{*,1/2}$ at slightly lower stellar masses. This may explain the systematic shift without invoking differences in underlying physics.

\section{Properties of host galaxies in the MW zoom-in simulations\label{App:Host}}

\begin{figure*}[ht]
    \centering
     \includegraphics[width=1
    \textwidth]{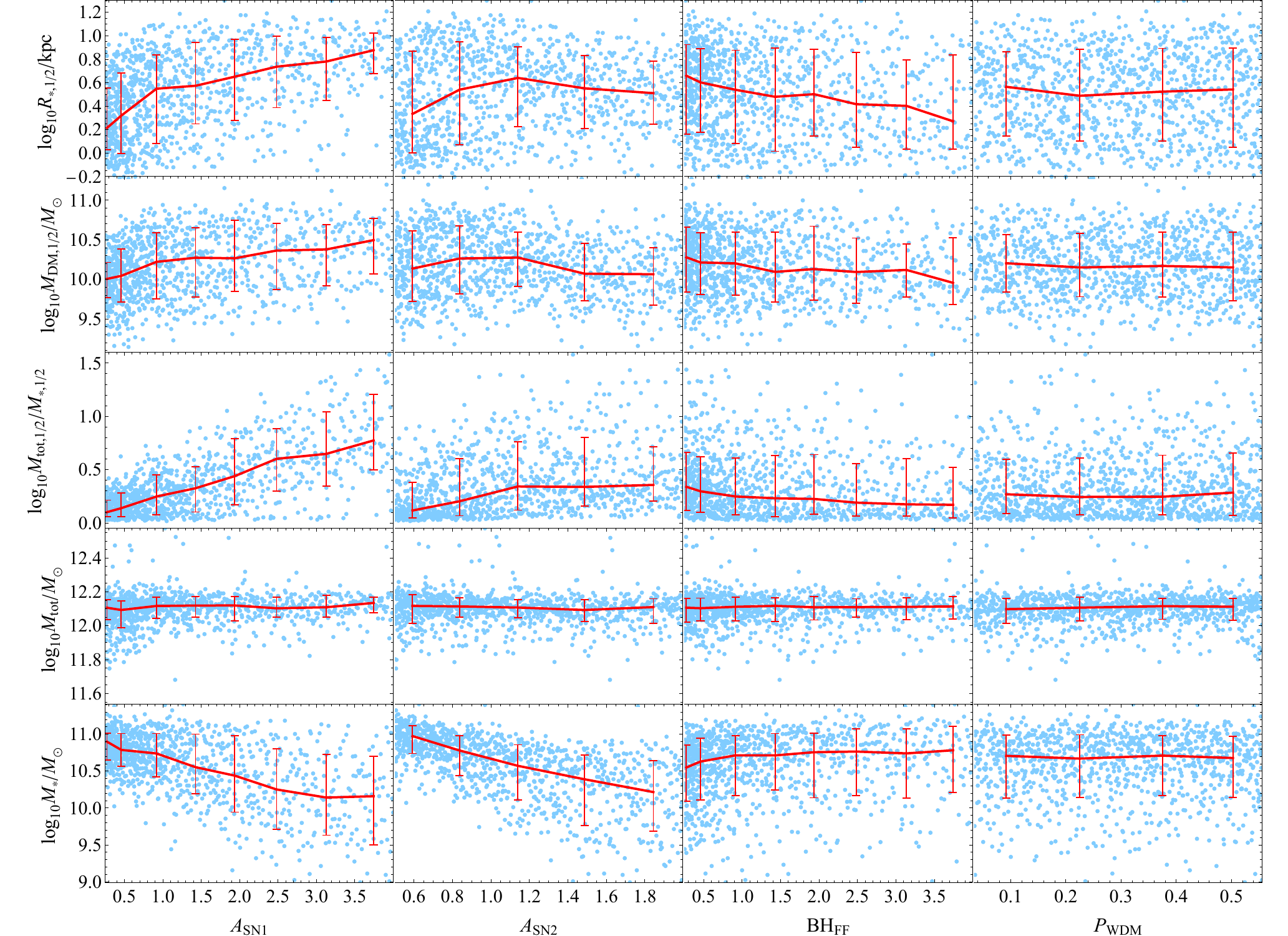}
        \caption{Trends of stellar half-mass radius, total-to-stellar mass ratio within the stellar half-mass radius, DM mass within the stellar half-mass radius, total galaxy mass and total stellar mass for the host galaxies in the MW zoom-in simulations, as a function of astrophysical parameters and the inverse of the WDM mass parameter. The light blue points represent individual galaxies; red lines indicate the median trends and their scatter.}   
  \label{fig:Host}
  \end{figure*}
  \begin{table*}[ht]
    \centering
    \small
      \caption{Maximum, minimum, and median values of astrophysical quantities for the host galaxies in the MW zoom-in simulations.}    
          \label{tab:hostprop}
    \begin{tabular}{cccccc}  
     \toprule
     \midrule
  
 &$\rm log_{10}R_{*,1/2}/kpc$  &$\rm log_{10}M_{\rm DM,1/2}/M_\odot$&$\rm log_{10} M_{\rm tot,1/2}/M_{*,1/2}$  &$\rm log_{10} M_{\rm tot}/M_\odot$&$\rm log_{10} M_*/M_\odot$\\
    \midrule
       Min &-0.22  &9.15  & 0.02  & 11.52 & 8.99  \\
     Max &1.30  & 11.20 &  1.58 &12.57  &  11.48 \\
    Median & 0.53 &10.17  & 0.26  & 12.11 & 10.69  \\
           \bottomrule
    \end{tabular}
\end{table*}
In this section, we summarize the main properties of the MW-like host galaxies from the MW zoom-in simulations of the DREAMS project. These simulations are constructed by selecting DM halos with masses comparable to that of the MW from a larger cosmological volume, which are then re-simulated at higher resolution using a zoom-in technique (see \citealt{rose2024introducingdreamsprojectdark}). As a result of this setup, each simulation contains a single, central galaxy—the host galaxy—that forms at the center of the high-resolution region. The host galaxies generally exhibit a contamination by low-resolution DM particles that remains below 2.5$\%$,  with a median value of 0$\%$. An analysis of the galaxy population in the MW zoom-in simulations shows that the level of contamination systematically increases with distance from the central galaxy, whereas the innermost regions, up to $\sim$ 1000 kpc (far beyond the virial radius), display a median contamination of 0$\%$ and only minimal scatter, increasing rapidly at larger distances. These tests confirm that the impact of low-resolution DM particle contamination is minimal for the host galaxies and the satellite populations considered in our analysis.

Table \ref{tab:hostprop} presents the minimum, maximum, and median values of the key astrophysical quantities analyzed throughout this work. Since each host galaxy evolves within a different realization of baryonic feedback parameters, its physical properties can vary 
significantly and often differ from those of the MW itself. 

In Fig. \ref{fig:Host}, we show the trends of $R_{*,1/2}$, $M_{\rm DM,1/2}$,  $M_{\rm tot,1/2}/M_{* ,1/2}$, $M_{\rm tot}$ and $M_{*}$ as functions of the astrophysical parameters and the inverse WDM particle mass. From Fig. \ref{fig:Host} we observe that increasing $A_{\rm SN1}$ leads to larger stellar half-mass radii, higher enclosed DM masses, and lower stellar masses, resulting in a steep increase in the ratio $M_{\rm tot,1/2} / M_{\rm *,1/2}$. The trends with $A_{\rm SN2}$ are non-monotonic: both $R_{\rm *,1/2}$ and $M_{\rm DM,1/2}$ rise mildly up to $A_{\rm SN2}$ $\sim$ 1.1 and decline thereafter, while the stellar mass decreases more steadily. This produces a moderate, but continuous, increase in $M_{\rm tot,1/2} / M_{\rm *,1/2}$. Higher $\rm BH_{\rm FF}$ values are associated with a slight increase in stellar mass, a decrease in $M_{\rm DM,1/2}$, and a mild reduction in $R_{\rm *,1/2}$, leading to a lower mass ratio. No significant variations are observed as a function of the inverse WDM particle mass across any of the plotted quantities. Finally, the total mass of the host galaxies does not appear to be significantly affected by any of the simulation parameters. Overall, the qualitative influence of astrophysical and WDM parameters on the host galaxies in the MW zoom-in simulations is consistent with the trends observed in Sect.~\ref{sec: MW parameter dep.} for the broader galaxy population. The only notable difference concerns the total galaxy mass, which shows no clear dependence on simulation parameters for the hosts alone, whereas a significant trend is observed when considering the full galaxy sample.

\section{Consistency and robustness tests
\label{Appendix: table recovery12}}

\begin{figure*}[ht]
    \centering
     \includegraphics[width=0.95
    \textwidth]{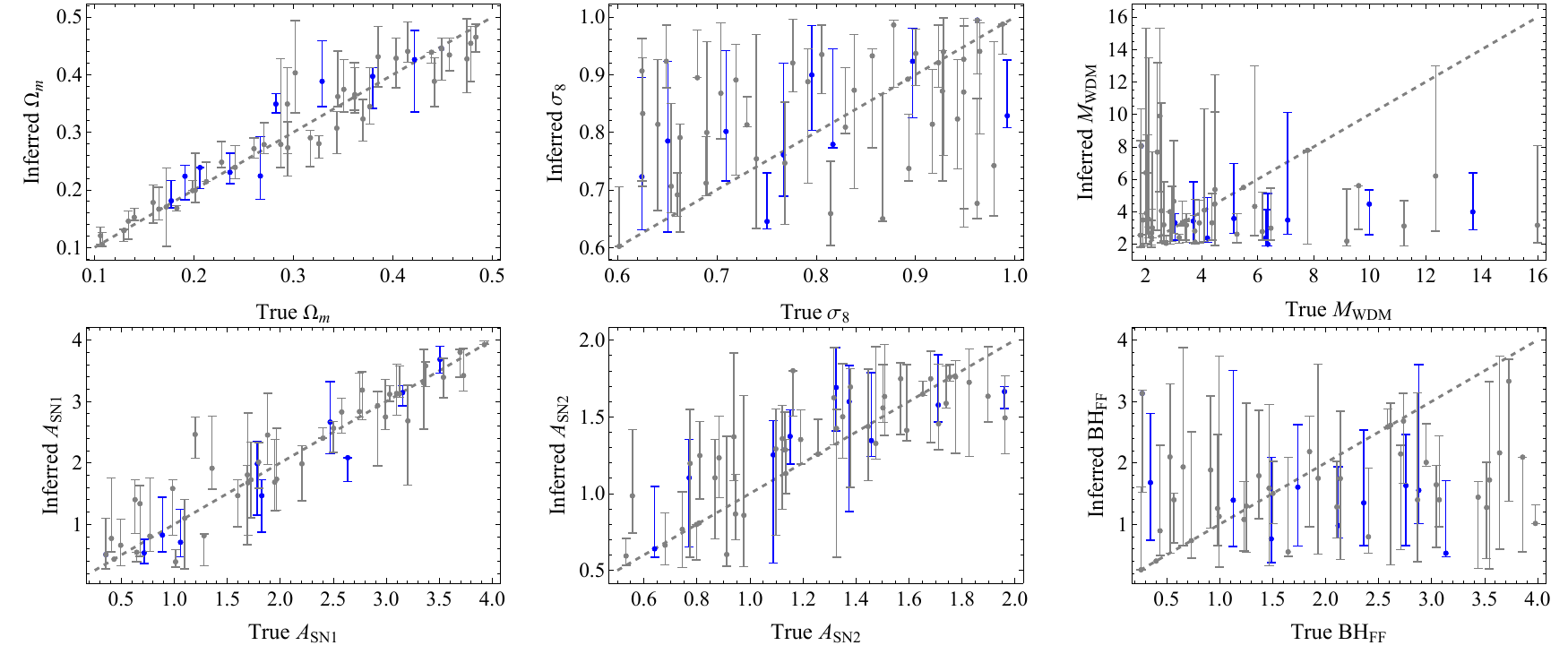}
        \caption{ Recovery plot for the uniform-box simulations, showing the bootstrap results with uncertainties for the estimated parameters:
$\Omega_{\rm m}$, $\sigma_{8}$, $M_{\rm WDM}$, $A_{\rm SN1}$, $A_{\rm SN2}$ and $\rm BH_{\rm FF}$ against their true values.
These parameters are obtained by comparing the simulations with the corresponding simulation outputs mimicking the SPARC catalog.
The dashed gray line indicates the ideal one-to-one relation between estimated and true parameter values. Blue points correspond to the reference analysis with $N=100$, while grey points show the test performed with $N=10$.}
  \label{fig:recovery10boot}
  \end{figure*}
In this section, we present a set of complementary validation tests aimed at assessing the overall reliability and robustness of our fitting procedure based on the bootstrap resampling method. We first examine the internal consistency of the recovered parameter correlations (\App\ref{Appendix: table recovery}), and then we perform dedicated tests to verify that the bootstrap approach does not introduce biases or underestimate uncertainties (\App\ref{Appendix: boot10}).
\subsection{Consistency tests on parameter correlations\label{Appendix: table recovery}}

In this section, we present the results of the correlation tests described in Sect. \ref{sec:evaluation fit}. Specifically, Table \ref{tab:recovery p ub} reports, for both the uniform-box and MW zoom-in samples, the outcomes of the Pearson correlation test along with the estimated slope $m$ and intercept $c$ for the parameters exhibiting statistically significant (i.e., $\alpha < 0.05$) correlations.

As discussed in the main text, we consider the fitting procedure described in Sect. \ref{sec:fit bootstrap} to be effective for estimating a given parameter when the correlation is significant, the slope $m$ is consistent with 1, and the intercept $c$ is consistent with 0.

%%%%%%%%%%%%%%%%%%%%%%%%%%%%%%%%%%%%%%%%%%%%%%%%%%%%%%%%%%%%%%%%%%%%%%%%%%%%%%%%%%%%%%%%%%%%%%%%%%%%%%%%%%%%%%%%%%%%%%%%%%%%%%%%%%%%%%%%%%%%%%%%%%%%%%%%%%%%%%%%%%%%%
\begin{table}[ht]
   \centering
   \small
        \caption{Pearson $r$, $p$-values, slopes ($m$), and intercepts ($c$) from weighted linear fits comparing bootstrap-inferred parameters to true values in uniform-box and MW zoom-in simulations.}   
    \begin{tabular}{p{0.8cm} p{0.8cm} p{1.8cm} p{1.4cm} p{2.15cm}}
        \toprule
         \midrule  
            \multicolumn{5}{c}{Uniform boxes} \\
       Param. & $r$ & $p$-value& $m$&$c$ \\ \midrule
             $\Omega_{\rm m}$ &0.95&2.2$\times 10^{-5}$  &1.01$\pm$0.14&0.03$\pm$0.04  \\
        $\sigma_8$  &0.68& 3.0$\times 10^{-2}$  &0.82$\pm$0.28&0.13$\pm$0.24 \\ 
        $\rm{A_{\rm SN1}}$ &0.97 &4.7$\times 10^{-6}$  &1.05$\pm$0.08&-0.24$\pm$0.20  \\
        $\rm{A_{\rm SN2}}$  &0.85 &$1.9\times 10^{-3}$ &0.53$\pm$0.09&0.67$\pm$0.15 \\ 
        $\rm{\rm BH_{\rm FF}}$  &-0.61 &5.9$\times 10^{-2}$    &-&-     \\ 
        $\rm{M_{\rm WDM}}$  &-0.13 &0.72   &-&-  \\
        \toprule
        \midrule
         \multicolumn{5}{c}{MW zoom-in} \\
       Param. & $r_{\rm SPARC}$ & $p_{\rm SPARC}$-value & $m\;$&$c$\\ \midrule
        $\rm{A_{\rm SN1}}$ &0.99   &8.5$\times 10^{-9}$   &0.81$\pm$0.03&0.16$\pm$0.03 \\
        $\rm{A_{\rm SN2}}$ &0.38 & 0.25&-&- \\ 
        $\rm{\rm BH_{\rm FF}}$ &0.04&0.91 &-&- \\ 
        $\rm{M_{\rm WDM}}$ &0.08&8.4$\times 10^{-3}$&0.03$\pm$0.02 &3.30$\pm$0.13  [keV]\\
           \midrule  
       Param. & $r_{\rm dwarf}$ & $p_{\rm dwarf}$-value& $m$&$c$ \\ \midrule
        $\rm{A_{\rm SN1}}$ & 0.90 &1.8$\times 10^{-4}$& 0.57$\pm$0.09&0.49$\pm$0.10 \\
        $\rm{A_{\rm SN2}}$  &-0.28  &0.41  &-&-\\ 
        $\rm{\rm BH_{\rm FF}}$ &0.30  &0.38 &-& - \\
        $\rm{M_{\rm WDM}}$ & 0.02  & 0.38  &-&-   \\
        \midrule
    Param. & $r_{\rm dwarf,alt.}$ & $p_{\rm dwarf,alt.}$-value&  $m$&$c$ \\ \midrule
        $\rm{A_{\rm SN1}}$ &0.86&7.1$\times 10^{-4}$ & $0.55\pm0.09$&0.65 $\pm$0.13\\
    $\rm{A_{\rm SN2}}$  &-0.29   &0.39  &-&- \\ 
       $\rm{\rm BH_{\rm FF}}$ &-0.21 &0.54  &-&-  \\ 
        $\rm{M_{\rm WDM}}$ & -0.51 & 0.11 &-&-   \\
       \bottomrule
    \end{tabular}

    \label{tab:recovery p ub}
\end{table}

%%%%%%%%%%%%%%%%%%%%%%%%%%%%%%%%%%%%%%%%%%%%%%%%%%%%%%%%%%%%%%%%%%%%%%%%%%%%%%%%%%%%%%%%%%%%%%%%%%%%%%%%%%%%%%%%%%%%%%%%%%%%%%%%%%%%%%%%%%%%%%%%%%%%%%%%%%%%%%%%%%%%%
\subsection{Bootstrap resampling robustness\label{Appendix: boot10}}
To further validate the reliability of the bootstrap–based fitting method described in Sect. \ref{sec:fit bootstrap}, we performed additional tests aimed at verifying that the resampling procedure does not bias parameter recovery or underestimate the associated uncertainties, particularly in the presence of parameter degeneracies. These tests are carried out using the uniform-box simulation suite, and for computational reasons we combine bootstrap estimates obtained with $N=100$ (used in Sect. \ref{subsec:rec unif}) and with $N=10$, which allows us to extend the statistical sampling while keeping the analysis tractable. Although a smaller $N$ increases the scatter in individual realizations, it provides a useful stress test for the stability and statistical validity of the method.

Figure \ref{fig:recovery10boot} presents the parameter–recovery plots, similarly to Fig. \ref{fig: recovery UB} but including the $N$=10 points in gray. The parameters $\Omega_m$ and $A_{\rm SN1}$ closely follow the one–to–one relation, indicating accurate recovery and no systematic bias. $A_{\rm SN2}$ also follows the general trend but with larger dispersion, while $\sigma_8$ shows a noisier behavior. The WDM mass parameter and the AGN feedback efficiency are not well recovered, as expected given the limited sensitivity of the adopted observables to these quantities. We notice a mild systematic trend across several parameters: low true values tend to be slightly overestimated, whereas high true values are mildly underestimated.
This pattern is consistent with a boundary bias, which occurs when the true parameter lies near the edges of a finite grid, causing the recovered distribution to be compressed toward the center of the sampled range (see, e.g.,  \citealt{Anthony1997,Silverman}).
This behavior reflects the finite coverage of the uniform-box simulations rather than a bias intrinsic to the bootstrap resampling.

\begin{figure*}[ht]
    \centering
     \includegraphics[width=0.95
    \textwidth]{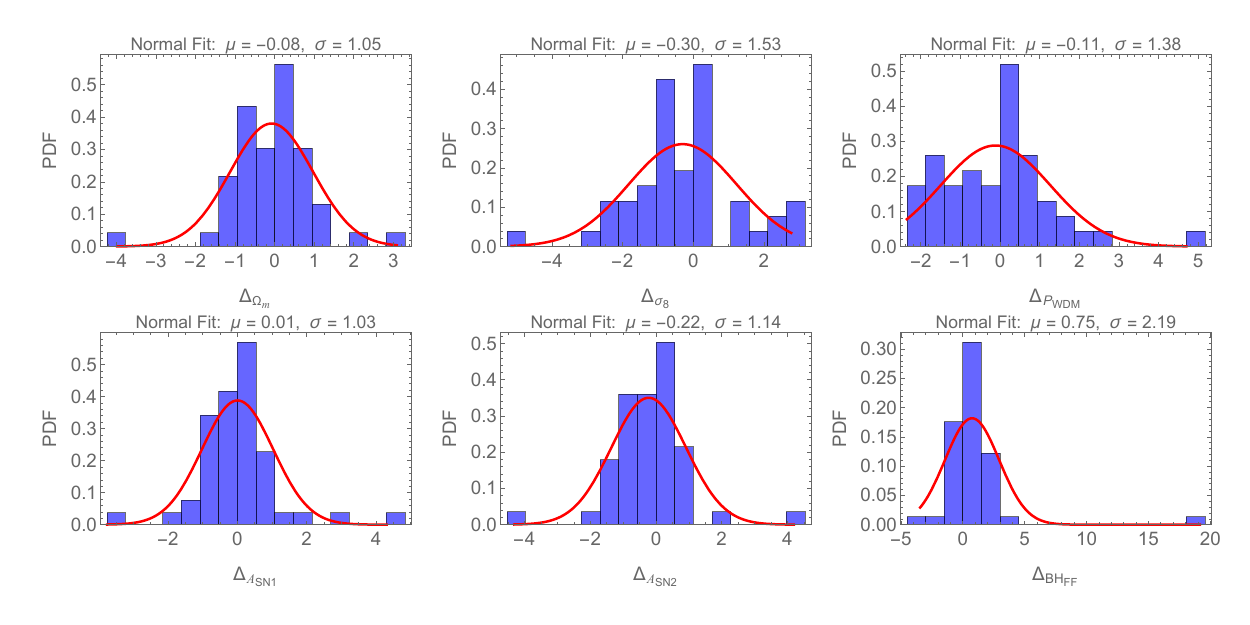}
        \caption{Normalized residual distributions ($\Delta_p$) for all parameters in the uniform-box suite.
Each panel shows the empirical distribution (histogram) together with a Gaussian fit (red solid curve); the fitted mean ($\mu$) and standard deviation ($\sigma$) are indicated in each panel. Histograms are normalized to unit area using Mathematica’s "PDF" option, allowing direct comparison with a standard normal.}
  \label{fig:Dist10boot}
  \end{figure*}

The normalized residual distributions, $\Delta_p = (p_{\mathrm{true}} - p_{\mathrm{rec}})/\sigma_p$, where $p_{\mathrm{true}}$ denotes the value of the parameter set in the simulation, $p_{\mathrm{rec}}$ is the recovered estimate, and $\sigma_p = (\sigma_+ + \sigma_-)/2$ is the mean bootstrap uncertainty from the 16th and 84th percentiles, are shown in Fig.~\ref{fig:Dist10boot} together with their Gaussian fits; the fitted mean ($\mu$) and standard deviation ($\sigma$) are indicated in each panel.
The parameters $\Omega_{\mathrm{m}}$ and $A_{\mathrm{SN1}}$ exhibit nearly Gaussian residuals ($\mu\simeq0$, $\sigma\simeq1$), confirming unbiased recovery and realistic uncertainty estimates. $A_{\mathrm{SN2}}$ and $\sigma_8$ display broader distributions ($\sigma>1$), consistent with the larger intrinsic scatter already visible in their recovery plots. In contrast, $M_{\mathrm{WDM}}$ and $\mathrm{BH_{FF}}$ show clearly non-Gaussian and asymmetric residuals ($\sigma\gg1$, $\mu\neq0$), reflecting the limited constraining power of the available observables rather than any bias introduced by the bootstrap procedure. Overall, these results demonstrate that the bootstrap resampling method yields statistically robust and unbiased estimates for parameters effectively constrained by the data, while correctly propagating the increased uncertainty for less constrained ones.

   \section{Fit results and parameter constraints}
    \begin{table*}[ht]
    \centering
    \small
      \caption{D-squared ($D^2$) and reduced D-squared ($\tilde{D}^2$) values for the best-fit simulations compared to the SPARC and dwarf galaxy catalogs, evaluated for the indicated scaling relations and cumulatively. }
    
    \begin{tabular}{ccccccc}  
     \toprule
     \midrule
  
  Relations  &$D^2_{\rm SPARC}$  &$\bar D^2_{\rm SPARC}$&$D^2_{\rm SPARC}$  &$\bar D^2_{\rm SPARC}$&$D^2_{\rm dwarf}$&$\bar D^2_{\rm dwarf}$\\
     &Uniform-box&Uniform-box & MW zoom-in &MW zoom-in &MW zoom-in&MW zoom-in\\
    \midrule
       $R_{*,1/2}-M_*$ &49.89&0.24  & 0.07  &  0.04&0.22 &  0.06  \\
       $f_{\rm DM,1/2}-M_*$&10.99&0.08&0.01  &0.01 &0.02 &0.02 \\ 
       $M_{\rm DM,1/2}-M_*$  & 24.42&0.11&0.01  & 0.01 & 0.33  &  0.33  \\ 
       $M_{\rm tot}-M_*$  &72.36&0.35& 0.05  &0.02   &  - & - \\
       Cumul.&159.2&1.09& 7.63  & 1.27 & 0.96   &0.96  \\

           \bottomrule
    \end{tabular}
  
    \label{tab:chi2SPARCdwarf}
\end{table*}
\begin{table*}[ht]
    \centering
    \small
\caption{Constraints on astrophysical parameters and WDM mass from MW zoom-in simulations.}
       \label{tab:bootstrapSPARC}
    \begin{tabular}{cccccc|cccc}   
        \toprule
        \midrule  
        & \multicolumn{5}{c}{SPARC} & \multicolumn{4}{c}{dwarf}\\
       Param. & $R_{*,1/2}-M_*$ & $f_{\rm DM,1/2}-M_*$ &$M_{\rm DM,1/2}-M_*$ &$M_{\rm tot}-M_*$&Cumul.& $R_{*,1/2}-M_*$ & $f_{\rm DM,1/2}-M_*$&$M_{\rm DM,1/2}-M_*$ &Cumul.
       \\ \midrule
        $\rm{A_{\rm SN1}}$  &$1.89^{+1.19}_{-0.91}$
        &$1.89^{+1.32}_{-1.03}$ &$1.45^{+1.49}_{-0.52}$ & $1.91^{+1.23}_{-0.91}$  &$1.31^{+0.83}_{-0.52}$  &$0.93_{-0.42}^{+1.23} $  &$1.72_{-0.98}^{+1.15} $   &$0.70_{-0.36}^{+0.40} $  &$0.68_{-0.34}^{+0.50} $  
        \\
        \\
        $\rm{A_{\rm SN2}}$  &$1.06^{+0.64}_{-0.38}$ &$0.87^{+0.67}_{-0.29}$  &  $1.03^{+0.65}_{-0.35}$  &$0.98^{+0.71}_{-0.39}$&$0.89^{+0.67}_{-0.32}$&$0.82_{-0.21}^{+0.72} $  &$1.02_{-0.39}^{+0.58} $   &$0.92_{-0.32}^{+0.80} $  &$0.91_{-0.34}^{+0.77} $ 
        \\ 
        \\
        $\rm{\rm BH_{\rm FF}}$  &$0.83^{+2.43}_{-0.47}$  &$0.65^{+1.69}_{-0.29}$ &$0.74^{+2.08}_{-0.38}$    &  $0.85^{+1.49}_{-0.49}$ &   $1.08^{+1.23}_{-0.72}$ &$0.93_{-0.48}^{+1.08} $  &$1.12_{-0.71}^{+1.39} $   &$0.70_{-0.33}^{+1.51} $  &$0.78_{-0.46}^{+1.14} $
        \\ 
        \\
         $\rm{P_{\rm WDM}}$  &$0.32^{+0.13}_{-0.25}$&$0.31^{+0.13}_{-0.24}$&$0.31^{+0.16}_{-0.19}$ &$0.34^{+0.12}_{-0.27}$ &$0.31^{+0.18}_{-0.20}$ &$0.34_{-0.23}^{+0.11} $  &$0.36_{-0.21}^{+0.15} $   &$0.39_{-0.20}^{+0.12} $  &$0.37_{-0.17}^{+0.10} $  
        \\
        \\
        $\rm{M_{\rm WDM}}$  &$3.08^{+10.36}_{-0.89}$&$3.20^{+10.24}_{-0.93}$&$3.23^{+5.24}_{-1.12}$ &$2.91^{+10.53}_{-0.76}$ &$3.56^{+6.08}_{-1.79}$ &$2.97_{-0.73}^{+6.14} $  &$2.79_{-0.83}^{+4.04} $   &$2.55_{-0.59}^{+2.74} $  &$2.73_{-0.65}^{+2.39} $  
        \\
        \\
        $D^2$        &  $0.05^{+0.07}_{-0.4}$ & $0.06^{+0.14}_{-0.04}$  &$0.06^{+0.18}_{-0.04}$ & $0.073^{+0.13}_{-0.06}$&$7.22^{+7.13}_{-4.58}$  & $0.10_{-0.08}^{+0.24} $  &$0.07_{-0.05}^{+0.20} $   &$0.76_{-0.62}^{+2.20} $  &$7.04_{-4.06}^{+10.52} $
        \\
        \\
        $\tilde{D}^2$& $0.04^{+0.05}_{-0.03}$  &$0.05^{+0.06}_{-0.03}$&$0.04^{+0.05}_{-0.03}$ &  $0.05^{+0.06}_{-0.03}$  &  $2.71^{+0.63}_{-0.70}$&$0.07_{-0.05}^{+0.10} $  &$0.05_{-0.03}^{+0.09} $   &$0.36_{-0.28}^{+0.32} $  &$2.82_{-0.92}^{+1.69} $ 
        \\
        \bottomrule
    \end{tabular}
  \tablefoot{ Values correspond to the 16th, 50th, and 84th percentiles from bootstrap fits (Sect.~\ref{sec:fit bootstrap}). Constraints are derived using the SPARC galaxy catalog and the dwarf galaxy catalog (LVDB).}
  \end{table*}

In this section, we present the results of the fits obtained through bootstrap resampling, as discussed in Sect. \ref{sec:bootstrap cataloghi}.
\begin{table*}[ht]
    \centering
    \small
      \caption{Constraints on astrophysical parameters and WDM mass from calibrated uniform-box simulations.}
        \label{tab:bootstrapSPARCub}
     \begin{tabular}{cccccc}  
        \toprule
         \midrule  
       Param. & $R_{*,1/2}-M_*$ & $f_{\rm DM,1/2}-M_*$ & $M_{\rm DM,1/2}-M_*$ & $M_{\rm tot}-M_*$ & Cumul. \\ \midrule
        $\Omega_{\rm m}$  & $0.32^{+0.08}_{-0.07}$  & $0.19 ^{+0.07}_{-0.07}$ & $ 0.21^{+0.05}_{-0.06}$  &  $ 0.36^{+0.10}_{-0.25}$  &  $ 0.28^{+0.08}_{-0.04}$ \\
        \\
        $\sigma_8$  & $ 0.88^{+0.07}_{-0.20}$  & $ 0.75^{+0.19}_{-0.10}$ & $ 0.76^{+0.18}_{-0.12}$  &  $ 0.95^{+0.04}_{-0.26}$  &  $ 0.89^{+0.10}_{-0.18}$ \\
        \\
        $S_8$  & $ 0.89^{+0.16}_{-0.18}$  & $ 0.62^{+0.18}_{-0.18}$ & $ 0.61^{+0.17}_{-0.09}$  &  $ 1.01^{+0.18}_{-0.58}$  &  $ 0.85^{+0.09}_{-0.12}$ \\
        \\
        $\rm{A_{\rm SN1}}$  & $2.77 ^{+0.55}_{-1.05}$  & $ 3.59^{+0.25}_{-0.83}$ & $ 2.76^{+0.70}_{-1.61}$  &  $ 1.75^{+1.81}_{-1.05}$  &  $ 1.67^{+0.49}_{-0.84}$ \\
        \\
        $\rm{A_{\rm SN2}}$  & $ 1.75^{+0.16}_{-0.30}$  & $ 1.73^{+0.20}_{-0.60}$ & $ 1.72^{+0.23}_{-0.38}$  &  $ 1.38^{+0.48}_{-0.72}$  &  $ 1.46^{+0.32}_{-0.21}$ \\
        \\
        $\rm{\rm BH_{\rm FF}}$  & $ 1.73^{+1.13}_{-0.96}$  & $ 1.74^{+1.38}_{-1.12}$ & $ 1.70^{+1.19}_{-1.17}$  &  $ 2.01^{+1.17}_{-1.11}$  &  $ 1.40^{+0.59}_{-0.66}$ \\
        \\
           $\rm{P_{\rm WDM}}$  & $ 0.36^{+0.12}_{-0.19}$  &  $0.29 ^{+0.14}_{-0.17}$ &  $ 0.30^{+0.15}_{-0.17}$  &   $ 0.26^{+0.18}_{-0.18}$  &   $ 0.37^{+0.10}_{-0.13}$ \\
        \\
        $\rm{M_{\rm WDM}}$  & $ 2.75^{+3.10}_{-0.68}$  & $ 35^{+47}_{-22}$ & $ 3.33^{+4.61}_{-1.09}$  &  $ 3.87^{+8.53}_{-1.61}$  &  $2.7^{+1.3}_{-0.6}$ \\
        \\
        $D^2$        &  $ 76^{+97}_{-29}$  & $ 0.75^{+0.10}_{-0.14}$ & $ 70^{+98}_{-43}$  &  $ 169^{+73}_{-137}$  &  $844 ^{+603}_{-175}$ \\
        \\
        $\tilde{D}^2$& $ 0.96^{+0.10}_{-0.12}$ & $ 0.75^{+0.10}_{-0.14}$   & $ 0.88^{+0.08}_{-0.11}$ &  $ 1.14^{+0.13}_{-0.24}$  &  $ 5.08^{+0.32}_{-0.36}$ \\
       \bottomrule
    \end{tabular}
  \tablefoot{Values correspond to the 16th, 50th, and 84th percentiles from bootstrap fits (Sect.~\ref{sec:fit bootstrap}). Cumulative constraints are shown in the last column.}  
\end{table*}

\begin{table*}[ht]
    \centering
    \small
\caption{Constraints on astrophysical parameters and WDM mass from uncalibrated uniform-box simulations.}
 \label{tab:bootstrapSPARCuncub}
     \begin{tabular}{cccccc}  
        \toprule
         \midrule  
      Param. & $R_{*,1/2}-M_*$ & $f_{\rm DM,1/2}-M_*$ & $M_{\rm DM,1/2}-M_*$ & $M_{\rm tot}-M_*$ & Cumul.\\ \midrule
               $\Omega_{\rm m}$  & $0.29^{+0.11}_{-0.07}$  &$0.19^{+0.06}_{-0.09}$ &$0.20^{+0.04}_{-0.03}$  & $0.41^{+0.05}_{-0.21}$  & $0.26^{+0.04}_{-0.04}$ \\
        \\
        $\sigma_8$   & $0.89^{+0.15}_{-0.07}$  &$0.73^{+0.16}_{-0.09}$ &$0.83^{+0.11}_{-0.16}$  & $0.93^{+0.06}_{-0.13}$  & $0.87^{+0.11}_{-0.16}$ \\
        \\
        $S_8$  & $0.83^{+0.23}_{-0.12}$  & $0.60^{+0.19}_{-0.19}$  &$0.70^{+0.06}_{-0.06}$ &$1.05^{+0.19}_{-0.30}$  & $0.80^{+0.11}_{-0.12}$   \\
        \\
        $\rm{A_{\rm SN1}}$   & $2.59^{+0.63}_{-0.25}$  &$3.59^{+0.29}_{-0.94}$ &$1.96^{+0.81}_{-1.21}$  & $0.72^{+1.33}_{-0.45}$  & $1.04^{+0.78}_{-0.56}$ \\
        \\
        $\rm{A_{\rm SN2}}$   & $1.76^{+0.18}_{-0.25}$  &$1.77^{+0.19}_{-0.53}$ &$1.72^{+0.19}_{-0.52}$  & $1.26^{+0.29}_{-0.61}$  & $1.27^{+0.47}_{-0.48}$ \\
        \\
        $\rm{\rm BH_{\rm FF}}$  & $1.71^{+0.18}_{-0.96}$  &$1.51^{+1.44}_{-0.90}$ &$1.83^{+1.12}_{-1.20}$  & $2.10^{+1.09}_{-1.46}$  & $1.87^{+1.38}_{-0.95}$ \\
        \\
        $\rm{M_{\rm WDM}}$  & $3.20^{+5.74}_{-1.00}$  &$3.72^{+7.31}_{-3.61}$ &$3.14^{+3.26}_{-1.06}$  & $3.78^{+6.55}_{-1.57}$  & $3.54^{+4.49}_{-1
        46}$ \\
        \\
        $D^2$        & $87^{+90}_{-43}$  &$37^{+57}_{-31}$ &$99^{+133}_{-41}$  & $219^{+141}_{-86}$  & $1251^{+474}_{-557}$ \\
        \\
        $\tilde{D}^2$ & $0.95^{+0.13}_{-0.13}$  &$0.78^{+0.10}_{-0.17}$ &$0.91^{+0.07}_{-0.08}$  & $1.12^{+0.10}_{-0.10}$  & $5.08^{+0.28}_{-0.34}$ \\
       \bottomrule
    \end{tabular}  
   \tablefoot{ Values correspond to the 16th, 50th, and 84th percentiles from bootstrap fits (Sect.~\ref{sec:fit bootstrap}). Cumulative constraints are shown in the last column.}
\end{table*}

%%%%%%%%%%%%%%%%%%%%%%%%%%%%%%%%%%%%%%%%%%%%%%%%
%%%%%%%%%%%%%%%%%%%%%%%%%%%%%%%%%%%%%%%%%%%%%%%%

\end{appendix}

\end{document}